%% file: letter.tex
\documentclass[%
 twocolumn,
 superscriptaddress,
 preprintnumbers,
 nofootinbib,
 amsmath,amssymb,
 aps,
 prl,
]{revtex4-2}
\usepackage{psfrag,slashed,cancel,array,graphicx,hyperref}
\usepackage{subfigure}
\usepackage[utf8]{inputenc}
\usepackage{mathtools}
\usepackage{mathrsfs}
\usepackage{multirow}
\usepackage{braket}
\usepackage{booktabs} 
\usepackage[normalem]{ulem}
\hypersetup{pdftitle={},pdfcreator={},linkcolor=[rgb]{0.15,0.35,0.75},colorlinks=true,citecolor=[rgb]{0.675,0,0.2},urlcolor=[rgb]{0.15,0.35,0.65}}

\input{numbers.tex}

\newcommand{\G}{\mathcal G}
\newcommand{\M}{\mathcal M}

\newcommand{\bb}{\boldsymbol b}
\newcommand{\DD}{\Delta_C}
\newcommand{\DG}{D_\gamma}
\newcommand{\Kt}{\widetilde{\mathcal K}}

\begin{document}

\title{Double parton scattering singularities in two-loop gluon-fusion amplitudes}

\author{Arunima Bhattacharya\,\href{https://orcid.org/0000-0003-4409-0901}{\includegraphics[scale=0.05]{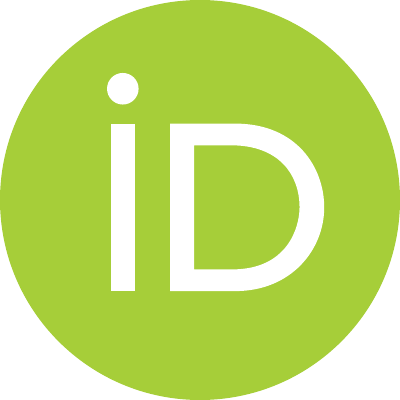}}}
\affiliation{Theory Division, IFIC, University of Valencia-CSIC,\\
E-46980 Paterna, Valencia, Spain}
%
\email{arunimab@ific.uv.es}

\begin{abstract}
Gluon fusion to two massive colour singlets can develop a double parton scattering (DPS) singularity as $p_T\to0$: each gluon splits, and its daughters fuse pairwise. We determine two-loop logarithms using a common kernel and derive finite matching functions analytically for gluon and quark splittings. Normalised to one loop, the cubic coefficient changes sign between channels. Complete $gg\to HH$, $gg\to AA$ and $gg\to ZZ$ amplitudes confirm the predicted colour, helicity and parity dependence and phase. The kernel gives leading DPS logarithms for other pairs with this DPS configuration.
\end{abstract}

\maketitle

\begin{figure*}[t]
\centering
\includegraphics[width=0.72\textwidth]{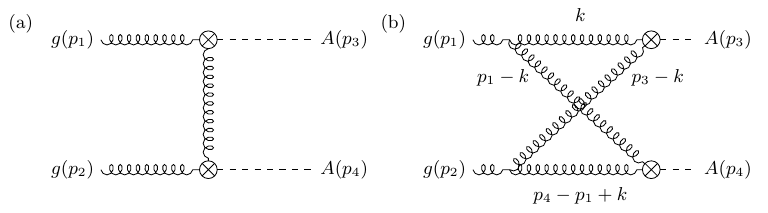}
\caption{(a) Tree diagram for $gg\to AA$ with effective vertices
(crossed circles). (b) Crossed box at the DPS pinch, where the four
internal partons are on shell and $k=xp_1$, $p_3-k=(1-x)p_2$.
For $HH$ pairs replace $A$ by $H$; for $ZZ$ pairs the loop contains
massless quarks.}
\label{fig:dps}
\end{figure*}

\textit{Introduction:} In gluon-fusion amplitudes with massless internal
partons, production of two massive colour singlets can develop a double
parton scattering (DPS) singularity as the transverse momentum $p_T$
vanishes~\cite{Gaunt:2011xd,Diehl:2011yj}: each incoming gluon splits
into two nearly on-shell partons, which fuse pairwise in two hard
subprocesses [Fig.~\ref{fig:dps}(b)]. This occurs in a loop amplitude and has the kinematic structure of double parton scattering. Such amplitudes arise, for example, in calculations of Higgs and electroweak boson pairs at higher orders in QCD~\cite{Glover:1988rg,Banerjee:2018lfq,vonManteuffel:2015msa}.
At two loops, the colour factor $\DD$ controls
the cubic DPS logarithm, whose coefficient
changes sign between gluon and quark splittings.
The same colour factor determines the phase
in Eq.~(\ref{eq:gammaK}), tested against
complete amplitudes below.
These analytic coefficients provide benchmarks
at small $p_T$, where direct numerical evaluation
requires high precision.
These splitting configurations also overlap with single hard scattering
and require a consistent treatment to avoid double
counting~\cite{Gaunt:2011xd,Manohar:2012pe,Gaunt:2012dd,Diehl:2017kgu}.
Perturbative parton splitting and parton
correlations in double hard scattering,
including four-jet production, have been
studied in Refs.~\cite{BDFS2011,BDFS2012,BDFS2014}.
At one loop, the DPS singularity is at most logarithmic and is described
by two splitting amplitudes and two hard subprocesses~\cite{Gaunt:2011xd}.
At two loops, the dependence of this singularity on the daughter partons,
their helicities and the hard vertices is not fixed by the one-loop analysis.

We determine the leading two-loop DPS logarithms and finite matching
functions for gluon and quark splittings at fixed momentum fractions
away from the endpoints. The coefficients agree with the two-loop
$gg\to HH$, $gg\to AA$ and $gg\to ZZ$ amplitudes across the tested
colour and helicity channels. The scalar and pseudoscalar channels use
effective gluon vertices, while the vector channel uses massless
quark loops.
We organise these results in a common tensor kernel contracted with
the two hard subprocess amplitudes. We derive the finite matching
functions from the collinear and soft contributions and parton
self-energy corrections, including their helicity and parity dependence.
The derivation requires only the one-loop corrections to these
subprocess amplitudes and to the kernel, with no analytic expansion
of a two-loop amplitude at small $p_T$.
Complete two-loop amplitudes test the predicted coefficients;
numerical integration separately checks the region integrals.

Our earlier $gg\to AA$ threshold analysis identified
the gluon-channel cubic and quadratic DPS logarithms
and the scalar and pseudoscalar helicity weights~\cite{P1}.
Here we extend the logarithmic structure to quark
splittings and derive the finite matching functions
analytically for both splitting channels. Two-loop splitting kernels have also been calculated for double parton distributions~\cite{Diehl:2019rdh,Diehl:2021non,Diehl:2026quant}.
These describe parton correlations in DPS cross
sections, whereas the present calculation concerns
singularities of scattering amplitudes.
The Glauber region produces a phase in the infrared (IR) subtracted amplitude in the prescription used here. This is consistent with the cancellation of Glauber contributions in the inclusive double Drell--Yan cross section~\cite{Diehl:2015bca}.
For $gg\to ZZ$, the predicted logarithmic coefficients
agree with complete two-loop amplitudes to $\zzDevMax$
in the normalisations of Table~\ref{tab:tests}, providing
analytic benchmarks at small $p_T$~\cite{Caola:2015zz}.
The Supplemental Material~\cite{SM} presents the analytic derivation,
subtraction conventions and numerical validation.

\textit{Two-loop DPS structure:} Let $\bb$ be the transverse position conjugate to $\boldsymbol p_T$,
$b=|\bb|$. For a given helicity projection and at fixed momentum fractions, the DPS term of the
IR-subtracted amplitude, the part not analytic in $\boldsymbol p_T$ at $\boldsymbol p_T=0$,
is represented as
\begin{equation}
\widetilde{\mathcal A}_{\rm DPS}(b)=\mathcal H_1\mathcal H_2\,\Kt(b),
\label{eq:factorised}
\end{equation}
with $\mathcal H_{1,2}$ the IR-subtracted amplitudes of the two $2\to1$ subprocesses, of
masses $m_1$ and $m_2$, and $\Kt$ a kernel that starts at order $a_s=\alpha_s/(4\pi)$ with one
splitting of each incoming parton. The product denotes contraction over the daughter parton
colour and helicity indices. At one loop this is the factorisation of
Ref.~\cite{Gaunt:2011xd}. At two loops we expand about
the DPS pinch, retaining the leading power in
$\lambda=p_T/\sqrt{s}$ at fixed momentum fractions
away from the endpoints. Hard corrections confined
to either fusion subprocess enter $\mathcal H_i$.
The factorised description assumes the cancellation
of leading hard exchanges between the subprocesses
and the eikonal reduction of soft attachments.
We calculate the collinear, soft, Glauber and ultrasoft
contributions to the kernel and test the resulting
description against complete two-loop amplitudes
in both splitting channels.

The one-loop anomalous dimensions of IR-subtracted amplitudes are determined by the
external partons~\cite{Becher:2009cu}; subtracting those of $\mathcal H_1$ and $\mathcal H_2$ from
that of the $2\to2$ amplitude gives $d\Kt/d\ln\mu=\gamma_{\mathcal K}\Kt$ with
\begin{equation}
\gamma_{\mathcal K}=a_s\Big[-4\DD\ln\frac{Q_*^2}{\mu^2}+4i\pi\DD+2\DG\Big],
\label{eq:gammaK}
\end{equation}
$\DD=C_{R_1}+C_{R_2}-C_a$ and $\DG=\gamma_a-\gamma_{R_1}-\gamma_{R_2}$, where $C_a$ is the Casimir
of the incoming partons, $C_{R_i}$ that of the two partons entering subprocess $i$,
$\gamma_g=-\beta_0$ and $\gamma_q=-3C_F$ are the one-loop non-cusp anomalous dimensions,
$\beta_0=\frac{11}3C_A-\frac43T_Fn_f$.
Colour conservation at the singlet hard vertices gives
$2T_k\cdot T_{\bar r}=\DD$ for exchanges between the
hard vertices, relating this colour factor to
crossed Wilson lines~\cite{Korchemskaya:1994qp}.
The scale $Q_*$ is fixed by
the two outgoing masses and the incoming energy:
$\DD\ln Q_*^2=C_{R_1}\ln m_1^2+C_{R_2}\ln m_2^2-C_a\ln s$.
This combines the mass and energy logarithms
with the corresponding colour factors.
Solving the renormalisation group equation with the
anomalous dimension in Eq.~(\ref{eq:gammaK}) from
$\mu_b=b_0/b$ to $\mu$, where $b_0=2e^{-\gamma_E}$,
and evaluating the coupling of the lowest-order kernel
at $\mu_b$, gives
\begin{equation}
\Kt=a_s\Kt_0\big\{1+a_s\big[\DD X_\mu^2-2\DD XX_\mu+BX_\mu\big]\big\}+a_s^2\widetilde R,
\label{eq:kernel}
\end{equation}
Here $X=\ln(Q_*^2e^{-i\pi}b^2/b_0^2)$,
$X_\mu=\ln(\mu^2b^2/b_0^2)$ and $B=\beta_0+\DG$,
up to $O(a_s^3)$. The renormalisation group determines the scale
dependence of the kernel, while the $\mu$-independent remainder
$\widetilde R$ contains matching information at $\mu_b$.
We parametrise its two-loop boundary as \(\widetilde R=\widetilde{\mathcal K}_0(r_0+r_1X+r_2X^2)\), where \(r_0\) is the finite matching coefficient and \(r_1,r_2\) represent additional logarithms not fixed by renormalisation group evolution.
At $\mu^2=Q_*^2e^{-i\pi}$, the RG terms
in Eq.~(\ref{eq:kernel}) reduce to
$-\Delta_C X^2+BX$.

The lowest-order kernel is proportional to $e^{i\nu\phi_b}/b^2$.
We denote the $\nu=0$ and $|\nu|=2$ projections by $\M_1$ and $\M_2$;
for scalar and pseudoscalar pairs these are the equal- and opposite-helicity
amplitudes. For spin-one pairs, $\nu=\Delta J_z$ also includes the outgoing
helicities~\cite{SM}. In momentum space, we define \(\mathcal G=\ln(p_T/Q_*)+i\pi/2\). At one loop, the DPS terms are \(C_1\mathcal G\) in \(\mathcal M_1\) and \(K\) in \(\mathcal M_2\). The difference arises from the Fourier transforms: the isotropic (\(\nu=0\)) integral is logarithmically divergent at small \(b\), whereas the \(|\nu|=2\) integral converges (\hyperref[sec:endmatter]{End Matter}).
The region analysis supports $r_1=r_2=0$ for the DPS
contributions studied. Independent two-loop amplitude fits
find both coefficients consistent with zero within numerical
uncertainties estimated from fit-window variations~\cite{SM}.
With this boundary condition, the terms $-\DD X^2+BX$ give the two-loop logarithms
\begin{equation}
\begin{aligned}
\M_1^{(2)}&\supset C_1\Big[-\tfrac43\DD\,\G^3-B\,\G^2\Big],\\
\M_2^{(2)}&\supset K\Big[-4\DD\,\G^2+(4\DD-2B)\,\G\Big].
\end{aligned}
\label{eq:rg}
\end{equation}
For $g\to gg$, $\DD=C_A$ and $B=2\beta_0$.
The two contributions to $B$ arise from the running
coupling of the lowest-order kernel and the non-cusp
anomalous dimensions. The four coefficients in
Eq.~(\ref{eq:rg}) are $-\frac43C_A$, $-2\beta_0$,
$-4C_A$ and $4C_A-4\beta_0$, respectively, with
$Q_*^2=m^4/s$ for equal masses.
For $g\to q\bar q$, $\DD=2C_F-C_A=-1/N_c$ and
$B=6C_F$. The cubic coefficient in $\M_1^{(2)}/C_1$
is $\frac4{3N_c}$, compared with $-\frac43C_A$
for $g\to gg$, reflecting the colour factors in
$\DD$; $\beta_0$ does not enter the quadratic
coefficient, and
$Q_*^2=s^{N_c^2}/m^{2(N_c^2-1)}$.
For $g\to q\bar q$, the Fourier transform is
performed order by order in $a_s$, since the
unexpanded large-$b$ exponent does not yield
a convergent real-$b$ Fourier integral.

The one-loop coefficients follow from the residues of the pinch~\cite{Gaunt:2011xd}.
For two effective vertices and equal final-state masses $m$, the daughter partons
carry longitudinal momentum fractions $x$ and $1-x$ at the DPS pinch,
with $x(1-x)=m^2/s$ [Fig.~\ref{fig:dps}(b)],
\begin{equation}
\begin{aligned}
C_1&=-4\pi iN_c\,x(1-x)\,F_0(x)\,\M_1^{(0)},\\
|K|&=2\pi N_c\,x(1-x)\,F_2(x)\,|\M_1^{(0)}|,
\end{aligned}
\label{eq:oneloop}
\end{equation}
with $\M_1^{(0)}$ the tree-level value of $\M_1$ [Fig.~\ref{fig:dps}(a)],
$F_0=\frac1{x(1-x)}+2x(1-x)$ for $gg\to HH$,
$F_0\to F_A=\frac1{x(1-x)}-2x(1-x)$ for $gg\to AA$ and $F_2=\frac{x^4+(1-x)^4}{x(1-x)}$ for both processes. These functions arise from the helicity overlaps of the
two splittings~\cite{SM}.
The second relation gives the magnitude of $K$.
In the helicity conventions of the Supplemental Material,
$K>0$ for both $gg\to HH$ and $gg\to AA$.
For $gg\to ZZ$ through a massless quark loop,
$C_1$ is proportional to $x^2$ or $(1-x)^2$,
while $K$ is proportional to $x(1-x)$, depending
on the helicities~\cite{Gaunt:2011xd,SM}.

\textit{Boundary term and phase:}
The potential $X^2$ contribution (Eq.~(\ref{eq:kernel})) comes from ultrasoft exchange. An ultrasoft gluon cannot resolve the transverse separation of the
two hard vertices and couples to the combined colour charge of each
splitting.
The calculation must therefore include diagrams in which the ultrasoft gluon couples to both the daughter partons and their parent gluon.
With the same rapidity regulator throughout, the daughter-daughter and parent-parent exchanges cancel in the static limit, while the mixed parent-daughter contributions vanish.
The calculated soft subtraction also has no finite rapidity logarithm
at $\mu_b=b_0/b$. Together with the two-loop amplitude comparisons,
these results support $r_1=r_2=0$ for the DPS contributions studied.
The remaining finite matching function $r_0$ is determined below.
The Supplemental Material~\cite{SM} gives the regulator and overlap
calculations.

For the leading DPS contribution in Eq.~(\ref{eq:factorised}),
Glauber exchange between partons from the two incoming
splittings generates a phase. Its coefficient is determined
by the colour factor $\DD$, independently of the hard
subprocess amplitudes. With the rapidity regulator and
IR subtraction used here, the poles of the parton
propagators produce an imaginary contribution to the
rapidity logarithm. At fixed momentum
fractions, the dependence of the regions at scale $\mu_b$
on the hard scales enters through $X$ in
Eq.~(\ref{eq:kernel}), with $\operatorname{Im}X=-\pi$.
The imaginary part of the bare kernel is therefore
\begin{equation}
i\,{\rm Im}\,\frac{\Kt^{(1)}_{\rm bare}}{\Kt_0}
=2\pi i\DD\Big[\frac1\epsilon+X_\mu\Big]+O(\epsilon),
\label{eq:imK}
\end{equation}
The $1/\epsilon$ pole is accounted for by the IR subtraction
of the full amplitude relative to the two hard
subprocesses~\cite{Catani:1998bh}. After subtraction, the
phase is given by the $2\pi i\DD X_\mu$ term in
Eq.~(\ref{eq:kernel}). The calculated regions leave no
additional imaginary matching constant at $\mu=\mu_b$.

\textit{Two-loop QCD amplitudes:}
We study the two-loop finite remainders for $gg\to HH$
with effective gluon vertices~\cite{Banerjee:2018lfq},
$gg\to AA$~\cite{Bhattacharya:2019oun},
and $gg\to ZZ$ through massless quark loops~\cite{vonManteuffel:2015msa}.
At fixed velocity $\beta=\sqrt{1-4m^2/s}$,
with $x=(1+\beta)/2$, we approach $p_T=0$
and evaluate the amplitudes using multiprecision
arithmetic. We extract the logarithmic coefficients
by fitting polynomials in $\G$ with $p_T^2$
corrections~\cite{SM}; for $gg\to HH$, omitting
these corrections increases the deviations
to approximately $\plainDevMax$.

For $gg\to HH$ and $gg\to AA$, the coefficients of
$\G^3$ and $\G^2$ in the equal-helicity amplitude,
and of $\G^2$ and $\G$ in the opposite-helicity
amplitude, agree with Eq.~(\ref{eq:rg}) for each of
the colour structures $N_c^2$, $N_cn_f$, $n_f/N_c$
and $n_f^2$.
The comparison covers $\beta=\frac14,\frac12,\frac23$,
with the renormalisation scale dependence of
$gg\to HH$ checked separately.
For $gg\to ZZ$, the predicted logarithms agree
with the two-loop amplitudes in eight helicity
configurations, including symmetry-related ones,
separately for the $C_F$ and
$C_A$ contributions. The fits also confirm the additional
$\ln[x(1-x)]$ terms arising from the different
scale $Q_*$ for quark splittings.
We further test Eq.~(\ref{eq:rg}) using the
massless quark $gg\to V_1^*V_2^*$ amplitudes
with different invariant masses for the two
off-shell vector bosons.
For $(m_1^2,m_2^2,s)=(1,2,15/2)$
and $(2,3,17)$, in common mass-squared units,
the $++++$ and $++--$ transverse-helicity
amplitudes reproduce the predicted cubic and
quadratic logarithms separately in the $C_F$
and $C_A$ terms. The fits include power
corrections and window variations, as detailed
in the Supplemental Material~\cite{SM}.
Figure~\ref{fig:threeprocess} compares the leading and next-to-leading
logarithms across $gg\to HH$, $gg\to AA$ and $gg\to ZZ$.
The normalised cubic coefficients have opposite signs for gluon and
quark splittings. The quadratic coefficient is common to $HH$ and $AA$,
while the $ZZ$ result depends on the splitting momentum fraction
through the scale $Q_*$. The extracted coefficients agree with both
predictions. Table~\ref{tab:tests} gives the corresponding numerical
deviations, fit-window tests and additional velocities.
Together, they support the absence of additional two-loop
boundary logarithms in both splitting channels.
\begin{table}[!htb]
\centering
\small
\setlength{\tabcolsep}{3pt}
\begin{tabular}{lccc}
\toprule
& $gg\to HH$ & $gg\to AA$ & $gg\to ZZ$\\
\midrule
Logarithms, Eq.~(\ref{eq:rg}) & $\devLeadMax$ & $\rzAAlead$ & $\zzDevMax$\\
Fit-window shift & $\shiftMax$ & $\rzAALogShift$ & $\zzShiftMax$\\
$r_0$, fitted points & $\rzFitHH$ & $\rzFitAA$ & $\rzFitZZ$\\
$r_0$, extra velocities & $\rzBlindHH$ & $\rzBlindAA$ & $\rzBlindZZ$\\
$r_0$, region numerics & $\rzRegG$ & --- & $\rzRegQ$\\
\bottomrule
\end{tabular}
\caption{Tests of Eqs.~(\ref{eq:rg}), (\ref{eq:r0g})
and~(\ref{eq:r0q}).
The first row gives the largest deviations of the fitted
logarithmic coefficients from Eq.~(\ref{eq:rg}) at
$\beta=\scanBetas$, relative to the nonzero predictions
for $gg\to HH$ and $gg\to AA$, and in units of $C_1$
or $K$ for $gg\to ZZ$.
The second row gives the largest changes upon shifting
the fit window, with the same normalisations as the first row.
The remaining rows give the absolute deviations of $r_0$
from Eqs.~(\ref{eq:r0g}) and~(\ref{eq:r0q}) at the
fitted velocities, at additional velocities
$\beta=\aaBlindBetas$ for $gg\to HH,AA$ and
$\beta=\frac45$ for $gg\to ZZ$, and from numerical integration of the same
region contributions, respectively.
For $gg\to AA$, the modifications described in the
text are applied.}
\label{tab:tests}
\end{table}

\begin{figure*}[!htb]
\centering
\includegraphics[width=0.94\textwidth]{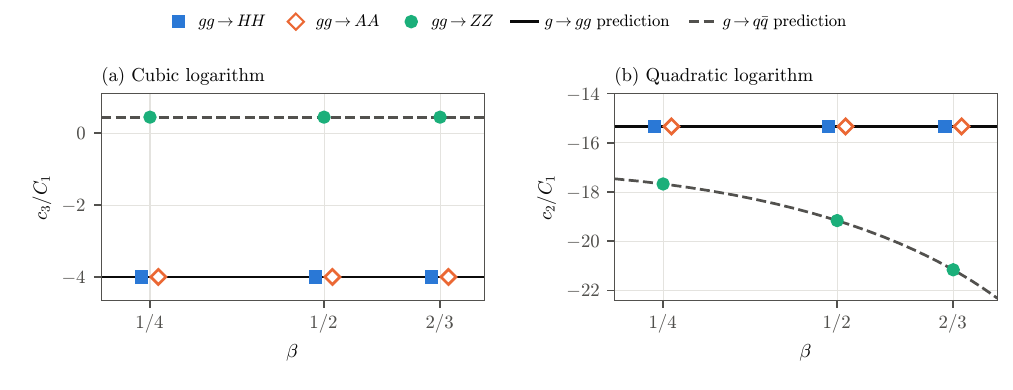}
\caption{Coefficients of the cubic and quadratic DPS logarithms in the
two-loop $gg\to HH$, $gg\to AA$ and $gg\to ZZ$ amplitudes with
$N_c=3$, $n_f=5$ and
$\mathcal M_1^{(2)}\supset c_3\mathcal G^3+c_2\mathcal G^2$.
Lines are the predictions of Eq.~(\ref{eq:rg}); symbols are coefficients
extracted from the complete amplitudes, divided by the corresponding
one-loop coefficient $C_1$. The $ZZ$ quadratic coefficient is plotted
in the gluon-scale logarithm, as in the amplitude fits; $HH$ and $AA$
share the same prediction. The $HH$ and $AA$ markers are displaced
slightly in $\beta$ to show both sets.}
\label{fig:threeprocess}
\end{figure*}

\textit{Finite matching and parity:}
The finite matching coefficient $r_0$ enters
the term proportional to $\mathcal G$ in
$\M_1^{(2)}$ and the constant term in
$\M_2^{(2)}$. High-precision fits to complete
two-loop amplitudes first identified the constants in $r_0$.
We subsequently derived the matching functions
analytically from the collinear and soft contributions
and parton self-energy corrections~\cite{SM}.
After conversion to a common IR subtraction scheme,
the complete amplitude comparisons confirm the
predicted colour and helicity dependence.
The imaginary part of the extracted $r_0$ is below
$\glauberImag$ in all \glauberValues{} fitted components,
testing the predicted phase in Eq.~(\ref{eq:gammaK}).
We obtain
\begin{align}
r_0^{g\to gg}&=C_A\Big[\tfrac{134}9-5\zeta_2-4\ln x\ln(1-x)\Big]-\tfrac{40}9T_Fn_f\notag\\
&\quad+\frac{2(C_A-2T_Fn_f)}{3F_\nu(x)},\label{eq:r0g}\\
r_0^{g\to q\bar q}&=C_F\Big[12-6\zeta_2+2\ln^2\tfrac{x}{1-x}\Big]\notag\\
&\quad+C_A\Big[2+\zeta_2-2\ln^2x-2\ln^2(1-x)\Big],\label{eq:r0q}
\end{align}
with $x$ the momentum fraction of the splitting and $F_\nu$
the overlap of the component, $F_0$ or $F_2$.
The collinear integrals determine the common finite terms,
while the one-loop splitting tensors and their contraction
with the hard amplitudes determine the helicity dependence.
The last term of Eq.~(\ref{eq:r0g}) carries this dependence,
whereas $r_0^{g\to q\bar q}$ is helicity independent.
For $gg\to AA$, $F_A$ replaces $F_0$ for $\nu=0$,
while the last term changes sign for $\nu=2$.

The amplitude reconstruction uses three velocities
and is tested at additional velocities
(Table~\ref{tab:tests}).
The collinear and soft rapidity poles cancel analytically
for $0<x<1$. Numerical integration of the same
region contributions at $x=0.1$, $0.3$ and $0.5$
confirms the matching functions in the tested colour
and helicity channels.
At $x=1/2$ and $\mu_R=m_A$, the $gg\to AA$ prediction
from Eq.~(\ref{eq:r0g}) reproduces the coefficient of
$\G$ in Eq.~(4.6) of Ref.~\cite{P1} exactly, including
the hard contribution and finite IR-scheme conversion.
The helicity-dependent term of Eq.~(\ref{eq:r0g}) originates from the
rational part of the one-loop $g\to gg$ splitting amplitudes~\cite{Bern:1999ry}, with colour factor $C_A-2T_Fn_f$.
For $\nu=2$, the helicity-flip contribution has the relative sign
of $M_{--}/M_{++}$ of the hard vertex.

\textit{Planar $\mathcal N=4$ checks:}
The known self-crossing limit of six-gluon
amplitudes, related to the evolution of
intersecting Wilson lines~\cite{Korchemskaya:1994qp,Dixon:2016epj},
provides a separate setting in which to test
the angular decomposition~\cite{CaronHuot:2019vjl}.
We use the BDS-like normalisation of the six-gluon amplitude,
which divides out a modified Bern-Dixon-Smirnov (BDS)
factor~\cite{Dixon:2016epj}. For $a=g^2N_c/(8\pi^2)$,
the Sudakov factor of Ref.~\cite{Dixon:2016epj} is
\begin{equation}
\begin{aligned}
\mathcal S(a,X)
&=\exp\left[-\frac{\gamma_K(a)}8X^2+B(a)X\right],\\
\gamma_K(a)
&=4a-4\zeta_2a^2+22\zeta_4a^3
+\gamma_K^{(4)}a^4+\ldots .
\end{aligned}
\label{eq:n4exp}
\end{equation}
Here $\gamma_K$ is the cusp anomalous dimension~\cite{Beisert:2006ez}, distinct from the kernel
anomalous dimension $\gamma_{\mathcal K}$ in
Eq.~(\ref{eq:gammaK}), and
$B(a)=\sum_{n\ge2}B^{(n)}a^n$.
The prefactor $a\Kt_0(1+\sum_n\kappa_na^n)$
multiplies $\mathcal S(a,X)$ and includes the
hard functions and BDS-like normalisation.
In $3\to3$ kinematics, the phase cancels and
$\G=\tfrac12\ln|\delta_*|$, with
$\delta_*=p_T^2/Q_*^2$.
The MHV singularity corresponds to $\nu=0$,
with $C_1=2\pi i$.
With the normalisation and boundary constants
fixed, the representation reproduces \nFourPass{}
additional logarithmic coefficients of the known
planar six-gluon amplitudes through five loops
(Table~\ref{tab:n4letter},
\hyperref[sec:endmatter]{End Matter}).

In the $v\to0^-$ overlap with multi-Regge kinematics, the selected
next-to-MHV component carries $|\Delta J_z|=2$. The transform with
$\nu=2$ reproduces \nFourNmhvChecked{} coefficients through four loops
and \nFourNmhvFiveCount{} five-loop coefficients obtained from the
published multi-Regge amplitude~\cite{CaronHuot:2019vjl}, without an
additional matching constant. For the leading nonanalytic terms,
$\M_2(\G)/K=\int_0^\infty ds\,e^{-s}C_1^{-1}
\partial_\G\M_1(\G-s/2)$ follows from the known helicity-flip
factor~\cite{Basso:2014bfkl,CaronHuot:2019vjl}. These are consistency
tests of the DPS interpretation, not a new all-order relation.

The region calculation in planar $\mathcal N=4$ theory,
combined with the hard amplitudes and BDS-like
normalisation, yields $\kappa_1=\zeta_2$, with
deviations below \kappaOneReg{} for \nFourRegPairs{}
pairs of splitting fractions in the four-dimensional
helicity scheme. The $\mathcal N=4$ field content
cancels the rational helicity-dependent term, giving
identical finite matching coefficients for the two
angular components. The Supplemental Material details
the normalisation and analytic continuation to $2\to4$
kinematics, including the phase in
Eq.~(\ref{eq:imK}).

\textit{Three loops:}
The kernel evolution also yields conditional predictions
for the leading three-loop DPS logarithms. If the
factorised description persists and the boundary
correction at relative order $a_s^2$ contains no
powers of $X$ higher than $X^2$, the two highest
logarithmic powers in each component are fixed.
The third is fixed if the $X^2$ coefficient also
vanishes. These boundary conditions remain to be
tested at three loops~\cite{SM}.

Under both conditions, the relative boundary correction
is $a_sr_0+a_s^2(r_0^{(3)}+r_1^{(3)}X)$, with $a_s$
evaluated at $\mu_b$.
The leading three-loop logarithms follow from the
renormalisation group evolution with the known
two-loop cusp anomalous dimension
$\gamma_{\rm cusp}(a_s)=4a_s+\Gamma_1a_s^2$,
where
$\Gamma_1=4[(67/9-2\zeta_2)C_A-20T_Fn_f/9]$.
Using the running coupling and the boundary
conditions stated above, we obtain in
minimal subtraction
\begin{equation}
\begin{aligned}
\M_1^{(3)}&\supset C_1\Big[\tfrac85\DD^2\G^5+c_4\G^4+c_3^{(0)}\G^3\Big],\\
\M_2^{(3)}&\supset K\Big[8\DD^2\G^4+(4c_4-16\DD^2)\G^3\\
&\qquad+\big(3c_3^{(2)}-4\DD(3B+2\beta_0-6\DD)\big)\G^2\Big].
\end{aligned}
\label{eq:threeloop}
\end{equation}
Here
$c_4=2\DD(B+\frac23\beta_0)$ and
$c_3^{(\nu)}=\frac23B(B+\beta_0)-\frac13\DD\Gamma_1-\frac83\DD h-\frac43\DD r_0^{(\nu)}-\frac43\DD L_*(B+\beta_0+\DD L_*)$.
Here $r_0^{(\nu)}$ is the finite matching coefficient
of the corresponding component, $h$ is the one-loop
hard correction relative to the tree amplitude
for either of the identical subprocesses, and
$L_*=\ln[\mu^2/(Q_*^2e^{-i\pi})]$.

For gluon splittings with $T_F=\frac12$, the
coefficients of the two highest powers are
$\frac85C_A^2$ and
$\frac{176}9C_A^2-\frac{32}9C_An_f$
in $\M_1^{(3)}/C_1$, and $8C_A^2$ and
$\frac{560}9C_A^2-\frac{128}9C_An_f$
in $\M_2^{(3)}/K$.
These coefficients are independent of the
renormalisation scale and IR subtraction
scheme. Table~\ref{tab:threeloop}
(\hyperref[sec:endmatter]{End Matter})
gives the third coefficient for $gg\to HH$.

\textit{Conclusion:}
We determine the two-loop DPS logarithms of
gluon-fusion amplitudes and derive the finite
matching functions analytically. The cubic coefficient,
normalised to the one-loop term, changes sign between
gluon and quark splittings. The helicity and parity
dependence follows from the one-loop splitting tensors.
The predictions agree with two-loop
$gg\to HH$, $gg\to AA$ and $gg\to ZZ$ amplitudes
across the tested colour and helicity channels.
Numerical integration of the same region contributions
confirms the analytical matching at the tested momentum
fractions, while the calculated Glauber contribution
reproduces the predicted phase after IR subtraction.
The predicted logarithms also agree with amplitudes
for off-shell vector bosons of unequal invariant
masses and with known planar $\mathcal N=4$
results through five loops. For massive colour-singlet
pair production with the DPS configuration studied here,
the kernel determines the leading two-loop logarithms
from the hard subprocess amplitudes. These predictions provide
analytic benchmarks for complete two-loop calculations.
Under the stated boundary conditions, the kernel
evolution also determines the leading three-loop
logarithms.

\begin{acknowledgments}
I thank IFIC, University of Valencia, and Prof. Francisco Campanario for providing the computing resources used in this work.
ChatGPT (OpenAI) and Claude (Anthropic) were used to
draft notes, draft and debug code, and organise the Git
repository.
The author takes full responsibility for the content of this Letter.
\end{acknowledgments}

\textit{Data availability:} The scripts, stored amplitude points, numerical
results and validation tests accompanying this work are supplied in the
\texttt{anc/} directory of the arXiv source archive and in the companion
repository at \url{https://github.com/abhattacharya1992/two-loop-dps-prl}.
The additional $gg\to V_1^*V_2^*$ scan records and precision checks are in \texttt{paper2/unequal\_virtuality/} of the companion repository.
Re-evaluation from the original published two-loop amplitudes requires
the external inputs identified in the repository README.

\input{sm_nocite.tex}
\bibliography{references,sm}

\section*{End Matter}\phantomsection\label{sec:endmatter}

\input{tables/table_n4_letter.tex}
\textit{Appendix A: Transforms:} The Fourier transforms of the two components of $\Kt_0$ with
$X^n$ inserted follow by differentiating
\begin{equation}
\begin{aligned}
\Phi_\nu(\alpha)&=\int\!\frac{d^2\bb}{2\pi}\frac{e^{i\boldsymbol p_T\cdot\bb+i\nu\phi_b}}{b^2}
\Big(\frac{Q_*^2e^{-i\pi}b^2}{b_0^2}\Big)^{\!\alpha}\\
&=\frac{i^\nu e^{i\nu\phi_q}}{2}\,e^{2(\gamma_E-\G)\alpha}\,
\frac{\Gamma(\frac\nu2+\alpha)}{\Gamma(1+\frac\nu2-\alpha)}
\end{aligned}
\label{eq:Fm}
\end{equation}
at $\alpha=0$, with $\phi_q$ the azimuth of
$\boldsymbol p_T$. For $\nu=0$, the pole of
$\Gamma(\alpha)$ is independent of $p_T$ and
corresponds to a contact term in impact-parameter
space, which is omitted from the nonanalytic
DPS part of the amplitude.
The finite part contains one additional power of $\G$.
No such pole occurs for $\nu=2$.

\input{tables/table_threeloop_letter.tex}

\textit{Appendix B: The boundary term from the regions:}
We apply the method of regions~\cite{Beneke:1997zp}
to the two-loop DPS term at leading power in $\lambda$.
We use dimensional regularisation and a common
rapidity regulator, with scaleless integrals set to
zero in the adopted prescription. Collinear, soft
and Glauber gluons have transverse momenta of order
$p_T$ and resolve the separation $b$. Their transverse
scale is $\mu_b$, while the hard-scale dependence
enters via a regulated rapidity logarithm at
this order~\cite{Chiu:2012ir}.

An ultrasoft gluon has light-cone and transverse
momentum components
$\ell^+\sim\ell^-\sim|\boldsymbol\ell_\perp|
\sim p_T^2/\sqrt{s}$.
Since $|\boldsymbol\ell_\perp|b\ll1$,
it does not resolve the transverse separation $b$.
However, $p_1^+\ell^-$ and $p_2^-\ell^+$
are of order $p_T^2$ and therefore contribute
to the denominators of the nearly on-shell
parton propagators.
Denoting the dimensionful
light-cone energy denominators of the two splittings
by $\sigma_1$ and $\sigma_2$, their product scales as
$\sigma_1\sigma_2\sim\mu_b^4s/(m_1^2m_2^2)$, or
$\mu_b^4/Q_*^2$ for gluon splittings.
An individual exchange between daughter partons
contains a factor
$\Gamma(\epsilon)^2\Gamma(1-\epsilon)
(\sigma_1\sigma_2e^{i\pi}/\mu^2)^{-\epsilon}$,
which can generate an $X^2$ boundary term.

At leading eikonal power, colour conservation
combines the daughter charges into their parent
charge, $T_k+T_{\bar k}=T_{p_1}$, and similarly
for the second splitting. With the same regulator
for all attachments, the static daughter-daughter
contribution $-2\pi^2$ cancels the parent-parent
term $+2\pi^2$; mixed terms vanish~\cite{SM}.
The Glauber phase is treated separately in Appendix C.

After combining the momentum-region contributions
with the required overlap subtractions, their
Laurent expansions have the form
$(\mu^2/\Lambda^2)^\epsilon
(e_2/\epsilon^2+e_1/\epsilon+e_0)$.
For the collinear, soft and Glauber contributions,
$\Lambda=\mu_b$. Their hard-scale dependence enters
through the rapidity logarithm
$(g/\epsilon+g_1)\ln(Q_*^2e^{-i\pi}/\mu_b^2)$;
ultrasoft terms use
$\Lambda^2=\mu_b^4/(\kappa Q_*^2e^{-i\pi})$,
with $\kappa$ independent of $b$.
Independence of the IR poles from $b$, followed
by subtraction, gives $r_2=e_2^{\rm us}/2$ and
$r_1=g_1+e_1^{\rm us}+e_2^{\rm us}\ln\kappa$.
The regulated ultrasoft cancellation gives
$r_2=0$ and $r_1=g_1$ for the calculated terms.

The soft and collinear rapidity poles cancel.
Their sum yields $2\DD$ times the Fourier
transform of $1/\boldsymbol\ell^2$ in
$2-2\epsilon$ dimensions at separation $b$,
\begin{equation}
\Big(\frac{\mu^2b^2}{b_0^2}\Big)^\epsilon e^{-\epsilon\gamma_E}\Gamma(-\epsilon)
=\Big(\frac{\mu^2}{\mu_b^2}\Big)^\epsilon\Big(-\frac1\epsilon-\frac{\zeta_2}2\epsilon+O(\epsilon^2)\Big).
\label{eq:rapcoef}
\end{equation}
At $\mu=\mu_b$ the finite part vanishes,
giving $g_1=r_1=0$. The resulting boundary
coefficients agree with the two-loop amplitudes;
the imaginary term in Eq.~(\ref{eq:imK}) follows
from the same coefficient multiplied by $-i\pi$.

\textit{Appendix C: Glauber exchange:}
In the rapidity-regulator and overlap-subtraction scheme
of Ref.~\cite{Chiu:2012ir}, the phase in
Eq.~(\ref{eq:imK}) arises from the Glauber region.
For an exchange between a daughter parton from each
incoming splitting, the off-shell propagators reduce
at leading power to eikonal denominators with positive
light-cone energy offsets $\sigma_1$ and $\sigma_2$.
Integrating over the longitudinal components of the
exchanged gluon gives
\begin{equation}
\frac{2\pi i}{\boldsymbol\ell^2+\sigma_1\sigma_2}\Big[\ln\frac{\boldsymbol\ell^2}{\sigma_1\sigma_2}-i\pi\Big],
\label{eq:glauberpair}
\end{equation}
where the pole at $\ell^-=\sigma_1$ lies within the
integration range and produces the $-i\pi$ term.
The logarithm is the rapidity logarithm, with
$\sigma_1\sigma_2\sim\mu_b^4/Q_*^2$.

Soft and collinear rapidity poles cancel; the Glauber
region supplies the phase in Eq.~(\ref{eq:glauberpair}).
Zero bins vanish in this prescription, while delta
and principal-value terms contribute equally~\cite{SM}.
The phase cancels for incoming--outgoing pairs
($\sigma_2<0$). Direct longitudinal integration checks
\glauberConfigs{} configurations in a two-loop graph~\cite{SM}.

\end{document}


\title{Supplemental Material for ``Double parton scattering singularities in two-loop gluon-fusion amplitudes''}

\author{Arunima Bhattacharya\,\href{https://orcid.org/0000-0003-4409-0901}{\includegraphics[scale=0.05]{orcidid.pdf}}}
\affiliation{Theory Division, IFIC, University of Valencia-CSIC,\\
E-46980 Paterna, Valencia, Spain}
\email{arunimab@ific.uv.es}

\maketitle

We develop the double parton scattering (DPS) description tested in the
Letter. Part~I derives the one-loop singularity, the two-loop
kernel evolution and the finite matching functions from the
momentum regions. Part~II tests the results against
$gg\to HH$, $gg\to AA$ and $gg\to ZZ$, including off-shell
vector bosons of different invariant masses, and presents
the independent reconstruction from amplitudes and numerical
region checks. Part~III contains scalar pinch checks, comparisons with
known planar $\mathcal N=4$ amplitudes and the conditional three-loop
predictions. We distinguish analytical derivations, numerical
reconstructions and consistency tests throughout. Equations, tables and
figures here carry the prefix~S; unprefixed references refer to the Letter.

\partheading{Part I. The DPS singularity and its two-loop structure}

\section{The one-loop DPS singularity}
\label{sec:oneloop}

The crossed box develops a DPS pinch when the four
internal partons become nearly on shell and collinear
to the incoming gluons~\cite{Gaunt:2011xd}.
We first consider effective scalar and pseudoscalar
vertices, followed by massless quark loops.

\subsection{The pinch and transverse integral}

Let $p_1$ and $p_2$ be the incoming gluon momenta,
with $s=2p_1\cdot p_2$. We consider two final-state
particles of equal mass $m$ and take $p_T\to0$ at
fixed $s$ and $m$. At $p_T=0$, the momentum of the
first final-state particle is
$p_3=xp_1+(1-x)p_2$ [Fig.~\Lref{fig:dps}(b)],
so that $sx(1-x)=m^2$, with $x=(1+\beta)/2$
and $\beta=\sqrt{1-4m^2/s}$.
We define $\lambda=p_T/\sqrt{s}$ and write
$k=\alpha p_1+\gamma p_2+k_\perp$, with
$k\simeq xp_1$ near the DPS pinch. The two
propagators adjacent to the first incoming gluon are
\begin{equation}
D_1=sx\gamma-\kk^2+i0,\qquad
D_2=-s(1-x)\gamma-\kk^2+i0.
\end{equation}
Their poles pinch the $\gamma$ contour as
$\kk\to0$. The propagators adjacent to the
second incoming gluon similarly pinch the
$\alpha$ contour as $\kk-\qq\to0$, where
$\qq$ is the transverse momentum of $p_3$.
Taking both residues and using
$d^4k=\frac s2\,d\alpha\,d\gamma\,d^2\kk$
gives the leading DPS integral,
\begin{equation}
L_{\rm DPS}=\frac{(2\pi i)^2}{2s}\int d^2\kk\,
\frac{\mathcal N_{\rm DPS}(\kk,\qq)}
{\kk^2(\kk-\qq)^2},
\label{eq:pinchmaster}
\end{equation}
where $\mathcal N_{\rm DPS}$ is the leading
numerator, bilinear in $\kk$ and $\kk-\qq$.
This reproduces Eq.~(3.11) of
Ref.~\cite{Gaunt:2011xd} in four dimensions.

\paragraph{Power counting:}
At the pinch, $\alpha-x\sim\gamma\sim\lambda^2$
and $|\kk|\sim\lambda\sqrt{s}$. The integration
measure scales as $\lambda^6$, the four
propagators together as $\lambda^{-8}$, and
the numerator as $\lambda^2$. The resulting
DPS contribution is therefore of leading
power $\lambda^0$, up to logarithms.
Both longitudinal integrations must be pinched:
a single nearby pole gives an integration of
order $\lambda^0$ rather than $\lambda^{-2}$
and cannot produce the same leading contribution.

\subsection{Scalar and pseudoscalar numerators}
\label{sec:numerator}

For $Y=H,A$, the crossed-box numerator in Feynman
gauge, with incoming gluon helicities
$\lambda_1,\lambda_2$, is
\begin{equation*}
\begin{aligned}
\mathcal N^{YY}_{\lambda_1\lambda_2}
={}&\epsilon_1^\mu\epsilon_2^\nu
V_{\mu\rho\sigma}(-p_1,k,\bar k)
V_{\nu\kappa\tau}(-p_2,r,\bar r)
\times V_Y^{\rho\kappa}(k,r)
V_Y^{\sigma\tau}(\bar k,\bar r).
\end{aligned}
\end{equation*}
Here $V$ denotes the three-gluon vertex. With the
Wilson coefficients $C_Y$ removed, the effective
vertices~\cite{Bhattacharya:2019oun,Banerjee:2018lfq}
for incoming momenta $k_a$ and $k_b$ are
\begin{equation*}
\begin{aligned}
V_A^{\rho\sigma}(k_a,k_b)
&=\epsilon^{\mu\rho\nu\sigma}k_{a\mu}k_{b\nu},\\
V_H^{\rho\sigma}(k_a,k_b)
&=(k_a\cdot k_b)g^{\rho\sigma}-k_b^\rho k_a^\sigma.
\end{aligned}
\end{equation*}

We route the momenta as $r=p_3-k$,
$\bar k=p_1-k$ and
$\bar r=p_2-r=p_4-p_1+k$
[Fig.~\Lref{fig:dps}(b)].
Expanding the numerators to quadratic order in the
transverse momenta, with $\sigma=x(1-x)$ and
$E=\sqrt{s}/2$ the incoming gluon energy in the
centre-of-mass frame, gives
\begin{align}
\mathcal N^{AA}_{++}&=8E^4\big[(1-2\sigma^2)\,\kk\cdot(\kk-\qq)+i(1+2\sigma^2)(k_yq_x-k_xq_y)\big],\nonumber\\
\mathcal N^{HH}_{++}&=-8E^4\big[(1+2\sigma^2)\,\kk\cdot(\kk-\qq)+i(1-2\sigma^2)(k_yq_x-k_xq_y)\big],\\
\mathcal N^{AA}_{+-}=\mathcal N^{HH}_{+-}&=-8E^4\big[x^4+(1-x)^4\big]\,k_+(k_+-q_+),\nonumber
\end{align}
where $k_\pm=k_x\pm ik_y$ and
$q_\pm=q_x\pm iq_y$.
The terms proportional to $k_yq_x-k_xq_y$
vanish upon angular integration. The remaining
transverse integrals are
\begin{equation}
\int_{|\kk|<R}d^2\kk\,\frac{\kk\cdot(\kk-\qq)}{\kk^2(\kk-\qq)^2}=\pi\ln\frac{R^2}{q^2}\quad(R>q),\qquad
\int d^2\kk\,\frac{k_+(k_+-q_+)}{\kk^2(\kk-\qq)^2}=-\pi e^{2i\phi_q},
\label{eq:transverse}
\end{equation}
with $q=|\qq|$ and $\phi_q$ its azimuthal angle.
The angular average of the first integrand vanishes
for $|\kk|<q$, while that of the second vanishes
for $|\kk|>q$. Thus the equal-helicity contribution
contains a logarithm, whereas the opposite-helicity
contribution approaches a finite limit.

\subsection{Helicity coefficients and normalisation}

The two effective vertices contribute $i^2$, the
four propagators $(-i)^4$, the two three-gluon
vertices $g_s^2$, the residues $(2\pi i)^2/(2s)$
and the loop measure $(2\pi)^{-4}$.
With the colour contraction
$f^{acd}f^{bcd}=C_A\delta^{ab}$, these factors
give $2a_sN_c/s$, where $a_s=\alpha_s/(4\pi)$.
With the effective couplings factored out, we
decompose the $gg\to YY$ amplitude as
$\M^{\mu\nu}_{ab}\propto
\delta_{ab}(T_1^{\mu\nu}\M_1+T_2^{\mu\nu}\M_2)$,
using the tensor basis of Ref.~\cite{P1},
\begin{equation}
\begin{aligned}
T_1^{\mu\nu}&=g^{\mu\nu}-\frac{p_2^\mu p_1^\nu}{p_1\cdot p_2},\\
T_2^{\mu\nu}&=g^{\mu\nu}+\frac{m^2p_2^\mu p_1^\nu-2(p_1\cdot p_3)\,p_2^\mu p_3^\nu-2(p_2\cdot p_3)\,p_3^\mu p_1^\nu
+2(p_1\cdot p_2)\,p_3^\mu p_3^\nu}{(p_1\cdot p_2)\,p_T^2},
\end{aligned}
\label{eq:tensors}
\end{equation}
where $p_T^2=(tu-m^4)/s$.
For physical polarisations, $T_1$ contributes only
to equal-helicity amplitudes and $T_2$ only to
opposite-helicity amplitudes. The corresponding
contractions are $-1$ for the $++$ projection
onto $T_1$ and $e^{2i\phi_q}$ for the $+-$
projection onto $T_2$.
In these conventions, the tree amplitudes are
$\M_1^{(0)}=is/2$ for $AA$ and $-is/2$ for $HH$.
Combining the numerators, transverse integrals
and normalisation factors, we obtain
\begin{align}
\M_1^{(1)}\big|_{\rm DPS}^{AA}&=\pi N_cm^2F_A(x)\ln\frac{p_T^2}{m^2}+\dots,&
F_A(x)&=\frac1{x(1-x)}-2x(1-x),\label{eq:FA}\\
\M_1^{(1)}\big|_{\rm DPS}^{HH}&=-\pi N_cm^2F_S(x)\ln\frac{p_T^2}{m^2}+\dots,&
F_S(x)&=\frac1{x(1-x)}+2x(1-x),\label{eq:FS}\\
\M_2^{(1)}\big|_{\rm DPS}&=m^2K_1(x),&
K_1(x)&=\pi N_cW(x),\quad W(x)=\frac{x^4+(1-x)^4}{x(1-x)},\label{eq:K1}
\end{align}
The dots in the first two lines denote terms
finite as $p_T\to0$. The opposite-helicity
result holds for both processes. In the notation
of the Letter, $F_S=F_0$ and $W=F_2$.

The dependence on $x$ follows from the helicity
sums of the gluon splitting amplitudes~\cite{Mangano:1990by,Dixon:1996wi,P1}.
For an incoming gluon of positive helicity,
the relevant weights are
$w_{++}=1/\sqrt{x(1-x)}$,
$w_{+-}=x^2/\sqrt{x(1-x)}$ and
$w_{-+}=(1-x)^2/\sqrt{x(1-x)}$.
The scalar and pseudoscalar vertices give
opposite signs for the mixed-helicity terms,
producing $F_S$ and $F_A$, respectively.
The combination
$W=w_{+-}(x)w_{-+}(1-x)+w_{-+}(x)w_{+-}(1-x)$
is the same for both processes. These terms
are products of amplitudes, not splitting
probabilities.

The logarithmic variable used in the Letter is
\begin{equation}
\G=\frac12\ln\frac{m^4-tu}{m^4}=\frac12\ln\frac{s\,p_T^2}{m^4}+\frac{i\pi}{2},
\label{eq:Gdef}
\end{equation}
where the principal branch is understood, since
$m^4-tu=-s\,p_T^2<0$.
At fixed $s$ and $m$,
$\ln(p_T^2/m^2)$ and $2\G$ differ only by a
constant. The one-loop DPS terms are therefore
$C_1\G$ and $K$, with
$C_1=2\pi N_cm^2F_A(x)$ for $AA$,
$C_1=-2\pi N_cm^2F_S(x)$ for $HH$,
and $K=m^2K_1(x)$ for both.
Relative to the tree amplitude,
$C_1/\M_1^{(0)}=-4\pi iN_cm^2F(x)/s$,
where $F=F_A$ or $F_S$, as in
Eq.~(\Lref{eq:oneloop}).
This ratio is purely imaginary in the stated
conventions; $\G$ also contains the constant
$i\pi/2$.
For comparison with Ref.~\cite{P1}, the threshold
variable satisfies $\LL=\G+\ln(1-\beta^2)$.
At threshold, $\beta=0$ and $x=1/2$, so that
$\LL=\G$, $C_1=7\pi N_cm^2$ for $AA$
and $K_1=\pi N_c/2$.
Direct one-loop checks of $AA$ reproduce the
logarithmic coefficient to $\pOneAALogDev$
and the opposite-helicity limit to
$\pOneKoneDev$~\cite{P1}.
For $HH$, the logarithmic coefficient agrees
to $\pOneHHLogDev$~\cite{P1}. An independent extraction
of the one-loop coefficients from the scale
dependence of the two-loop $HH$ amplitudes is
given in Sec.~\ref{sec:hhoneloop}.
For $p_T^2<\dpsOnset\,m^2$, the $AA$ and $ZZ$
scans reproduce the one-loop logarithmic
coefficients and finite opposite-helicity
limits to within one per cent at
$\beta=\zzBetas$.

\subsection{Massless quark loops}
\label{sec:quarkloops}

For massless quark loops, the residue calculation
reproduces the single crossed-box coefficients of
Eqs.~(3.16)--(3.18) in Ref.~\cite{Gaunt:2011xd}.
The full amplitude includes both orientations of the
fermion loop. Under charge conjugation, reversing the
fermion flow gives a factor $(-1)^{n_V}$, where $n_V$
counts the vector couplings, including those of the
two incoming gluons. Axial-vector and scalar
couplings do not change the sign.
Consequently, the two orientations add in the
$v_q^2$ and $a_q^2$ contributions to $gg\to ZZ$
and cancel in the $v_qa_q$ contribution. They
also add for scalar couplings.
The DPS coefficient of the full $gg\to ZZ$
amplitude is therefore twice the single-box
result of Eqs.~(3.17)--(3.18) in
Ref.~\cite{Gaunt:2011xd}.\footnote{
Eqs.~(2.6)--(2.7) of Ref.~\cite{Gaunt:2011xd}
have twice the prefactors of its
Eqs.~(3.17)--(3.18), despite the statement
following Eq.~(3.18) that they agree.
For $gg\to HH$, Eqs.~(2.5) and (3.16)
both give the single crossed-box result.}
We verify the normalisation independently using
a four-dimensional fermion trace, including
both loop orientations, $T_F=\frac12$ and
the loop measure $(2\pi)^{-4}$.
With the overall couplings stripped, the
coefficients in the expansion in
$a_s=\alpha_s/(4\pi)$ are
$C_1=-16\pi ix^2$ for $\hel++++$ and
$C_1=-16\pi i(1-x)^2$ for $\hel++--$,
with the same results for their parity partners.

\subsection{The forward limit of the opposite-helicity amplitude}
\label{sec:K1complete}

The opposite-helicity amplitude
$\mathcal A_{\pm\mp}=C_Y^2\M_2e^{\pm2i\phi_q}$
has $|J_z|=2$. A contribution analytic in
the transverse momentum must have the form
$(p_x\pm ip_y)^2g(p_T^2)$ and therefore
vanishes as $p_T^2$ in $\M_2$.
The nonzero forward limit of Eq.~(\ref{eq:K1})
must thus arise from a contribution nonanalytic
at $p_T=0$.

For massless internal lines, the leading Landau
conditions for the three box orderings give
\begin{equation}
\det\mathcal Y_{(1234)}=(su)^2,\quad
\det\mathcal Y_{(1243)}=(st)^2,\quad
\det\mathcal Y_{(1324)}=(tu-m^4)^2=(s p_T^2)^2,
\end{equation}
where $\mathcal Y_{ij}=-(q_i-q_j)^2$~\cite{Landau:1959fi,Eden:1966dnq}.
At fixed $\beta$, neither $t$ nor $u$
vanishes as $p_T\to0$. Only the crossed
box develops the additional Landau pinch,
corresponding to the two collinear splittings
in Fig.~\Lref{fig:dps}(b), as expected from
the Coleman-Norton interpretation~\cite{Coleman:1965xm}.
The reduced triangle and bubble integrals
have no additional singularity at $p_T=0$.
Infrared subtraction terms are proportional
to the tree amplitude,
$\M_2^{(0)}=O(p_T^2)$.
Spurious inverse Gram factors introduced
by tensor reduction~\cite{Passarino:1978jh}
cancel in the full amplitude.
An analytic hard contribution also vanishes
as $p_T^2$. The crossed-box DPS region
therefore supplies the constant in
Eq.~(\ref{eq:K1}).

The Landau analysis does not exclude an
additional physical-sheet singularity of
the second type at the vanishing Gram
determinant. We therefore check the result
against the exact one-loop amplitudes.
For $gg\to AA$, the forward limit agrees
with Eq.~(\ref{eq:K1}) to
$\pOneKoneDev$~\cite{P1}.
The ratio
$(\M_2^{(1)}-m^2K_1)/p_T^2$
remains stable to \aaApproachDigits{} digits
as $p_T^2$ decreases over
\aaApproachDecades{} orders of magnitude
below $\aaApproachBelow\,m^2$.
For $gg\to HH$, the same ratio is stable
to \kOneApproachDigits{} digits over
\kOneApproachDecades{} orders of magnitude
in $p_T^2$, with its slope stable to
$\kOneSlopeSpread$. The extrapolated
forward limit agrees with $m^2K_1$
to $\devKMax$.

\section{Two-loop DPS structure and momentum regions}
\label{sec:factorisation}

We expand the two-loop diagrams about the DPS pinch
of Sec.~\ref{sec:oneloop}, retaining the leading
power in $\lambda$ at fixed $m_i^2/s$ and momentum
fractions away from the endpoints. We first make
the helicity contractions in
Eq.~(\Lref{eq:factorised}) explicit, then identify
the contributions from the different momentum regions.
We use the DPS loop momentum
$k=\alpha p_1+\gamma p_2+k_\perp$ and write the
additional gluon momentum as
$\ell=a p_1+c p_2+\ell_\perp$, with
$k_\perp^2=-\kk^2$ and
$\ell_\perp^2=-\boldsymbol\ell^2$.
The daughter parton momenta are $k$,
$\bar k=p_1-k$, $r=p_3-k$ and
$\bar r=p_4-p_1+k$. The pairs $k,r$ and
$\bar k,\bar r$ enter the two hard vertices,
respectively [Fig.~\Lref{fig:dps}(b)].
The parent partons have quadratic Casimir $C_a$,
while the partons entering hard vertex $i$
have Casimir $C_{R_i}$. We retain these
colour factors for both gluon and quark splittings.

\paragraph{Nonanalytic DPS terms: }
We retain the leading terms that are nonanalytic
in $\boldsymbol p_T$ at $p_T=0$. These include
logarithms and, for $|J_z|=2$, the azimuthal
factor $e^{2i\phi_q}=p_{T+}^2/\boldsymbol p_T^2$,
where $p_{T\pm}=p_{Tx}\pm ip_{Ty}$.
Terms polynomial in $\boldsymbol p_T$ give
contact terms in impact parameter space. For
example, $(\boldsymbol p_T^2)^n$ transforms
into $(-\nabla_b^2)^n\delta^2(\bb)$, and
$e^{2i\phi_q}(\boldsymbol p_T^2)^n
=p_{T+}^2(\boldsymbol p_T^2)^{n-1}$
is polynomial for $n\ge1$.
The logarithms and the angular factor
$e^{2i\phi_q}$ instead contribute at nonzero
$b$. The large-$b$ expansion in
$1/(b\sqrt{s})$ isolates the DPS behaviour
near $p_T=0$; the full Fourier transform
also depends on the amplitude away from this
limit. Thus a one-loop constant in the
$\nu=0$ component is omitted, while the
$|\nu|=2$ constant $K$, which multiplies
$e^{2i\phi_q}$, is retained.

\paragraph{Helicity structure: }
The kernel carries the helicity indices of
the daughter partons. For incoming gluon
helicities $\lambda_1,\lambda_2$, its contraction
with the hard amplitudes is
\begin{equation}
\widetilde{\mathcal A}^{\lambda_1\lambda_2}_{\rm DPS}(\bb)
=\sum_{\lambda_k\lambda_{\bar k}\lambda_r\lambda_{\bar r}}
H_1^{\lambda_k\lambda_r}\,H_2^{\lambda_{\bar k}\lambda_{\bar r}}\,
\widetilde{\mathcal K}^{\lambda_1\lambda_2}_{\lambda_k\lambda_{\bar k}\lambda_r\lambda_{\bar r}}(\bb),
\qquad
\widetilde{\mathcal K}=a_s\widetilde{\mathcal K}_0+a_s^2\widetilde{\mathcal K}^{(1)}+O(a_s^3),
\label{eq:tensorfact}
\end{equation}
where $H_i^{\lambda\lambda'}$ are the
infrared-subtracted on-shell amplitudes
for the two $2\to1$ subprocesses. Colour
indices are also contracted.
Equation~(\Lref{eq:factorised}) suppresses
these indices.
The kernel depends on $\bb$, the momentum
fractions, the scale and the parton
representations, but not on the hard vertices.
The finite matching coefficient $r_0$ is
defined after contraction with the tree-level
hard amplitudes and normalisation to the
corresponding lowest-order contribution.
Its helicity projections can therefore depend
on the hard interaction. For $HH$ and $AA$,
this dependence enters through the overlaps
$F_\nu$ and the hard helicity ratio
$M_{--}/M_{++}=\pm1$.

For fixed parton representations and the
colour-singlet hard projection, the anomalous
dimensions are independent of helicity.
The kernel components therefore satisfy
Eq.~(\Lref{eq:gammaK}) at the order considered.
At order $a_s$, Eq.~(\ref{eq:tensorfact})
reduces to the one-loop DPS representation
of Ref.~\cite{Gaunt:2011xd}. At order $a_s^2$,
\begin{equation}
\widetilde{\mathcal A}^{(2)}_{\rm DPS}(\bb)=\big(H_1^{(1)}H_2^{(0)}+H_1^{(0)}H_2^{(1)}\big)\widetilde{\mathcal K}_0(\bb)
+H_1^{(0)}H_2^{(0)}\,\widetilde{\mathcal K}^{(1)}(\bb),
\label{eq:twoloopfact}
\end{equation}
with the helicity and colour contractions
understood as in Eq.~(\ref{eq:tensorfact}).
The first term contains the one-loop corrections
to the hard subprocesses, while the second
contains the correction to the DPS kernel.

\paragraph{Momentum regions: }
The additional loop momentum has the following
scalings in the method of regions~\cite{Beneke:1997zp}:
\begin{center}
\begin{tabular}{lccc}
\toprule
Region & $a$ & $c$ & $|\boldsymbol\ell|/\sqrt{s}$\\
\midrule
Hard & $1$ & $1$ & $1$\\
Collinear to $p_1$ & $1$ & $\lambda^2$ & $\lambda$\\
Collinear to $p_2$ & $\lambda^2$ & $1$ & $\lambda$\\
Soft & $\lambda$ & $\lambda$ & $\lambda$\\
Glauber & $\lambda^2$ & $\lambda^2$ & $\lambda$\\
Ultrasoft & $\lambda^2$ & $\lambda^2$ & $\lambda^2$\\
\bottomrule
\end{tabular}
\end{center}

At fixed momentum fractions, the squared momenta
scale as $\ell^2\sim s$ in the hard region,
$\ell^2\sim p_T^2$ in the collinear, soft and
Glauber regions, and $\ell^2\sim p_T^4/s$
in the ultrasoft region. The ultrasoft scale, including
its dependence on the momentum fractions,
is derived in Sec.~\ref{app:boundary}.
Glauber momentum is off shell and contributes
to the phase in the rapidity-regulator scheme
of Ref.~\cite{Chiu:2012ir}.
Intermediate longitudinal scalings must be
accounted for through the overlap subtractions.
Additional scalar pinch checks are given in
Sec.~\ref{sec:pinch}, and the region contributions
are compared with the full amplitudes in
Sec.~\ref{sec:r0regions}.

\paragraph{Hard contributions: }
We choose the loop routing so that a hard
correction confined to either fusion vertex
leaves the four DPS propagators near shell.
Both longitudinal pinches are then preserved.
Expanding the hard subgraph about the pinch
gives the one-loop correction to the corresponding
hard subprocess. Its infrared-subtracted part
enters $H_1^{(1)}$ or $H_2^{(1)}$.
A hard correction to an incoming splitting
can also preserve both pinches. Its external
momenta are $p_1$, $xp_1$ and $(1-x)p_1$,
so all external invariants vanish. In
covariant gauge, the bare hard integral is
scaleless in dimensional regularisation,
and the splitting correction is assigned
to the collinear region. This does not remove
its separate ultraviolet (UV) and infrared (IR)
counterterms. The argument does not apply
diagram by diagram in axial gauge, where
$(n\cdot k)^2/n^2$ introduces another scale.

Hard exchanges connecting the two subprocesses
are not included in the hard term of
Eq.~(\ref{eq:twoloopfact}). The pointwise
scaling checks find suppression by $\lambda$
for $ZZ$ in Feynman and axial gauges, and
for $HH,AA$ in axial gauges. Some individual
scalar-channel diagrams in covariant gauge
instead scale as $\lambda^0$. Their cancellation
after summing diagrams is not established
by these checks. The comparisons with the
complete two-loop amplitudes therefore provide
an essential test of the factorised
hard contribution.

\paragraph{Collinear contributions: }
For $\ell$ collinear to $p_1$, diagrams involving
$p_1,k,\bar k$ correct the first splitting,
including parton self-energy corrections.
Attachments to $r,\bar r$ and the hard vertices
also contribute at leading power.
The propagators
$(r+\ell)^2\simeq r^-\ell^+\sim s$
are hard. Integrating out these propagators
generates the Wilson lines $W_{n_1}$
in the gauge-invariant collinear fields
$\chi_{n_1}$ and $\mathcal B_{n_1\perp}$~\cite{Bauer:2002nz}.
For an effective $ggH$ or $ggA$ vertex,
the non-Abelian attachments complete
$\partial^+A_\perp$ to $\mathcal B_\perp$.
A collinear gluon with both ends attached
to hard lines is suppressed by $\lambda^2$.

The collinear correction is obtained from
the one-loop matrix element
$\langle0|O_1(0)O_2(\bb)|g(p_1)\rangle$.
We normalise the components present at tree
level to the corresponding tree splitting.
Helicity-flip contributions without a tree-level
counterpart are retained as separate tensor
components~\cite{Bern:1999ry}.
Their contraction with the tree hard amplitudes
gives the first-beam contribution to
$\widetilde{\mathcal K}^{(1)}$.
The second beam gives the corresponding
correction $J_2$. The collinear functions
$J_1$ and $J_2$ are evaluated in
Sec.~\ref{app:regions}.

\paragraph{Soft contributions: }
For soft $\ell$, the couplings to
$p_1,k,\bar k,p_2,r,\bar r$ are eikonal
at leading power, since
$(k+\ell)^2\simeq k^+\ell^-\sim\lambda s$
dominates $k^2\sim\lambda^2s$.
Individual diagrams involving an effective
hard vertex can also contribute at leading
power in covariant gauge. Their power
counting alone does not establish cancellation
after summing diagrams. We use the eikonal
soft function defined by the corresponding
Wilson-line matrix element.

An exchange within one hard vertex is
scaleless. An exchange between partons
entering different vertices changes the
DPS transverse momentum and produces
$e^{i\boldsymbol\ell\cdot\bb}$.
Colour conservation at the singlet vertices
and incoming splittings gives
$T_k+T_r=T_{\bar k}+T_{\bar r}=0$,
$T_k+T_{\bar k}=T_{p_1}$ and
$T_r+T_{\bar r}=T_{p_2}$. Hence
\begin{equation}
T_k\cdot T_{\bar r}=T_{\bar k}\cdot T_r=-T_k\cdot T_{\bar k}=\tfrac12\big(C_{R_1}+C_{R_2}-C_a\big)=\tfrac12\DD ,
\label{eq:softcolour}
\end{equation}
The parent-parent exchanges have colour
factor $-C_a$, while exchanges between
partons entering hard vertex $i$ have
factor $-C_{R_i}$. These terms are
independent of $b$ and enter the IR
poles of the full amplitude and hard
subprocess amplitudes, respectively.
The remaining soft function $S$ is a
Wilson-line matrix element at two
transverse positions. It depends on the
parton representations, not their helicities,
and contributes as $\widetilde{\mathcal K}_0S$.

\paragraph{Glauber and ultrasoft contributions: }
Glauber exchange between daughter partons
leaves the hard vertices unchanged at
leading power. The propagators carrying
$\ell$ reduce to eikonal denominators
containing the energy offsets
$\sigma_1,\sigma_2$ of the splittings.
The Glauber function $G$ depends on these
offsets, the parton directions and their
colour representations.
Exchanges within a hard vertex contribute
its Coulomb phase $i\pi C_{R_i}$.
Exchanges between the vertices produce
the $-i\pi$ term in
Eq.~(\ref{eq:glauberint}).
The calculation includes parent attachments
and uses a common rapidity regulator.
A vanishing longitudinal integral for an
individual diagram does not imply the
vanishing of the complete regulated overlap.

Individual ultrasoft exchanges contribute
at leading power, but their sum cancels in
the static eikonal limit, as shown in
Sec.~\ref{app:boundary}.
In the calculated regions, the
hard-collinear overlap is scaleless,
the soft and collinear rapidity poles cancel,
and the Glauber-soft overlap is subtracted
as in Eq.~(\ref{eq:glauberregions}).
The remaining contributions have the
structure of Eq.~(\ref{eq:twoloopfact}).

\paragraph{Helicity and colour representations: }
Angular momentum along the beam determines
the harmonic $e^{i\nu\phi_b}$.
For scalar and pseudoscalar pairs, equal and
opposite incoming gluon helicities correspond
to $\nu=0$ and $\nu=\pm2$, respectively.
For $ZZ$, the outgoing helicities also enter,
as discussed in Sec.~\ref{sec:zz}.
The kernel is block diagonal in $\nu$,
with the leading-power angular dependence
$e^{i\nu\phi_b}/b^2$.
Each block can contain more than one
helicity component. At one loop, a
helicity-flip splitting generates a
daughter configuration absent at tree
level~\cite{Bern:1999ry}.
Its projection gives the rational
helicity-dependent term of
Eq.~(\Lref{eq:r0g}), with the relative
sign determined by $M_{--}/M_{++}$.
This accounts for the different finite
terms in $HH$ and $AA$ without changing
their leading logarithms.
The parton representations enter through
their Casimirs, non-cusp anomalous dimensions
and splitting amplitudes. The leading-power
counting is the same for $g\to gg$ and
$g\to q\bar q$.

\paragraph{Scope: }
The region calculation identifies the leading
DPS contributions and their colour and helicity
structure. The complete two-loop amplitudes
provide independent tests of the resulting
factorised description in the channels studied.
A systematic classification of all pinch
surfaces and complete renormalised overlap
matching would extend this analysis~\cite{Rothstein:2016glauber}.
The kernel here describes scattering amplitudes
and is distinct from the splitting kernels
of double parton distributions~\cite{Diehl:2011yj,Buffing:2017mqm,Diehl:2019rdh}.
The collinear function $J_1$ retains open colour
and helicity indices. A comparison with DPD splitting
kernels requires the corresponding operator
projections and subtraction conventions.

\section{Evolution and Fourier transforms}
\label{sec:rg}

We derive the two-loop logarithms of
Eq.~(\Lref{eq:rg}) from the evolution and
Fourier transforms of the kernel defined in
Sec.~\ref{sec:factorisation}. The boundary
terms not fixed by evolution are examined
in Sec.~\ref{app:boundary}.

\paragraph{Anomalous dimension: }
For the parton representations and colour-singlet
hard projections considered here, the anomalous
dimensions of IR-subtracted amplitudes
depend only on the external coloured partons~\cite{Becher:2009cu,Becher:2009qa,Gardi:2009qi}.
Subtracting the anomalous dimensions of the two
hard subprocesses from that of the full $2\to2$
amplitude gives Eq.~(\Lref{eq:gammaK}).
The contributions from the effective operators
cancel in this difference. Since the anomalous
dimensions are independent of $p_T$, the DPS
term obeys the same evolution equation as the
full amplitude. For $g\to gg$,
$Q_*^2=m_1^2m_2^2/s$.

\paragraph{Boundary condition: }
The lowest-order kernel contains one splitting
from each incoming gluon and starts at
$a_s(\mu_b)$, where $\mu_b=b_0/b$.
Using
$a_s(\mu_b)=a_s(\mu)[1+\beta_0a_s(\mu)X_\mu]+O(a_s^3)$
and evolving Eq.~(\Lref{eq:gammaK}) from
$\mu_b$ gives Eq.~(\Lref{eq:kernel}).
Evolution fixes the scale dependence but leaves
the matching boundary $\widetilde R$ undetermined.
The region calculations and amplitude comparisons
support $r_1=r_2=0$ for the DPS terms studied,
leaving the finite matching coefficient $r_0$
(Sec.~\ref{app:boundary}).
The separation between evolution and matching
depends on the scale convention.
At $\mu^2=Q_*^2$, $X_\mu=X+i\pi$, so the
RG terms in Eq.~(\Lref{eq:kernel}) differ
from $-\DD X^2+BX$ only by terms
independent of $b$.
Expressing the lowest-order coupling at $Q_*$
instead of $\mu_b$ requires a corresponding
change in the boundary:
$a_s(\mu_b)=a_s(Q_*)[1+\beta_0a_s(Q_*)(X+i\pi)]+O(a_s^3)$.
Imposing a boundary without logarithms at
$Q_*$ would instead replace $B$ by
$B-\beta_0$. Throughout this work, the
boundary is defined at $\mu_b$.

\paragraph{Transforms: }
We use the generating integral
$\Phi_\nu(\alpha)$ of Eq.~(\Lref{eq:Fm})
to transform the kernel to momentum space.
The angular integration gives
$2\pi i^\nu e^{i\nu\phi_q}J_\nu(p_Tb)$,
and the radial integration is a Mellin
transform of the Bessel function.
Differentiating $\Phi_\nu$ $n$ times
with respect to $\alpha$ at $\alpha=0$
inserts $X^n$.

For $\nu=0$, the pole of $\Gamma(\alpha)$
at $\alpha=0$ is independent of $p_T$
and corresponds to a contact term in
impact parameter space. We omit this term
and retain the finite parts of
$\Phi_0$, $\Phi_0'$ and $\Phi_0''$:
$-\G$, $\G^2$ and
$-\frac43\G^3-\frac23\zeta_3$, respectively.
For $\nu=2$, there is no pole, and
$\Phi_2'(0)/\Phi_2(0)=1-2\G$ and
$\Phi_2''(0)/\Phi_2(0)=4\G^2-4\G+2$.
With $r_1=r_2=0$, applying these transforms
to $-\DD X^2+BX$ gives the logarithms
of Eq.~(\Lref{eq:rg}), in units of the
one-loop coefficients $C_1$ and $K$.
The pole at $\alpha=0$ gives the $\nu=0$
component one additional power of $\G$
and accounts for the factor $1/3$ in
the cubic coefficient.

\paragraph{IR scheme: }
Equation~(\Lref{eq:gammaK}) uses minimally
subtracted amplitudes, whereas the $HH$
remainders of Sec.~\ref{sec:hh} use Catani's
subtraction operators~\cite{Catani:1998bh}.
With two coloured incoming partons and
colour-singlet final states, a finite
scheme change multiplies lower-order
amplitudes by functions of $s$ and $\mu$
only. It therefore leaves the
$\G^3$ and $\G^2$ coefficients of
$\M_1^{(2)}$ unchanged, as well as the
$\G^2$ and $\G$ coefficients of
$\M_2^{(2)}$. The latter follows because
$\M_2^{(1)}$ has no DPS logarithm.

\section{Boundary logarithms and regulated overlaps}
\label{app:boundary}

The evolution equation leaves the boundary
$\widetilde R=\widetilde{\mathcal K}_0(r_0+r_1X+r_2X^2)$
undetermined. The two-loop amplitude comparisons support
$r_1=r_2=0$. The fitted $gg\to ZZ$ coefficients also permit the
additional boundary logarithms to be extracted rather than fixed.
After the coupling and logarithmic-scale conversions,
the largest fitted residuals at $N_c=3$ are
$|r_1|<1.5\times10^{-9}$ and
$|r_2|<2.3\times10^{-11}$.
The maximum fit-window changes give the conservative
numerical estimates
$|r_1|\lesssim1.1\times10^{-7}$ and
$|r_2|\lesssim1.7\times10^{-9}$
for the tested helicities and velocities.
Both coefficients are consistent with zero within these
uncertainties, which are not rigorous error bounds.
The calculation is reproduced by the accompanying
boundary-coefficient check. We examine the momentum regions that contribute
to these boundary logarithms: ultrasoft exchange for the
$X^2$ term and the finite rapidity logarithm for the
$X$ term. We work at leading power in $\lambda$ in
$d=4-2\epsilon$ dimensions, using the rapidity regulator
of Ref.~\cite{Chiu:2012ir} and a prescription in which
bare scaleless integrals vanish. Appendix B of the
Letter summarises the calculation.

\paragraph{Scales and energy offsets: }
Corrections confined to either hard subprocess enter
$\mathcal H_1$ or $\mathcal H_2$. The collinear, soft
and Glauber regions have transverse momenta of order
$p_T$ and resolve the separation $b$, with characteristic
scale $\mu_b$. At one exchange, their dependence on the
hard scales enters through a regulated rapidity logarithm.
After cancellation of the regulator dependence, its
coefficient is given by Eq.~(\Lref{eq:rapcoef})~\cite{Chiu:2012ir,Echevarria:2011epo,Buffing:2017mqm}.

An ultrasoft gluon does not change the leading transverse
or large longitudinal momenta of the daughter partons,
but is sensitive to their small off-shellness. At the
DPS pinch, $\alpha\simeq x$, $\gamma\sim\lambda^2$
and $|\kk|\sim\lambda\sqrt{s}$.
For equal final-state masses, the dimensionless
longitudinal energy offsets are
\begin{equation}
\sigma_1=\frac{\kk^2}{x(1-x)s},\qquad
\sigma_2=\frac{(\kk-\qq)^2}{x(1-x)s}.
\end{equation}
For unequal masses, write
$p_3=xp_1+yp_2$ at $p_T=0$, with
$sxy=m_1^2$ and $s(1-x)(1-y)=m_2^2$.
The second denominator is then $y(1-y)s$,
while $y=1-x$ for equal masses. Thus
\[
s\sigma_1\sigma_2
=\frac{\kk^2(\kk-\qq)^2}{x(1-x)y(1-y)s}
=\frac{\kk^2(\kk-\qq)^2s}{m_1^2m_2^2}.
\]
For gluon splittings this equals
$\kk^2(\kk-\qq)^2/Q_*^2$, which scales as
$\mu_b^4/Q_*^2$ in impact parameter space.
For quark splittings, $Q_*$ differs from
$m_1m_2/\sqrt{s}$ by a factor depending on
the momentum fractions, without changing
the power counting.

The dimensionful light-cone energy offsets used
in the Letter are $\sigma_1p_2^-$ and
$\sigma_2p_1^+$. Since $p_1^+p_2^-=s$,
their product is $s\sigma_1\sigma_2$.
We use the same conversion in the Glauber
integral of Sec.~\ref{app:regions}.

\paragraph{Eikonal attachments: }
An ultrasoft gluon exchanged between the two
splittings couples at leading power through
$p_2^\mu$ on the second beam.
Writing
$p_2^\mu=\ell^\mu/c+R^\mu$, the contraction
of $R^\mu$ with a momentum collinear to $p_1$
is suppressed by $\lambda$.
The first splitting
therefore sees a longitudinally polarised gluon.
Ward identities reduce the sum of its couplings
to eikonal factors multiplying the tree splitting.
For a quark,
$S(k)\slashed\ell S(k+\ell)=S(k)-S(k+\ell)$,
where $S(k)=1/\slashed k$. The corresponding
three-gluon Ward identity includes the contact
vertices.

The same result follows from the field
redefinition of Ref.~\cite{Bauer:2001yt},
which expresses the ultrasoft couplings through
Wilson lines along the incoming directions.
At leading power, the ultrasoft gluon cannot
resolve the separation between the hard vertices.
Their transverse and longitudinal separations
scale as $1/(\lambda\sqrt{s})$ and $1/\sqrt{s}$,
respectively. Colour conservation combines the
charges of the daughter partons into that of
their parent,
$T_k+T_{\bar k}=T_{p_1}$,
and similarly for the second splitting.
The complete bare exchange between the
resulting Wilson lines is scaleless in the
adopted prescription.
This cancellation can be seen explicitly at
one exchange. Relative to the lowest-order
residue $2\pi i/(s\kk^2)$, integration over
$\gamma$ gives the following longitudinal
factors for the attachments to $k$,
$\bar k$ and $p_1$, respectively~\cite{Gaunt:2011xd}:
\begin{equation}
\frac1{c-\sigma_1+i0},\qquad
\frac1{c-\sigma_1+i0},\qquad
\frac1{c+i0}-\frac1{c-\sigma_1+i0}.
\label{eq:usattach}
\end{equation}
Multiplying by the respective colour charges
and summing gives $T_{p_1}/(c+i0)$.
The second splitting similarly gives
$T_{p_2}/(-a+i0)$. The combined expression is
\begin{equation}
\widetilde{\mathcal K}_0\,g_s^2\,T_{p_1}\!\cdot T_{p_2}
\int\frac{d^d\ell}{(2\pi)^d}
\frac{4p_1\cdot p_2\,\mu^{2\epsilon}/s^2}{(\ell^2+i0)(c+i0)(-a+i0)},
\qquad T_{p_1}\!\cdot T_{p_2}=-C_a,
\label{eq:usscaleless}
\end{equation}
with the $i0$ prescriptions appropriate to
two incoming Wilson lines.

The complete bare integral is scaleless,
but this does not imply that individual
diagrams or their separate UV
and IR poles vanish. For example,
the exchange between $k$ and $\bar r$
contains $T_k\cdot T_{\bar r}$ times
\begin{equation}
\int\frac{da\,dc\,d^{d-2}\boldsymbol\ell}{(sac-\boldsymbol\ell^2+i0)(c-\sigma_1+i0)(-a-\sigma_2+i0)}
\propto\Gamma(\epsilon)^2\Gamma(1-\epsilon)
\big(\sigma_1\sigma_2e^{i\pi}\big)^{-\epsilon}.
\end{equation}
This contribution contains the scale dependence
that can generate $X^2$. Its cancellation
requires the sum over parent and daughter
attachments.

\paragraph{Regulated static overlap: }
To evaluate this sum with a common rapidity
regulator, define
\begin{equation}
J(u,v)=\int da\,dc\,
\frac{|a-c|^{-\eta}}
{(c-u+i0)(-a-v+i0)},
\label{eq:jointoverlap}
\end{equation}
for positive $u$ and $v$.
In the analytic scaleless prescription,
$J(0,0)=0$. This is the regulated
homogeneous integral, not the limit
obtained by taking $u,v\to0$ in the
expression for positive offsets.

Including exchanges between daughter partons,
between parents, and between parents and
daughters gives
\begin{equation}
\begin{split}
&J(\sigma_1,\sigma_2)
+[J(0,\sigma_2)-J(\sigma_1,\sigma_2)]+[J(\sigma_1,0)-J(\sigma_1,\sigma_2)]
\\&+[J(0,0)-J(0,\sigma_2)-J(\sigma_1,0)
+J(\sigma_1,\sigma_2)]
=J(0,0).
\end{split}
\label{eq:fullstaticcancel}
\end{equation}
As $\eta\to0$ at fixed positive offsets,
the daughter-daughter contribution approaches
$-2\pi^2$ and the parent-parent contribution
approaches $+2\pi^2$, while the mixed terms
vanish. The regulated static overlap therefore
cancels. Taking the longitudinal offsets to
zero before summing the diagrams would miss
the parent contribution.
The subtraction associated with the transverse
separation is proportional to
$\DD[\cos(\boldsymbol\ell\cdot\bb)-1]$.
For ultrasoft momentum,
$|\boldsymbol\ell|b\sim\lambda$,
so this term is suppressed.

\paragraph{Boundary coefficients: }
We now relate the region calculation to
$r_1$ and $r_2$. After normalisation to the
lowest-order kernel, the pole terms of a
region have the form
\[
\Big(\frac{\mu^2}{\Lambda^2}\Big)^\epsilon
\Big(\frac{e_2}{\epsilon^2}
+\frac{e_1}{\epsilon}+e_0\Big).
\]
For collinear, soft and Glauber contributions,
$\Lambda=\mu_b$. For ultrasoft contributions,
we write
$\Lambda^2=\mu_b^4/(\kappa Q_*^2e^{-i\pi})$,
where $\kappa$ depends on the momentum fractions
but not on $b$.
The IR poles of the normalised kernel
are determined by those of the full amplitude
and the two hard subprocesses~\cite{Catani:1998bh,Becher:2009cu}.
Their combined pole coefficients cannot depend
on $b$. Expanding in $X$ and $X_\mu$ and
performing minimal subtraction gives
\begin{equation}
\tfrac12(e_2^{\rm b}+e_2^{\rm us})X_\mu^2-(e_2^{\rm b}+e_2^{\rm us})XX_\mu+\tfrac12e_2^{\rm us}X^2
+(g_1+e_1^{\rm us}+e_2^{\rm us}\ln\kappa)X+\dots,
\end{equation}
where the superscripts b and us refer to
the regions at scale $\mu_b$ and the ultrasoft
regions, respectively. Here $g_1$ is the finite
coefficient of the rapidity logarithm at $\mu_b$.
Comparison with Eq.~(\Lref{eq:kernel}) gives
$e_2^{\rm b}+e_2^{\rm us}=2\DD$ and
\[
r_2=\tfrac12e_2^{\rm us},\qquad
r_1=g_1+e_1^{\rm us}+e_2^{\rm us}\ln\kappa.
\]
Thus the quadratic boundary term is associated
with ultrasoft exchange in this region analysis.
The regulated static cancellation removes the
ultrasoft terms in the evaluated one-exchange
sum, giving $r_2=0$ and $r_1=g_1$.

The remaining coefficient $g_1$ is fixed by
the one-exchange soft subtraction.
The rapidity divergences cancel between the
soft and collinear regions, and the coefficient
of the remaining logarithm is given by the
Fourier transform in Eq.~(\Lref{eq:rapcoef}).
At $\mu=\mu_b$, this transform has no finite
term at order $\epsilon^0$, so $g_1=0$.
The region calculation therefore supports
$r_1=r_2=0$ in the stated prescription.

This result depends on the convention
$b_0=2e^{-\gamma_E}$ used in $X$, $X_\mu$
and the Fourier transform. Replacing $b_0$
by $b_0e^{-\kappa'/2}$ moves
$2\DD\kappa'X$ between the evolution terms
and the matching boundary without changing
the amplitude. The statement $r_1=0$ therefore
refers to this choice of convention.

\paragraph{Additional attachment and amplitude checks: }
We also check the eikonal colour structure
for two ordered ultrasoft gluons.
The twelve leading attachments to a splitting,
summed over both gluon orderings, combine through
colour conservation,
$(T_k+T_{\bar k})\mathcal C_1
=\mathcal C_1T_{p_1}$, into
\begin{equation}
\mathcal C_1\left[
\frac{T_{p_1}^bT_{p_1}^a}{(c_1+i0)(c_1+c_2+i0)}+
\frac{T_{p_1}^aT_{p_1}^b}{(c_2+i0)(c_1+c_2+i0)}
\right],
\label{eq:ustwo}
\end{equation}
which is the ordered two-gluon Wilson-line
vertex of the parent. The second splitting
gives the corresponding expression.
This check covers parent and daughter
attachments, but not gluon self-interactions,
self-energies or diagrams involving mixed
momentum regions.

An explicit two-loop diagram with a gluon
exchanged between daughter partons, with
numerator $\kk\cdot(\kk+\boldsymbol\ell-\qq)$
at $x=\kernX$, confirms a leading ultrasoft
contribution. For
$|\boldsymbol\ell|\sim\lambda\sqrt{s}$,
the coefficient of the single logarithm agrees
with its eikonal expression to $\kernSlopeDev$
over \kernSlopePoints{} configurations.
At $\lambda=\kernUsLam$, the complete
one-exchange sum agrees with
$T_{p_1}/(c+i0)$ to $\kernUsDev$,
although the individual terms remain nonzero.

We verify the identities in
Eqs.~(\ref{eq:usattach}) and~(\ref{eq:ustwo})
using exact arithmetic, with open spinor indices
and explicit SU(3) colour matrices.
The checks cover vector and axial
$g\to q\bar q$ vertices and $g\to gg$
with effective scalar vertices.
For the latter, \usTwoGluonLeading{} of
\usTwoGluonGraphs{} two-gluon diagrams
contribute at leading power.
The calculations were repeated using Lorentz
algebra in \usDimLow{} and \usDimHigh{}
dimensions. The gluon contractions depend
linearly on $d$, and the quark contractions
contain no closed $d$-dimensional traces.
These calculations establish the stated
attachment identities and static cancellation
with the second splitting kept at tree level.
A complete renormalised treatment of the
unexpanded Glauber diagrams and all overlaps
remains outside these checks.

The two-loop $gg\to HH$ and $gg\to ZZ$
amplitudes independently test $r_1=r_2=0$.
In the numerical region calculation, the
collinear terms $2\DD/(\eta\epsilon)$ cancel
against the soft contribution, with no finite
rapidity logarithm. The remaining finite
constant agrees with $r_0$
(Sec.~\ref{sec:r0regions}).
As a separate check, the same pole analysis
reproduces the nonzero ultrasoft $X^2$ term
of a Sudakov triangle with two off-shell legs.

\section{One-loop correction to the DPS kernel}\label{app:regions}

We calculate $\widetilde{\mathcal K}^{(1)}$ in impact parameter space at
$\mu=\mu_b$, where
$\widetilde{\mathcal K}=a_s\widetilde{\mathcal K}_0+
 a_s^2\widetilde{\mathcal K}^{(1)}+O(a_s^3)$.
The evolution terms vanish at this scale. With the boundary condition
supported in Sec.~\ref{app:boundary}, the finite real correction in
minimal subtraction is $r_0\widetilde{\mathcal K}_0$.
We calculate the collinear, soft and Glauber contributions and
parton self-energy corrections separately, checking the cancellation of their poles
and the resulting finite terms. With $p_1^+=p_2^-=\mu=1$ and $b=b_0$,
the bare correction, normalised to the lowest-order kernel, is
\begin{equation}
\dfrac{\widetilde{\mathcal K}^{(1)}_{\rm bare}}{\widetilde{\mathcal K}_0}=J_1+J_2+S+\sum_{\rm daughters}\Sigma+G,
\label{eq:Kregions}
\end{equation}
for helicity structures present at tree level. Contributions that
first appear at one loop are retained separately. In the prescription
used here, $J_1,J_2,S$ and $\Sigma$ are real after the regulator
limits, whereas $G$ is purely imaginary. The real finite part
fixes $r_0$; $G$ has no finite matching constant at $\mu_b$.

\paragraph{Collinear matrix elements: }
The first collinear function is obtained from the coefficient of
$a_s$ in $\langle0|O_1(0)O_2(\bb)|g(p_1)\rangle$ at momentum
fractions $x$ and $1-x$. We normalise structures present at tree
level to the corresponding tree splitting and retain helicity
structures that first appear at one loop separately.
For $g\to q\bar q$, the fields are
$\chi_n=W_n^\dagger\xi_n$ and
$\bar\chi_n=\bar\xi_nW_n$, projected onto their good components.
For $g\to gg$, we use
$\mathcal B_{n\perp}^\mu=g_s^{-1}[W_n^\dagger iD_\perp^\mu W_n]$~\cite{Bauer:2002nz}. The Wilson lines account for collinear
gluons along the conjugate light-cone direction.

We work in Feynman gauge. Taking residues in the minus components
of the collinear and DPS loop momenta reduces $J_1$ to a one-loop
light-front wave function. We combine the transverse denominators
with Feynman parameters and integrate in $2-2\epsilon$ dimensions.
A term linear in $\kk$ acquires, relative to its tree transform,
the factor
\begin{equation}
F_\epsilon=\frac{e^{-\epsilon\gamma_E}\Gamma(1-2\epsilon)}{\Gamma(1+\epsilon)\Gamma(1-\epsilon)}
\label{eq:Feps}
\end{equation}
For $g\to q\bar q$, we include the Abelian and non-Abelian vertex
corrections and Wilson-line couplings to the quark, antiquark and
parent gluon. For $g\to gg$, the diagrams comprise gluon, ghost
and quark triangles, four-gluon vertex bubbles on either daughter,
and Wilson line couplings to the parent and daughter gluons.
Both orientations of the ghost and quark loops are included.
Diagrams containing both a Wilson line and a four-gluon vertex
from $\mathcal B_{n\perp}$ vanish in Feynman gauge because
$\bar n$ is orthogonal to itself and to transverse polarisations.

\paragraph{Rapidity prescription: }
We regulate the eikonal denominators using
$\nu_{\rm r}^\eta|\ell^+|^{-\eta}$ at each Wilson-line vertex
in $J_1$, $\nu_{\rm r}^\eta|\ell^-|^{-\eta}$ in $J_2$, and
$\nu_{\rm r}^\eta|\ell^+-\ell^-|^{-\eta}$ in the soft function~\cite{Chiu:2012ir}. Here $\nu_{\rm r}$ is the rapidity scale,
distinct from the angular harmonic $\nu$. The zero-bin integrals
are scaleless in this prescription, and we take $\eta\to0$ before
$\epsilon\to0$. Charge conjugation relates $J_2$ to $J_1(1-x)$;
for $g\to q\bar q$, the two quark helicity structures are interchanged.

\paragraph{Soft contribution: }
Only gluon exchanges between Wilson lines entering different
hard vertices depend on the transverse separation $b$. The two
cross pairs have total colour factor $\DD$. We integrate the
longitudinal component carrying the rapidity regulator along the
real axis. At the gluon pole, the regulator becomes
$|\boldsymbol\ell^2/\ell^--\ell^-|^{-\eta}$, with $\ell^->0$.
Near the eikonal endpoint, the integrand behaves as
$(\ell^-)^{\eta-1}$; the prescription removes the
$\delta$-function contribution at that endpoint. The soft
function is therefore real in this scheme. At $\mu=\mu_b$,
\begin{equation}
S=2\DD\,e^{-\epsilon\gamma_E}\Gamma(-\epsilon)\Big[\frac2\eta+\ln\frac{\nu_{\rm r}^2b^2}{b_0^2}-\gamma_E-\psi(-\epsilon)\Big].
\label{eq:softeta}
\end{equation}
The separation of the off-shell daughter exchange into momentum
regions below provides a further check of the phase assignment.

\paragraph{Parton self-energy corrections: }
At the DPS residue, a daughter carrying fraction $x$ has
squared momentum $-\kk^2/(1-x)$. In Feynman gauge, its
self-energy multiplies the tree contribution in impact parameter
space by
\begin{equation}
\Sigma=F_\epsilon\,\frac{\Gamma(\epsilon)\Gamma(1-\epsilon)^2}{\Gamma(2-2\epsilon)}\,(1-x)^\epsilon\times
\begin{cases}-(1-\epsilon)\,C_F&\text{quark},\\[2pt]
\big[(5-3\epsilon)C_A-4(1-\epsilon)T_Fn_f\big]/(3-2\epsilon)&\text{gluon}.
\end{cases}
\label{eq:sigma}
\end{equation}
The incoming-gluon self-energy is scaleless. Renormalisation of
the coupling in $a_s\widetilde{\mathcal K}_0$ contributes
$-\beta_0/\epsilon$ to Eq.~(\ref{eq:Kregions}).

\paragraph{Longitudinal Glauber integral: }
Consider a gluon exchanged between $k$ and $\bar r$. We choose
$r=p_3-k-\ell$ and $\bar r=p_2-p_3+k+\ell$, so the two hard
vertices receive $k+\ell,r$ and $\bar k,\bar r-\ell$,
respectively, with $\bar k=p_1-k$. Taking the DPS residues at
the $\bar k^2$ pole in $k^-$ and the $r^2$ pole in $k^+$ gives,
at leading power in the Glauber region,
\begin{equation}
\begin{aligned}
(k+\ell)^2&\to x(\ell^--\sigma_1)+i0,&\sigma_1&=\frac{\kk^2}{1-x}+\frac{(\kk+\boldsymbol\ell)^2}x,\\
(\bar r-\ell)^2&\to-x(\ell^++\sigma_2)+i0,&\sigma_2&=\frac{(\kk+\boldsymbol\ell-\qq)^2}{1-x}+\frac{(\kk-\qq)^2}x,
\end{aligned}
\end{equation}
where both energy offsets are positive for two incoming parents.
The offsets here include the exchanged transverse momentum and are
dimensionful in the units of this calculation. At
$\boldsymbol\ell=0$, they correspond, after restoring dimensions,
to $\sigma_1p_2^-$ and $\sigma_2p_1^+$, with the $\sigma_i$ in
Sec.~\ref{app:boundary} defined as dimensionless quantities.
Taking the $\ell^+$ residue restricts $\ell^->0$.
The remaining integration gives
\begin{equation}
\int\frac{d\ell^+d\ell^-}{(\ell^+\ell^--\boldsymbol\ell^2+i0)(\ell^--\sigma_1+i0)(-\ell^+-\sigma_2+i0)}
=\frac{2\pi i}{\boldsymbol\ell^2+\sigma_1\sigma_2}\Big[\ln\frac{\boldsymbol\ell^2}{\sigma_1\sigma_2}-i\pi\Big],
\label{eq:glauberint}
\end{equation}
The pole at $\ell^-=\sigma_1$ lies within the integration
range and produces the $-i\pi$ term. Appendix C of the Letter
relates this term to the kernel phase. With one incoming and one
outgoing parton, $\sigma_2<0$ and the imaginary contributions from
the two poles cancel. This is the $3\to3$ configuration considered
in Sec.~\ref{sec:n4regions}.

\paragraph{Separation into regions: }
The assignment of the phase to the Glauber region depends on the
rapidity regulator and overlap subtraction. To check it, we expand
Eq.~(\ref{eq:glauberint}) at fixed
$A=\boldsymbol\ell^2$. In the $n$-collinear region, the gluon
and $k$ propagators are kept without expansion, while the $\bar r$
propagator becomes $1/(-\ell^++i0)$ and the regulator is
$|\ell^+|^{-\eta}$. The $\bar n$-collinear region follows by
interchanging the two splittings. In the soft region both daughter
propagators are eikonal, with regulator
$|\ell^+-\ell^-|^{-\eta}$. In the Glauber region, the gluon
propagator reduces to $-1/A$, while both offsets and the symmetric
regulator are retained. The zero-bin integrals are scaleless.

The longitudinal contributions at fixed $A$ are
\begin{equation}
\begin{aligned}
C_n=-\frac{2\pi i}A\Big(\frac{\sigma_1}A\Big)^{\eta}\frac\pi{\sin\pi\eta},
\hspace{4pt}
C_{\bar n}=C_n\big|_{\sigma_1\to\sigma_2},
\hspace{4pt}
S=-4\big[Y_sR_s+Y_cR_c\big],
\hspace{4pt}
G=\frac{(\sigma_1+\sigma_2)^{-\eta}}A\Big[\pi^2+\pi^2-2i\pi^2\tan\frac{\pi\eta}2\Big],
\end{aligned}
\label{eq:glauberregions}
\end{equation}
where, for the soft integral, we set
$\ell^+=\pm\rho e^y$ and $\ell^-=\pm\rho e^{-y}$.
Here $C_n,C_{\bar n},S$ and $G$ denote the
longitudinal contributions at fixed
$\boldsymbol\ell$, before the transverse
integration. Their integrated counterparts
enter Eq.~(\ref{eq:Kregions}).
The rapidity factors are
$Y_s=\Gamma(\eta/2)\Gamma(1-\eta)/\Gamma(1-\eta/2)$ and
$Y_c=2^{-\eta}\sqrt\pi\,\Gamma(\eta/2)/\Gamma((1+\eta)/2)$,
for the $\sinh y$ and $\cosh y$ regulators in the regions with
$\ell^+\ell^->0$ and $\ell^+\ell^-<0$, respectively. The
radial factors are
$R_s=\tfrac12A^{-1-\eta/2}[\pi\cot(\pi\eta/2)-i\pi]$ and
$R_c=-\tfrac12A^{-1-\eta/2}\pi/\sin(\pi\eta/2)$.
For each collinear region, we take the residue in the component
not regulated and integrate the other component along the real axis.
The two $\pi^2$ terms in $G$ arise from different parts of the
eikonal denominators. Using
$1/(x+i0)=\mathrm{PV}(1/x)-i\pi\delta(x)$, one comes from the
product of delta functions and the other from the principal-value
cross term. The symmetric regulator retains the latter through
$[-\pi\tan(\pi\eta/2)][\pi\cot(\pi\eta/2)]$. A product
regulator $|\ell^+|^{-\eta}|\ell^-|^{-\eta}$ would miss this
term unless the overlap subtraction were changed.

The rapidity poles cancel between the collinear and soft regions:
in units of $2\pi i/(A\eta)$ their contributions are $-2$ and
$+2$, respectively. As $\eta\to0$, the sum of regions agrees
with Eq.~(\ref{eq:glauberint}) to $\glauberSum$ at
$A/(\sigma_1\sigma_2)=\glauberRatio$, consistent with the
neglected power correction. The Glauber term alone approaches
$2\pi^2/A$ to $\glauberG$. The collinear and soft longitudinal
integrals are imaginary, but become real after division by the
lowest-order factor $2\pi i$. For $\sigma_2<0$, the outgoing
collinear contribution cancels $2\pi^2/A$ to $\glauberCancel$.
We take $\eta\to0$ before expanding the transverse integral in
$\epsilon$. The Glauber contribution to the normalised kernel is
$-i\pi$ times $2\DD$ times Eq.~(\Lref{eq:rapcoef}), reproducing
Eq.~(\Lref{eq:imK}) without an additional finite matching constant
at $\mu_b$. In the explicit two-loop graph of Sec.~\ref{app:boundary},
exact integration over $\ell^\pm$ reproduces the same leading term
at fixed $\boldsymbol\ell=\lambda\hat{\boldsymbol\ell}$.
The result agrees with $-\pi/2$ times the coefficient of
$\ln\lambda$ to $\glauberGraph$ over \glauberConfigs{} configurations
of $x,\kk,\hat{\boldsymbol\ell}$; the hard scale enters as
$se^{-i\pi}$.

\paragraph{Helicity structures: }
For $g\to q\bar q$, projection onto the good spinor components
leaves $\slashed\varepsilon_\perp\slashed k_\perp$ and
$\slashed k_\perp\slashed\varepsilon_\perp$, corresponding to
the two quark helicity structures. For $g\to gg$, we use
\[
\begin{aligned}
S_1&=(\varepsilon\cdot\kk)(e_1\cdot e_2),&
S_2&=(\varepsilon\cdot e_1)(e_2\cdot\kk),\\
S_3&=(\varepsilon\cdot e_2)(e_1\cdot\kk),&
T_4&=\frac{(\varepsilon\cdot\kk)(e_1\cdot\kk)(e_2\cdot\kk)}{\kk^2}.
\end{aligned}
\]
Here $e_1$ and $e_2$ denote the transverse polarisations of the
daughter gluons. The $T_4$ structure first appears at one loop and is finite.
The combination $T_4-(S_1+S_2+S_3)/4$ is symmetric and
traceless in its three transverse polarisation indices
and carries the third angular harmonic, whereas the
tree splitting carries the first.
For their common radial dependence $1/|\kk|$,
the Fourier transforms differ by $i^3/i=-1$.
Thus, at $\epsilon=0$, $T_4\to(S_1+S_2+S_3)/2-T_4$,
with $\kk$ replaced by $\bb$.
At one loop, the corrected splitting from either incoming gluon
is contracted with the lowest-order splitting from the other.
Combining these splittings with the hard amplitudes gives
the factors $F_0$ and $F_2$ and the helicity-dependent
finite terms discussed below.

\paragraph{Analytic collinear contributions: }
We evaluate the collinear diagrams at fixed $0<x<1$,
retaining the dependence on $\epsilon$ and $\eta$.
Taking residues in the minus components reduces the
transverse integrals in $2-2\epsilon$ dimensions to
beta functions and parameter integrals of the form
\begin{equation}
I_{ab}(x,\epsilon)=
\int_0^1 d\zeta\,\zeta^a
\int_0^\zeta ds\,
\frac{s^{b-1-\epsilon}}
{[1-x\zeta-(1-x)s]^{1+\epsilon}}.
\label{eq:r0master}
\end{equation}
For example,
\begin{equation}
I_{-1,0}=
\frac1{\epsilon^2}
+\frac{\ln(1-x)}{\epsilon}
+\zeta_2-2\operatorname{Li}_2(x)
-\frac12\ln^2(1-x)+O(\epsilon).
\label{eq:r0masterex}
\end{equation}

For $g\to q\bar q$, the eight diagrams include vertex
corrections and Wilson-line contributions.
The additional Dirac structure cancels separately
in the $C_F$ and $C_A$ parts, leaving the tree splitting
multiplied by
\begin{equation}
\begin{aligned}
J_1^{g\to q\bar q}
={}&C_F\Big[
\frac4{\eta\epsilon}
+\frac{5-2L}{\epsilon}
+8-2\zeta_2+L
+\ln^2\frac{x}{1-x}\Big]
+C_A\Big[
-\frac2{\eta\epsilon}
+1-\ln^2x-\ln^2(1-x)\Big],
\end{aligned}
\label{eq:r0J1q}
\end{equation}
where $L=\ln[x(1-x)]$. The second collinear function
follows by interchanging the two incoming splittings.
Combining their finite parts with the soft contribution
and parton self-energy corrections gives
\begin{equation}
r_0^{g\to q\bar q}
=2\left.J_1^{g\to q\bar q}\right|_{\eta^0\epsilon^0}
-\zeta_2(2C_F-C_A)+2C_F(-2-L),
\label{eq:r0assembleq}
\end{equation}
which reproduces Eq.~(\Lref{eq:r0q}).

For $g\to gg$, we evaluate the gluon, ghost and quark
triangles, four-gluon vertex bubbles and Wilson-line
contributions. Including the terms related by exchange
of the two outgoing partons, the 23 diagram entries give
\begin{equation}
\begin{aligned}
c_s={}&C_A\Big[
\frac2{\eta\epsilon}
+\frac{1/3-2L}{\epsilon}
+\frac59-2\zeta_2-\frac53L
-2\ln x\ln(1-x)
+\frac{\rho_s}{3}\Big]
+T_Fn_f\Big[
\frac4{3\epsilon}
+\frac{20}9+\frac43L
-\frac{2\rho_s}{3}\Big],\\
c_4={}&\frac23(C_A-2T_Fn_f),
\end{aligned}
\label{eq:r0J1g}
\end{equation}
where $c_s$, $s=1,2,3$, are normalised to the
corresponding tree coefficients, $c_4$ is the
coefficient of $T_4$ in units of the tree coefficient
of $S_1$, and
$(\rho_1,\rho_2,\rho_3)=(0,1-x,x)$.
Contracting the two lowest-order splitting amplitudes
with the hard vertices gives
\[
\begin{aligned}
gg\to HH:\quad&
\frac{F_0}{2x(1-x)},\qquad
\frac{F_2}{x(1-x)},\\
gg\to AA:\quad&
-\frac{F_A}{2x(1-x)},\qquad
\frac{F_2}{x(1-x)},
\end{aligned}
\]
for $\nu=0$ and $\nu=2$, respectively.
The terms proportional to $\rho_s$ and $c_4$,
together with the Fourier transform of $T_4$,
determine the helicity dependence.
Combining the one-loop corrections with the soft
contribution and parton self-energy terms gives
Eq.~(\Lref{eq:r0g}) for both angular components.
For $gg\to AA$, $F_A$ replaces $F_0$ for $\nu=0$,
while the last term changes sign for $\nu=2$.
The rapidity poles cancel between the collinear
and soft contributions, and the remaining single
poles reproduce the coefficient $B$ after accounting
for the momentum-fraction dependence.

\paragraph{Numerical evaluation: }
We also evaluate the same collinear integrals numerically,
using a separate integration procedure.
After taking the residues and integrating over transverse
momenta, each diagram contains an integral over the loop momentum
fraction and at most two Feynman parameters. The integrands are
products of polynomials raised to powers linear in $\epsilon$
and $\eta$. We factor the polynomials, resolve endpoint zeros
using iterated sector decomposition~\cite{Binoth:2000ps},
subtract the endpoint singularities and use Gauss-Legendre
quadrature. We extract the Laurent coefficients from discrete
Fourier sums on $|\epsilon|=0.2$ and
$|\eta|=0.05|\epsilon|$, using 32 points on the
$\epsilon$ circle.

We evaluate the kernel at \rzXvalues; symmetry under $x\to1-x$
also gives the results at $x=0.7$ and $0.9$. At $x=1/2$,
reducing the number of Fourier points from 32 to 16 allows
aliasing from the pole at $\epsilon=1/2$ into $r_0$ at the level
of $\rzAlias$. Increasing the quadrature orders from
64, 48 and 28 to 96, 72 and 40 for integrals in one, two and
three variables, respectively, changes $r_0$ by less than
$\rzQuad$ and reduces the difference from
Eqs.~(\Lref{eq:r0g}) and~(\Lref{eq:r0q}) below $\rzQuadDev$
for every colour contribution. We also check the parameter
integrals directly at real $\epsilon$ and $\eta$.

For $g\to q\bar q$, charge conjugation maps each diagram in
$J_1(x)$ to one at $1-x$, with the quark helicity structures
interchanged. Their finite parts agree to $\rzJspread$ over
the five momentum fractions, and the four projected values of
$r_0$ agree to $\rzSpread$.
In both splitting channels, the rapidity poles are proportional
to the tree splitting, with coefficient
$2\DD/(\eta\epsilon)$ for each collinear function and no
$\eta^{-1}\epsilon^0$ term. The helicity structures first
appearing at one loop are kept separately. These poles cancel
against the soft contribution in Eq.~(\ref{eq:softeta}).
The finite results are in Table~\ref{tab:r0}.


\partheading{Part II. QCD amplitudes and finite matching}

\section{Two-loop \texorpdfstring{$gg\to HH$}{gg -> HH} amplitudes}
\label{sec:hh}

\subsection{Amplitudes}
\label{sec:hhamp}

We use the two-loop class-B amplitudes of
Ref.~\cite{Banerjee:2018lfq}, in which each Higgs boson is
produced through a separate effective gluon vertex.
The tensor basis and finite remainder conventions follow
Ref.~\cite{P1}, with $d=4-2\epsilon$.
The coupling and both effective operators are renormalised
in the $\overline{\rm MS}$ scheme. The two-loop finite
remainder is
$\M^{(2),{\rm fin}}=\M^{(2)}-2I^{(1)}\M^{(1)}-4I^{(2)}\M^{(0)}$,
where $I^{(1)}$ and $I^{(2)}$ are Catani's IR
subtraction operators~\cite{Catani:1998bh}.

We write $\M_i=m_H^2A_i$, with
$A_i=i\,{\rm Gauge}_i/(4m_H^2)$ in terms of the
expressions ${\rm Gauge}_1$ and ${\rm Gauge}_2$
provided with Ref.~\cite{Banerjee:2018lfq}.
The remainders contain four colour structures:
\begin{equation}
A_i=N^2A_i^{[1]}+Nn_f\,A_i^{[2]}+\frac{n_f}{N}\,A_i^{[3]}+n_f^2A_i^{[4]},
\label{eq:colourdec}
\end{equation}
where $N=N_c$. Their dependence on the renormalisation
scale is at most quadratic in
$L=\ln(m_H^2/\mu_R^2)$. Evaluating each colour
coefficient at \lValueCount{} values of $L$
therefore determines its dependence on $\mu_R$.

We use the kinematic variables $x,y,z$ of
Refs.~\cite{Banerjee:2018lfq,Bhattacharya:2019oun},
renaming their $x$ as $v$ to distinguish it
from the splitting fraction $x$. Thus
$s=m_H^2(1+v)^2/v$, $t=-m_H^2y$ and
$u=-m_H^2z$, with $z=v+1/v-y$. It follows that
$s\,p_T^2=m_H^4(yz-1)$ and
$\G=\frac12\ln(1-yz)$, with the principal
branch understood. At fixed $\beta$, we set
$v=(1-\beta)/(1+\beta)$ and approach the
forward limit along $y=v+\delta$, with
$\delta\to0$ and $p_T^2\propto\delta$.

\subsection{Method}
\label{sec:method}

We evaluate the amplitudes at $\beta=\scanBetas$,
with $\delta=\scanDeltaFirst,\ldots,\scanDeltaLast$.
Cancellations involving inverse Gram determinants
lead to a loss of approximately $3k$ decimal digits
at $\delta=10^{-k}$. We therefore use
$\digitsBase+\digitsPerDecade k$ digits.
Repeating the calculation at higher precision
for $(\beta,\delta)=(\ctrlBeta,\ctrlDelta)$
changes the physical colour sum by less
than $\precisionCtrl$.
We extract the forward coefficients by fitting
the $J+P+2$ points closest to $\delta=0$ to
\begin{equation}
\sum_{j=0}^{J}f_j\G^j+p_T^2\sum_{l=0}^{P}h_l\G^l,
\label{eq:fit}
\end{equation}
with $(J,P)=(3,2)$ for $\M_1$ and
$(2,2)$ for $\M_2$. We assess the stability
of the fitted coefficients by shifting the fit
window one point towards larger $\delta$
(Sec.~\ref{app:fits}). Omitting the
$p_T^2$ terms increases the deviations of
the leading coefficients to approximately
$\plainDevMax$.

\subsection{Leading logarithms by colour factor}

Let $c_{ij}$ denote the coefficient of $\G^j$
in $A_i^{(2)}$ at $\mu_R=m_H$.
The four leading DPS logarithmic coefficients
are $c_{13}$, $c_{12}$, $c_{22}$ and $c_{21}$.
The one-loop coefficients are
$c_1=C_1/m_H^2=-2\pi NF_S(x)$ and
$K_1(x)=K/m_H^2=\pi NW(x)$.
For gluon splittings, $\DD=C_A=N$ and
$B=2\beta_0$, with
$\beta_0=\frac{11}{3}N-\frac23n_f$.
Equation~(\Lref{eq:rg}) then gives the
following predictions for the separate
colour structures:
\begin{equation}
\begin{aligned}
&\G^3\ (A_1):\ \ N^2\!:\ \tfrac83\pi F_S;\qquad
\G^2\ (A_1):\ \ N^2\!:\ \tfrac{44}3\pi F_S,\quad Nn_f\!:\ -\tfrac83\pi F_S;\\
&\G^2\ (A_2):\ \ N^2\!:\ -4\pi W;\qquad
\G\ (A_2):\ \ N^2\!:\ -\tfrac{32}3\pi W,\quad Nn_f\!:\ \tfrac83\pi W,
\end{aligned}
\label{eq:colourpred}
\end{equation}
All other colour coefficients at these
logarithmic powers vanish.
The ratios to the one-loop coefficients agree
with the $gg\to AA$ threshold results of
Eq.~(4.5) in Ref.~\cite{P1}; the corresponding
equal-helicity forward relation is given in
Eq.~(4.16) of that reference.

Table~\ref{tab:colour} compares the predictions
with the extracted coefficients at
$\beta=\colourTableBeta$.
Across all \velocityCount{} velocities,
the largest relative deviation for a
nonzero coefficient is $\devLeadMax$,
smaller than the maximum fit-window change
of $\shiftMax$. Coefficients predicted
to vanish are below $\vanishMax$ of the
leading coefficients.
In particular, neither the $n_f/N$ nor
the $n_f^2$ colour structure contributes
to these leading logarithms. A term
proportional to $C_Fn_f$, with
$C_F=(N^2-1)/(2N)$, would generate an
$n_f/N$ contribution. Its absence confirms
that $n_f$ enters the leading logarithms
only through $\beta_0$.
Since the colour structures are tested
separately, the agreement is established
at the level of the $N_c$ and $n_f$
dependence, rather than for a single
numerical choice of these parameters.

\input{tables/table_colour.tex}

\subsection{The one-loop amplitude from the scale dependence}
\label{sec:hhoneloop}

Renormalisation group invariance of
$C_H^2(\M^{(0)}+a_s\M^{(1)}+a_s^2\M^{(2)})$,
where $C_H$ starts at order $a_s$, gives
(cf.\ Eqs.~(B.1) and (B.3) of Ref.~\cite{P1})
\begin{equation}
\frac{\partial\M^{(1)}}{\partial\ln\mu_R^2}=2\beta_0\M^{(0)},\qquad
\frac{\partial\M^{(2)}}{\partial\ln\mu_R^2}=3\beta_0\M^{(1)}+\varrho\,\M^{(0)},
\end{equation}
where $\varrho$ is a scale-independent coefficient
whose expression for $gg\to AA$ is given in
Eq.~(B.3) of Ref.~\cite{P1}.
Integrating these equations with
$L=\ln(m_H^2/\mu_R^2)$ yields
\begin{equation}
\M^{(2)}=\M^{(2)}\big|_{L=0}
-\big[3\beta_0\M^{(1)}\big|_{L=0}
+\varrho\,\M^{(0)}\big]L
+3\beta_0^2\M^{(0)}L^2,
\label{eq:Ldep}
\end{equation}
The coefficient of $L$ thus determines
the one-loop amplitude up to a term
proportional to the tree amplitude, while
the coefficient of $L^2$ is fixed entirely
by the tree amplitude. For $\M_1$, the
tree amplitude is independent of $\G$
at fixed $s$; for $\M_2$, it vanishes as
$p_T^2$. Consequently, the coefficient
of $\G$ in the $L$ term determines the
one-loop DPS logarithm, while the forward
limit of the $L$ term in $\M_2$ determines
the one-loop constant.
The extracted coefficients agree with
Eqs.~(\ref{eq:FS}) and~(\ref{eq:K1})
to $\devOneLoopMax$, with fit-window
changes of at most $\oneLoopShiftMax$.
The $n_f$-dependent part of the one-loop
amplitude contributes to neither coefficient.
The coefficient of $L^2$ reproduces
$3\beta_0^2\M_1^{(0)}$, with
$\M_1^{(0)}=-is/2$, point by point
to $\treeMax$.
Using these independently extracted one-loop
coefficients, we recover the logarithmic
ratios of Eq.~(\Lref{eq:rg}) at all three
velocities. The numerical checks are
in the released code.

\subsection{Scale independence of the leading logarithms}
\label{sec:scale}

Eq.~(\ref{eq:Ldep}) implies that
the coefficients of $\G^3$ and $\G^2$
in $\M_1^{(2)}$, and of $\G^2$ and
$\G$ in $\M_2^{(2)}$, are independent
of $\mu_R$. Neither the $L$ nor the
$L^2$ term contributes to these
logarithmic powers. The largest
coefficient predicted to vanish is
$\absentMax$ relative to the
corresponding leading coefficient.
We verify this directly at
$\mu_R^2=m_H^2/2$, $m_H^2$ and $4m_H^2$.
The extracted leading coefficients
agree across these scales to
$\scaleSpread$.
At leading DPS power, among terms containing
positive powers of $\G$, only the coefficient
of $\G$ in $\M_1^{(2)}$ depends on $\mu_R$.
Eq.~(\ref{eq:Ldep}) gives its slope
$-3\beta_0c_1$ with respect to $L$,
in agreement with the fits to $\scaleSlopeDev$.
The same relation holds for $gg\to AA$
at threshold. In Eq.~(4.6) of
Ref.~\cite{P1}, the terms proportional
to $L_\mu=\ln(m_A^2/\mu_R^2)$ in
$\widehat A_{11}$ combine to give
$-3\beta_0\cdot7\pi N_c\,L_\mu$.

\section{Renormalisation and IR subtraction}
\label{sec:subtraction}

We distinguish the DPS logarithms present in the
unrenormalised two-loop amplitude from those generated
by UV counterterms and IR subtraction.
At leading DPS power, the one-loop contribution is
homogeneous in $p_T$. In $d=4-2\epsilon$, using the
conventions of Ref.~\cite{Banerjee:2018lfq}, it has
the form
\begin{equation}
\M_1^{(1)}\supset
\Big(\frac{p_T^2}{\mu_R^2}\Big)^{-\epsilon}
\Big[-\frac{C_\ell}{\epsilon}+a_0
+a_1\epsilon+a_2\epsilon^2\Big],\qquad
\M_2^{(1)}\supset
\Big(\frac{p_T^2}{\mu_R^2}\Big)^{-\epsilon}
\big[K+k_1\epsilon+k_2\epsilon^2\big],
\label{eq:regionexp}
\end{equation}
where higher orders in $\epsilon$ are omitted.
With $\ell=\ln(p_T^2/\mu_R^2)=2\G+\text{const}$,
the finite term in $\M_1^{(1)}$ is
$C_\ell\ell+a_0$, giving $C_1=2C_\ell$.
The coefficients $a_i$ and $k_i$ depend on the
continuation of the helicity amplitudes to
$d$ dimensions, including the Levi-Civita
prescription for $AA$. We leave them symbolic.
The UV counterterm acting on the
one-loop amplitude is
$-(2r_{\mathcal O}+\beta_0)\M^{(1)}/\epsilon
=-3\beta_0\M^{(1)}/\epsilon$,
where $r_{\mathcal O}=\beta_0$ for each
effective operator [Eq.~(2.23) of
Ref.~\cite{Banerjee:2018lfq}].
The two-loop finite remainder also contains
the subtraction $-2I^{(1)}\M^{(1)}$, with~\cite{Catani:1998bh,Banerjee:2018lfq}
\begin{equation}
I^{(1)}=-\frac{e^{\epsilon\gamma_E}}{\Gamma(1-\epsilon)}
\Big(\frac{C_A}{\epsilon^2}+
\frac{\beta_0}{2\epsilon}\Big)
\Big(-\frac{\mu_R^2}{s}\Big)^\epsilon.
\label{eq:I1}
\end{equation}
Terms proportional to $\M^{(0)}$, including
$4I^{(2)}\M^{(0)}$, do not generate leading
DPS logarithms: $\M_1^{(0)}$ is independent
of $p_T$, while $\M_2^{(0)}=O(p_T^2)$.
Expanding
$(p_T^2/\mu_R^2)^{-\epsilon}
=\sum_n(-\epsilon\ell)^n/n!$
and retaining the finite terms from the
indicated poles gives
\begin{align}
\frac{2C_A}{\epsilon^2}
\cdot\frac{C_\ell\epsilon^2\ell^3}{6}
&=\frac{C_A}{3}C_\ell\ell^3
\ \to\ \frac43C_AC_1\G^3,\nonumber\\
\Big(-\frac{3\beta_0}{\epsilon}
+\frac{\beta_0}{\epsilon}\Big)
\Big(-\frac{C_\ell\epsilon\ell^2}{2}\Big)
&=\beta_0C_\ell\ell^2
\ \to\ 2\beta_0C_1\G^2,\nonumber\\
\frac{2C_A}{\epsilon^2}
\cdot\frac{K\epsilon^2\ell^2}{2}
&=C_AK\ell^2
\ \to\ 4C_AK\G^2,\nonumber\\
\Big(-\frac{3\beta_0}{\epsilon}
+\frac{\beta_0}{\epsilon}\Big)
(-K\epsilon\ell)
&=2\beta_0K\ell
\ \to\ 4\beta_0K\G.
\end{align}
The terms $2C_A/\epsilon^2$ and
$\beta_0/\epsilon$ arise from $-2I^{(1)}$,
while $-3\beta_0/\epsilon$ comes from
UV renormalisation.

Table~\ref{tab:stage2a} compares these
contributions with the two-loop finite
remainder. The calculation isolates the
highest-power $C_A$ terms in $\M_1$ and
$\M_2$, together with the $\beta_0$ terms
in the next logarithmic powers. In all
four cases, the counterterm and subtraction
contributions are equal in magnitude and
opposite in sign to the finite-remainder
coefficients. The corresponding terms in
the unrenormalised, unsubtracted two-loop
amplitude at $O(\epsilon^0)$ therefore
have twice the finite-remainder coefficients.
For example, the UV counterterm
contributes $+3\beta_0C_1$ to the
$\G^2$ coefficient of $\M_1$, while
IR subtraction contributes
$-\beta_0C_1$. Their sum,
$+2\beta_0C_1$, has the opposite sign
to the finite-remainder contribution
$-2\beta_0C_1$. Thus the quadratic
logarithm is already present in the
unrenormalised two-loop amplitude;
it is not generated solely by subtraction.
The remaining $C_A$ contributions to
$\G^2$ in $\M_1$ and $\G$ in $\M_2$
also involve the higher-order
$\epsilon$ terms of $I^{(1)}$,
the coefficients $a_0,k_1$, and the
constant in $\ell=2\G+\text{const}$.
These terms are not isolated in
Table~\ref{tab:stage2a}. The separation
of bare-amplitude and subtraction
contributions is distinct from a finite
change of IR subtraction scheme,
which leaves the four complete leading
logarithmic coefficients unchanged
(Sec.~\ref{sec:rg}).

The same analysis applies to $gg\to AA$,
where the coupling and effective operators
have the same renormalisation structure~\cite{P1}. Converting the convention
$d=4+\epsilon'$ used there through
$\epsilon'=-2\epsilon$ reproduces
the finite contributions above.

\input{tables/table_stage2a.tex}

\section{Quark splittings and two-loop \texorpdfstring{$gg\to ZZ$}{gg -> ZZ}}
\label{sec:quarks}

\subsection{Prediction}
\label{sec:quarkpred}

For $g\to q\bar q$, each incoming gluon splits into a
quark-antiquark pair, and partons from the two splittings
produce the final colour singlets
[Fig.~\Lref{fig:dps}(b)]. The relevant colour and
non-cusp anomalous dimension combinations are
$\DD=2C_F-C_A=-1/N_c$ and
$B=\beta_0+\DG=\beta_0-\beta_0+6C_F=6C_F$.
For equal final-state masses, the scale
$Q_*^2=s^{N_c^2}/m^{2(N_c^2-1)}$ differs from
the gluon-splitting scale $m^4/s$.
The logarithm in Eq.~(\Lref{eq:rg}) is therefore
\begin{equation}
\begin{aligned}
\G_q&=\ln\frac{p_T}{Q_*}+\frac{i\pi}{2}
=\G-\frac{C_A-C_F}{\DD}\,\ln[x(1-x)]\\
&=\G+\frac{N_c^2+1}{2}\,\ln[x(1-x)],
\end{aligned}
\label{eq:Gq}
\end{equation}
where $\G$ is defined in Eq.~(\ref{eq:Gdef})
and $x=(1+\beta)/2$.

At leading power, the sum of ultrasoft gluon
attachments to the parent and daughter partons
reduces to the eikonal coupling of the parent,
as shown in Sec.~\ref{app:boundary}.
This follows from Ward identities and colour
conservation for both $g\to gg$ and
$g\to q\bar q$.
Together with the two-loop amplitude comparisons,
the result supports $r_1=r_2=0$ for quark splittings.
With this boundary condition, a helicity amplitude
whose one-loop DPS term is $C_1\G$ contains
\begin{equation}
\begin{aligned}
\M^{(2)}&\supset
C_1\Big[\frac{4}{3N_c}\,\G_q^3
-6C_F\,\G_q^2\Big]\\
&=C_1\Big[\frac{4}{3N_c}\,\G^3
+\Big(\frac{2(N_c^2+1)}{N_c}
\ln[x(1-x)]-6C_F\Big)\G^2\Big]+O(\G).
\end{aligned}
\label{eq:quarks}
\end{equation}
Thus the quadratic logarithmic coefficient,
when expressed in terms of $\G$, depends on
the velocity through $x$.
The coefficient $C_1$ includes both orientations
of the quark loop, as established in
Sec.~\ref{sec:quarkloops}.

For helicity amplitudes with a one-loop DPS
forward constant $K$, the second relation of
Eq.~(\Lref{eq:rg}) gives
$K[\frac4{N_c}\G_q^2-
(\frac4{N_c}+12C_F)\G_q]$.
Expressed in $\G$, the coefficient of the
single logarithm is
$\frac{4(N_c^2+1)}{N_c}\ln[x(1-x)]
-\frac4{N_c}-12C_F$.
For spin-one final states, these expressions
apply to helicity projections with definite
angular-momentum difference $\Delta J_z$,
as required by Eq.~(\Lref{eq:Fm}).
The quark channel provides a direct test of
the colour structure of the kernel.
For gluon splittings, $\DD=C_a=C_A$, so
replacing $\DD$ by the incoming-parton Casimir
would not change the prediction. For quark
splittings, however, it would give a cubic
coefficient $-4C_A/3$ instead of
$4/(3N_c)$.

Analytic two-loop amplitudes are available
for $gg\to ZZ$ and $gg\to\gamma^*\gamma^*$~\cite{vonManteuffel:2015msa,Caola:2015ila},
and DPS singularities occur in some of the
corresponding master integrals~\cite{Henn:2014lfa}.
Under the same boundary condition,
Eq.~(\ref{eq:quarks}) also predicts the
leading logarithms for
$gg\to\gamma^*\gamma^*$ with equal
invariant masses. We do not test that
channel separately. The extension to
different invariant masses, for which
$Q_*$ and the splitting fractions change,
is tested using $gg\to V_1^*V_2^*$
amplitudes in Sec.~\ref{sec:unequal}.

\subsection{Test in two-loop \texorpdfstring{$gg\to ZZ$}{gg -> ZZ}}
\label{sec:zz}

We evaluate the massless quark loop two-loop
$gg\to ZZ$ amplitudes of
Ref.~\cite{vonManteuffel:2015msa}, using
the published $q_T$-scheme finite remainders
at $\mu^2=s$. The 20 tensor coefficients
are reconstructed from nine independent
coefficients using crossing relations.
We check their evaluation against the
compiled C++ expressions and a separate
FORM implementation (Sec.~\ref{app:eval}).
We retain the $C_F$ and $C_A$ contributions
separately. To use the expansion parameter
$a_s=\alpha_s/(4\pi)$, each published
$n$-loop coefficient in the
$\alpha_s/(2\pi)$ convention is multiplied
by $2^n$. There is no additional $n_f$
contribution at this order.
Finite changes of IR subtraction
scheme or renormalisation scale add terms
proportional to lower-order amplitudes,
with coefficients independent of $p_T$.
They do not affect the leading logarithms
considered here (Sec.~\ref{sec:rg}).

For a helicity amplitude
$\M_{\lambda_1\lambda_2\lambda_3\lambda_4}$,
the angular momentum difference is
$\Delta J_z=(\lambda_1-\lambda_2)
-(\lambda_3-\lambda_4)$.
We test two groups: $\hel++++$ and
$\hel++--$ with their parity partners,
and $\hel+-++$ and $\hel+---$
with their parity partners.
We use $\beta=\zzBetas$ and approach the
forward limit along
$-t/m_Z^2=v+\delta$, where
$v=(1-\beta)/(1+\beta)$, with
$\delta=\zzDeltaFirst,\ldots,\zzDeltaLast$.
The amplitudes are evaluated with
$\zzDigitsBase+\zzDigitsPerDecade k$
decimal digits at $\delta=10^{-k}$.
Increasing the precision at
$(\beta,\delta)=(\zzCtrlBeta,\zzCtrlDelta)$
by \zzCtrlExtraDigits{} digits changes
the result by less than $\zzPrecisionCtrl$.
We use the fit of Eq.~(\ref{eq:fit}),
with $(J,P)=(1,2)$ at one loop
and $(3,3)$ at two loops, requiring
five and eight forward points,
respectively. The fit-window shift
tests numerical stability.
For the first helicity group, reducing
$P$ to 2 at two loops increases the
leading-logarithm deviations to
$\zzLowerModelDev$ in units of $C_1$.
For the second group, $P=2$ gives a
smaller window shift. We retain $P=3$
for a uniform comparison and examine
the model dependence in Sec.~\ref{app:fits}.
\input{tables/table_zz.tex}
At one loop, $\hel++++$ and $\hel++--$
and their parity partners have
$|C_1|=16\pi x^2$ and
$16\pi(1-x)^2$, respectively.
The forward constants of $\hel+-++$
and $\hel+---$, together with their
parity partners, have
$|K|=8\pi x(1-x)$.
These results agree with the crossed-box
calculation including both fermion-loop
orientations (Sec.~\ref{sec:quarkloops})
and Ref.~\cite{Gaunt:2011xd}.
The closed forms agree with the evaluated
amplitudes to $\zzOneLoopDev$ at all
\zzVelocityCount{} velocities.
The numerical results also show the approach
to the forward limit from $90^\circ$:
$\partial\M_{\hel++++}/\partial\G\to C_1$
and $\M_{\hel+-++}\to K$.
For the latter, the ratio
$(\M^{(1)}-K)/(Kp_T^2)$ remains
constant to \zzFwdDigits{} digits
over \zzFwdDecades{} orders of magnitude
in $p_T^2$.

The first and second helicity groups have
$\Delta J_z=0$ and $\Delta J_z=\pm2$,
corresponding to the $\nu=0$ and
$\nu=\pm2$ components of
Eq.~(\Lref{eq:Fm}).
The amplitudes $\hel++00$ and
$\hel+-+-$, together with their
parity partners, approach nonzero
constants but have no DPS logarithm.
Other even-$\Delta J_z$ DPS terms
predicted to vanish are below
$\zzVanishMax$ relative to the largest
retained coefficient. Amplitudes with
odd $\Delta J_z$ vanish as powers of
$p_T$, up to logarithms.
For the first helicity group,
$\M^{(1)}\supset C_1\G$; for the second,
$\M^{(1)}\to K$. Substituting
$\DD=2C_F-C_A$, $B=6C_F$ and
$\G_q$ from Eq.~(\ref{eq:Gq}) into
Eq.~(\Lref{eq:rg}) gives the
following two-loop predictions:
\begin{equation}
\begin{aligned}
\M^{(2)}&\supset C_1\Bigl\{C_F\bigl[-\tfrac83\G^3-(4L_x+6)\G^2\bigr]
+C_A\bigl[\tfrac43\G^3+4L_x\G^2\bigr]\Bigr\},\\
\M^{(2)}&\supset K\Bigl\{C_F\bigl[-8\G^2-(8L_x+4)\G\bigr]
+C_A\bigl[4\G^2+(8L_x-4)\G\bigr]\Bigr\},
\end{aligned}
\label{eq:zzpred}
\end{equation}
where $L_x=\ln[x(1-x)]$.
The factors of $1/\DD$ introduced
by the shift from $\G$ to $\G_q$
cancel in these coefficients.
Table~\ref{tab:zz} compares these predictions
with the fitted coefficients at all
\zzVelocityCount{} velocities.
The maximum deviations, in units of
the respective one-loop coefficients,
are $\zzLogDevMax$ for the first group
and $\zzKDevMax$ for the second,
both smaller than the corresponding
fit-window changes. The coefficient
of $\G^3$ in the second group is
predicted to vanish and is below
$\zzKcubicMax\,|K|$.
For $\hel++00$, $\hel+-+-$
and their parity partners, which have
no one-loop DPS logarithm, the
extracted coefficients of $\G^3$,
$\G^2$ and $\G$ are below
$\zzNoLogMax\,|K|$.
Their absence confirms the predicted
helicity dependence of the DPS terms.
The separate agreement of the $C_F$
and $C_A$ contributions tests the
predicted colour structure.
In particular, the cubic coefficient
is $4/(3N_c)$ rather than $-4C_A/3$,
distinguishing the quark-splitting
kernel from a dependence on the
incoming gluon Casimir alone.
The $-6C_F$ term in the quadratic
coefficient confirms $B=6C_F$,
including the cancellation of
$\beta_0$ against
$\gamma_g=-\beta_0$.
The terms proportional to
$\ln[x(1-x)]$ test the different
scale $Q_*$ for quark splittings;
using the gluon value
$Q_*^2=m^4/s$ does not reproduce
the fitted coefficients.

Finally, the agreement in the
$\Delta J_z=\pm2$ amplitudes
supports identifying their one-loop
forward constants with the DPS
contribution, as discussed in
Sec.~\ref{sec:K1complete}.

\begin{figure}[t]
\centering
\includegraphics[width=0.55\textwidth]{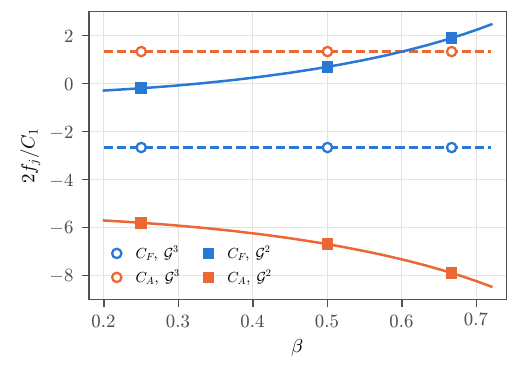}
\caption{The $C_F$ and $C_A$ contributions to the two-loop
$gg\to ZZ$ DPS logarithms in the $\Delta J_z=0$ component.
Lines are the predictions of Eq.~(\Lref{eq:rg}); symbols are
coefficients extracted from the complete amplitudes at the
three fitted velocities. The ratios use the one-loop coefficient $C_1$.}
\label{fig:zz_prediction}
\end{figure}

\subsection{Off-shell vector bosons with different invariant masses}
\label{sec:unequal}

We extend the massless quark loop test to
$gg\to V_1^*V_2^*$ with independent squared
invariant masses $a=m_1^2$ and $b=m_2^2$,
at $s>(\sqrt a+\sqrt b)^2$.
With $s+t+u=a+b$, the forward endpoint
and transverse momentum are
\begin{equation}
t_+=\frac{a+b-s+\sqrt{(s-a-b)^2-4ab}}{2},
\qquad
p_T^2=\frac{tu-ab}{s}.
\end{equation}
We define
\begin{equation}
\G_0=\frac12\ln\frac{sp_T^2}{ab}
+\frac{i\pi}{2},
\qquad
L_V=\ln\frac{\sqrt{ab}}{s}.
\end{equation}
For quark splittings, the logarithm entering
the kernel is
$\G_q=\G_0-(C_A-C_F)L_V/\DD$.

We use the $\alpha_s/(2\pi)$ expansion
of Ref.~\cite{vonManteuffel:2015msa}.
At the forward pinch, define momentum fractions
$x_1,x_2$ by $a=sx_1x_2$ and
$b=s(1-x_1)(1-x_2)$. The one-loop splitting
weights give
\[
c_1^{++++}=-8\pi i\,x_1(1-x_2),\qquad
c_1^{++--}=-8\pi i\,(1-x_1)x_2.
\]
We extract the one-loop coefficient of $\G_0$
independently from each helicity amplitude and
find agreement with these expressions to
$\uvCOneOne$ and $\uvCOneTwo$ in the
respective configurations. At two loops,
$f_{j,F}$ and $f_{j,A}$ denote the
coefficients of $C_F\G_0^j$ and
$C_A\G_0^j$, respectively.
With the boundary condition $r_1=r_2=0$,
Eq.~(\Lref{eq:rg}) predicts
\begin{equation}
\begin{aligned}
\frac{f_{3,F}}{c_1}&=-\frac43,&
\frac{f_{2,F}}{c_1}&=-2L_V-3,\\
\frac{f_{3,A}}{c_1}&=\frac23,&
\frac{f_{2,A}}{c_1}&=2L_V.
\end{aligned}
\label{eq:unequalpred}
\end{equation}

The factors differ from those in the
$a_s=\alpha_s/(4\pi)$ convention of the
Letter because the relative two-loop
normalisation is smaller by a factor of two.

We test two configurations,
$(a,b,s)=(1,2,15/2)$ and $(2,3,17)$,
using the published two-loop amplitudes
of Ref.~\cite{vonManteuffel:2015msa}.
For each, we evaluate nine forward points,
\[
t=t_+-\sqrt{ab}\,10^{-k},
\qquad k=4,\ldots,12,
\]
including all four crossing configurations
required to reconstruct the tensor
coefficients. We retain the $C_F$ and
$C_A$ contributions separately and fit
the complex $++++$ and $++--$ helicity
amplitudes without imposing the predicted
coefficients.
The two-loop fits contain a cubic
polynomial in $\G_0$ and a correction
proportional to
$\frac{p_T^2}{\sqrt{ab}}
\sum_{l=0}^{P}h_l\G_0^l$.
For $P=3$, the largest absolute deviations
of the fitted complex ratios from
Eq.~(\ref{eq:unequalpred}) are
$\uvLeadOne$ and
$\uvLeadTwo$ for the two
configurations, respectively.
With $P=2$, the corresponding deviations
reach $\uvPtwoOne$ and $\uvPtwoTwo$.
The $P=3$ fit-window changes are
$\uvWindowOne$ and $\uvWindowTwo$.
Repeating one point in each configuration
at 95-digit precision checks the
numerical accuracy of the amplitudes.
The central $P=3$ fit interpolates eight
points. We also compare shifted windows
and an overdetermined fit using all nine
points. These comparisons test numerical
stability but do not provide rigorous
error bounds.

The agreement in both configurations,
for two helicities and separately for
$C_F$ and $C_A$, supports the predicted
two-loop logarithms for vector bosons
with different invariant masses.
We also extract the single logarithm at two loops.
After converting the $q_T$ remainder to minimal
subtraction, subtracting the two hard-current
corrections at their respective invariant masses,
and removing the known kernel logarithms, the
finite matching agrees with the average
\[
r_0^{g\to q\bar q}(x_1,x_2)
\simeq\frac12\big[r_0^{g\to q\bar q}(x_1)
+r_0^{g\to q\bar q}(x_2)\big].
\]
The maximum complex deviations are
$\uvMatchingOne$ and $\uvMatchingTwo$.
The two independent mass configurations provide
numerical tests of this form. Its general validity
at arbitrary invariant masses requires further
analysis. The associated scripts, amplitude points
and fit-window comparisons are provided with the
numerical results.

\section{Finite matching from the two-loop amplitudes}
\label{sec:r0}

The analytic collinear calculation in Sec.~\ref{app:regions}
determines the finite boundary coefficient $r_0$.
It enters the coefficient of $\G$ in
$\M_1^{(2)}$ and the constant term
in $\M_2^{(2)}$.
We independently extract this coefficient from the complete
two-loop amplitudes by converting to a common IR subtraction
scheme and subtracting the known hard contributions.
The extracted values test Eqs.~(\Lref{eq:r0g})
and~(\Lref{eq:r0q}) at several momentum fractions.

\paragraph{Hard matching and scale dependence: }
Let $c_{11}$ denote the coefficient of $\G$
in $\M_1^{(2)}$, divided by $C_1$, and
$c_{20}$ the constant term of $\M_2^{(2)}$,
divided by $K$. In minimal subtraction,
Eqs.~(\Lref{eq:factorised}),
(\Lref{eq:kernel}) and~(\Lref{eq:Fm}) give
\begin{equation}
\begin{aligned}
c_{11}&=2h+\DD L_*^2+BL_*+r_0^{(\nu=0)},\\
c_{20}&=B-2\DD+2h+\DD L_*^2+BL_*+r_0^{(\nu=2)},
\end{aligned}
\qquad L_*=\ln\frac{\mu^2}{Q_*^2e^{-i\pi}},
\label{eq:r0read}
\end{equation}
where $\DD L_*^2+BL_*$ is the contribution
from the evolved kernel, and $B-2\DD$ arises
from the constant terms in the $\nu=2$
Fourier transforms.

For identical subprocesses,
$h=H^{(1)}/H^{(0)}$ is the one-loop
hard coefficient. For different subprocesses,
$2h$ is replaced by $h_1+h_2$.
The required hard functions are
$h=C_A(-L_m^2+\zeta_2)$ for an effective
$ggH$ vertex and
$h=C_F(-L_m^2+3L_m-8+\zeta_2)$
for $q\bar q\to Z$, where
$L_m=\ln(m^2/\mu^2)-i\pi$.
The $ZZ$ fits use $\G$ rather than
the quark-splitting logarithm $\G_q$.
To extract $r_0$ consistently, we express
the kernel in terms of the gluon-scale
variables $X_g,L_g$, defined with
$m^4/s$ in place of $Q_*^2$.
Writing $L_x=\ln[x(1-x)]$, the kernel
contribution becomes
\[
-\DD X_g^2+
[B-4(C_A-C_F)L_x]X_g+
\DD L_g^2-4(C_A-C_F)L_xL_g+BL_g.
\]
The constant from the $\nu=2$ Fourier
transform is then
$B-2\DD-4(C_A-C_F)L_x$.
We account for this change of logarithmic
variable before extracting the individual
colour coefficients of $r_0$.

\paragraph{IR-scheme conversion: }
The $HH$ amplitudes use Catani's
IR subtraction operator,
Eq.~(\ref{eq:I1}). The corresponding
finite remainders differ from the
minimally subtracted amplitudes by
$-[2I^{(1)}]_{\epsilon^0}\M^{(1)}$, with
\begin{equation}
[2I^{(1)}]_{\epsilon^0}=-C_AL_s^2+C_A\zeta_2-\beta_0L_s,\qquad L_s=\ln\frac{\mu^2}{s}+i\pi,
\label{eq:I1finite}
\end{equation}
At $\mu^2=s$, this coefficient is
$C_A(\pi^2+\zeta_2)-i\pi\beta_0$.

The published $ZZ$ amplitudes use the
$q_T$ subtraction scheme with
$\delta_{q_T}^{(0)}=0$~\cite{Catani:2013tia}.
Equation~(4.10) of
Ref.~\cite{vonManteuffel:2015msa}
gives the conversion to Catani's scheme.
In the published $\alpha_s/(2\pi)$
normalisation, the correction is
$(-\pi^2C_A/2+i\pi\beta_0/2)$ times
the one-loop amplitude, where $\beta_0$
is defined in our convention.
Converting to $a_s=\alpha_s/(4\pi)$
gives $-\pi^2C_A+i\pi\beta_0$.
Combining this conversion with
Eq.~(\ref{eq:I1finite}) gives
\[
-\pi^2C_A+i\pi\beta_0+
C_A(\pi^2+\zeta_2)-i\pi\beta_0
=C_A\zeta_2.
\]
Thus the $q_T$ finite remainder
equals the minimally subtracted
amplitude minus $C_A\zeta_2$
times its one-loop contribution.

\paragraph{Extracted finite coefficients: }
After the scheme conversion and subtraction
of the known terms in Eq.~(\ref{eq:r0read}),
the extracted imaginary parts are consistent
with zero for every tested colour and helicity
component, including the eight $ZZ$
helicity configurations.
The largest residual is $\glauberImag$
over \glauberValues{} extracted values
across \glauberProcs, approximately
$\glauberImagRatio$ of the corresponding
fit-window variation.
The extracted complex values agree with the closed forms
of Eqs.~(\Lref{eq:r0g}) and
~(\Lref{eq:r0q}) to $\rzFitHH$ for
$HH$ and $\rzFitZZ$ for $ZZ$.
The corresponding maximum fit-window
changes are $\rzBoundHH$ and
$\rzBoundZZ$, respectively.
Here $T_F=1/2$, $F_0=F_S$ and $F_2=W$.
The eight $ZZ$ helicity projections,
including parity partners, agree with
one another to $\rzZZspread$.
The results are in
Tables~\ref{tab:r0zz} and~\ref{tab:r0amp}.

\paragraph{Independent reconstruction and numerical tests: }
We reconstruct the analytic matching functions independently
from the complete two-loop amplitudes at the three
momentum fractions \rzScanX{}.
The ansatz is motivated by the one-loop
collinear functions and contains logarithms
of $x$ and $1-x$ of degree at most two,
together with the helicity overlaps.
For gluon splittings, we use
$c+a\ln x\ln(1-x)+b/F_\nu(x)$
for the $C_A$ contribution and
$c'+b'/F_\nu(x)$ for the $T_Fn_f$
contribution. The constant $c$ is common
to the two helicity components.
The three fractions determine $a,b$
to $\rzIdentRes$, while the third
fraction also tests the two-parameter
$T_Fn_f$ expression.
For quark splittings, the $C_F$
contribution is parametrised as
$c+a\ln^2[x/(1-x)]$ and the $C_A$
contribution as
$c+a[\ln^2x+\ln^2(1-x)]$.
Each contains two unknown coefficients,
with the third fraction providing
an additional check.

We use PSLQ with the basis
$\{1,\zeta_2\}$, integer coefficients
bounded by 200 and tolerance $10^{-8}$,
to identify the four constant terms.
The reconstructed values agree to
$\rzIdentC$. The chosen tolerance
exceeds the estimated numerical errors.
The basis assumes transcendental weight
at most two for these one-loop
matching functions. Enlarging it to
include $\ln2$ and $\zeta_3$ produces
near-relations with coefficients of
order 100 among the basis constants.
The reconstruction reproduces the matching functions
derived in Sec.~\ref{app:regions}.
Additional momentum fractions test the reconstruction
against complete amplitudes; numerical integration
separately checks the same region contributions.

We test the reconstructed functions at
kinematic points excluded from the fit:
$\beta=1/10,4/5$ for $HH$ and
$\beta=4/5$ for $ZZ$.
The deviations are $\rzBlindHH$ and
$\rzBlindZZ$, respectively, compared with
fit-window changes of $\rzBlindHHunc$
and $\rzBlindZZunc$.
These tests use additional kinematics
of the published amplitude implementations,
not independently calculated Feynman integrals.

\input{tables/table_r0_zz.tex}

\paragraph{Pseudoscalar amplitudes: }
The two-loop $gg\to AA$ amplitudes~\cite{Bhattacharya:2019oun}
provide a test of the parity dependence
of the finite matching.
We evaluate the forward kinematics of
Sec.~\ref{sec:method} using the code of
Ref.~\cite{P1}, with separate colour
contributions and the same fit degrees
as for $HH$ (Sec.~\ref{app:eval}).
At one loop, the amplitudes reproduce
$C_1=2\pi N_cm^2F_A$ and
$K=\pi N_cm^2W$ to $\rzAAoneLoop$,
with the $n_f$ contributions below
$\rzAAoneLoopNf$.
At two loops, the four leading
logarithmic coefficients agree with
Eq.~(\Lref{eq:rg}) to $\rzAAlead$
for the $N^2$ and $Nn_f$ colour
structures. The $n_f^2$ and $n_f/N$
coefficients are below $\rzAAleadVanish$
relative to the leading contribution.

The pseudoscalar hard function is
$h=C_A(-L_m^2+4+\zeta_2)$.
At $\mu^2=Q^2$, the renormalised
one-loop $G\widetilde G$ form factor
in $d=4+\epsilon$ is
$C_A\{-8/\epsilon^2+22/(3\epsilon)
+4+\zeta_2\}-8T_Fn_f/(3\epsilon)$~\cite{Ahmed:2015qpa}.
The corresponding $GG$ form factor
has $\zeta_2$ in place of
$4+\zeta_2$, with the same operator
renormalisation at this order.
The finite hard functions therefore
differ by $4C_A$.
Using the pseudoscalar hard function
in Eq.~(\ref{eq:r0read}) cancels the
extracted imaginary parts to
$\rzAAimag$ and gives
\begin{align}
r_0^{AA}\big|_{\nu=0}&=C_A\Big[\frac{134}9-5\zeta_2-4\ln x\ln(1-x)\Big]-\frac{40}9\,T_Fn_f
+\frac{2(C_A-2T_Fn_f)}{3F_A(x)},\notag\\
r_0^{AA}\big|_{\nu=2}&=r_0^{HH}\big|_{\nu=2}-\frac{4(C_A-2T_Fn_f)}{3W(x)}.
\label{eq:r0aa}
\end{align}
These expressions agree with the
amplitudes to $\rzFitAA$ at the three
scan velocities and to $\rzBlindAA$
at $\beta=\aaBlindBetas$.
Using the scalar hard function instead
would increase the extracted matching
coefficients in both components by $8C_A$.

For $\nu=0$, the rational correction
has the same numerator in the scalar
and pseudoscalar projections; the
difference arises from normalisation
by $F_S$ or $F_A$.
For $\nu=2$, the helicity-flip
contribution has the relative sign
of the hard helicity ratio
$M_{--}/M_{++}$, positive for
the scalar and negative for the
pseudoscalar interaction.
These rational contributions follow
from the one-loop splitting current
and its contraction with the hard
vertices, independently of the
$AA$ amplitude fits.
The five-velocity comparison tests
the difference
$-4(C_A-2T_Fn_f)/(3W)$
in the $\nu=2$ component to
$\rzAAparity$, and
$2(C_A-2T_Fn_f)(1/F_A-1/F_S)/3$
in the $\nu=0$ component to
$\rzAAoverlap$.

\paragraph{Analytic rational terms: }
The finite rational part of the
one-loop $g\to gg$ splitting current
contains spin-one and spin-three
transverse structures.
For $0<x<1$, its impact parameter
representation is
\begin{equation}
J_b^{\mathrm{rat}}
=\frac{C_A-2T_Fn_f}{3}(S_1-2T_4),
\label{eq:r0rational}
\end{equation}
where $S_1$ and $T_4$ are defined
in Sec.~\ref{app:regions}, with
$\kk$ replaced by $\bb$.
Contracting this current with the
tree splitting of the other incoming
gluon and the scalar or pseudoscalar
hard amplitudes gives
\begin{equation}
\begin{array}{c|cc}
&\nu=0&\nu=2\\ \hline
HH&2c/(3F_S)&2c/(3W)\\
AA&2c/(3F_A)&-2c/(3W)
\end{array},
\quad c=C_A-2T_Fn_f,
\label{eq:r0paritymatrix}
\end{equation}
with
$F_S=1/[x(1-x)]+2x(1-x)$,
$F_A=1/[x(1-x)]-2x(1-x)$
and $W=[x^4+(1-x)^4]/[x(1-x)]$.
These rational contributions follow
analytically from the one-loop
splitting amplitudes~\cite{Bern:1999ry,Kosower:1999xi}.
The remaining collinear contributions are evaluated
analytically in Sec.~\ref{app:regions}.
Together with the soft and parton self-energy terms,
they determine the complete finite matching functions.
Comparisons with complete amplitudes and numerical
integrations of the region contributions are
presented in Secs.~\ref{sec:r0} and~\ref{sec:r0regions}.
\input{tables/table_r0_amp.tex}
The common gluon constant obeys
\[
C_A\Big(\frac{134}9-5\zeta_2\Big)
-\frac{40}9T_Fn_f
=
2\frac{\gamma_{\rm cusp}^{(1)}}
{\gamma_{\rm cusp}^{(0)}}-C_A\zeta_2,
\]
where
$\gamma_{\rm cusp}
=\sum_n\gamma_{\rm cusp}^{(n)}a_s^{n+1}$.
The gluon constant can also be expressed
in terms of the two-loop cusp anomalous dimension.
The analytical matching coefficients are checked against
numerical region integrations in Sec.~\ref{sec:r0regions}.

\section{Finite matching from the momentum regions}
\label{sec:r0regions}

The analytic calculation of Sec.~\ref{app:regions}
determines the real finite matching coefficient $r_0$
from the collinear and soft contributions and parton
self-energy corrections at $\mu=\mu_b$.
Here we compare the analytical results with numerical
integrations of the same region contributions and
with coefficients extracted from complete two-loop
amplitudes in Sec.~\ref{sec:r0}.
Hard corrections are included in $\mathcal H_1$ and
$\mathcal H_2$, while the regulated static ultrasoft
contribution cancels as described in
Sec.~\ref{app:boundary}.

The collinear and soft functions and the daughter
self-energies are real to $\rzImag$. The Glauber
contribution produces the phase of
Eq.~(\Lref{eq:imK}). Its pole agrees with the
IR subtraction $2I^{(1)}_{\mathcal K}$,
obtained by subtracting the Catani operators of
the two $2\to1$ subprocesses from that of the
full $2\to2$ amplitude~\cite{Catani:1998bh}.
For each $q\bar q\to Z$ subprocess, the
corresponding operator uses $C_F$ and $3C_F/2$
in place of $C_A$ and $\beta_0/2$.
After IR subtraction, the imaginary part of
$\widetilde{\mathcal K}^{(1)}/\widetilde{\mathcal K}_0$
is $2\pi\DD X_\mu$. It vanishes at
$\mu=\mu_b$, giving $\operatorname{Im}r_0=0$.

The soft and collinear rapidity poles cancel analytically
for $0<x<1$. Numerical integrations of the same contributions at the
momentum fractions and their symmetry partners specified
in Sec.~\ref{app:regions} confirm this cancellation
to $\rzEta$. The remaining poles in $\epsilon$
agree with $2I^{(1)}_{\mathcal K}$ to $\rzPole$.
The finite terms reproduce
Eqs.~(\Lref{eq:r0g}) and~(\Lref{eq:r0q})
within $\rzRegG$ for gluon splittings and
$\rzRegQ$ for quark splittings, separately
for each colour contribution and angular
component (Table~\ref{tab:r0}).
The amplitude-based extraction independently
gives imaginary parts consistent with zero
to $\glauberImag$ over \glauberValues{}
values from \glauberProcs. The largest
residual is approximately $\glauberImagRatio$
of the corresponding fit-window variation
(Sec.~\ref{sec:r0}).
For quark splittings, the finite coefficient
is independent of the four combinations of
daughter helicities, in agreement with the
amplitude results. For gluon splittings,
the helicity dependence in
Eq.~(\Lref{eq:r0g}) arises from two rational
terms of the one-loop splitting current,
both proportional to $C_A-2T_Fn_f$.

For $\nu=0$, the contribution is
$\frac13x(1-x)(C_A-2T_Fn_f)$ times the
tree splitting in the helicity configuration
where both daughters have the helicity of
the parent gluon. For $\nu=2$, it comes
from a helicity-flip splitting absent at
tree level. Its sign relative to the tree
contribution is fixed by $M_{--}/M_{++}$,
positive for scalar and negative for
pseudoscalar hard vertices
(Sec.~\ref{sec:r0}).
These contributions agree with the rational
terms of the one-loop $g\to gg$ splitting
amplitudes in Eqs.~(4.32) and (4.33) of
Ref.~\cite{Bern:1999ry}. In that reference's
matter conventions, their coefficient at
$\epsilon=0$ is proportional to
$1-n_f/N_c+n_s/N_c$, with $n_f$ quark
flavours and $n_s$ complex scalars in the
fundamental representation.
For the four adjoint Weyl fermions and three
complex adjoint scalars of $\mathcal N=4$
super Yang--Mills theory, the corresponding
leading-colour contributions cancel as
$1-4+3=0$.
Consequently, the rational helicity-dependent
terms vanish in planar $\mathcal N=4$, and
the two angular components have the same
finite matching constant. This is consistent
with the amplitude checks in Sec.~\ref{sec:n4};
Sec.~\ref{sec:n4regions} gives the
corresponding region calculation.

\input{tables/table_r0.tex}

\section{Numerical evaluation and fits}
\label{app:eval}

We evaluate the published two-loop amplitudes
for each colour contribution and helicity
component separately. We describe the numerical
implementations and the extraction of the
forward coefficients, including tests of
precision, fit-window stability and dependence
on the fit model.

\subsection{\texorpdfstring{$gg\to HH$ and $gg\to AA$}{gg -> HH and gg -> AA}}

We evaluate the two-loop $gg\to HH$ finite remainders
of Ref.~\cite{Banerjee:2018lfq} from the published
ancillary expressions using \texttt{mpmath}~\cite{mpmath}.
The expressions contain logarithms, classical
polylogarithms and
\begin{equation}
\mathrm{Li}_{2,2}(a,b)=\sum_{m>n\ge1}\frac{a^mb^n}{m^2n^2}
=-a\int_0^1dt\,\frac{\ln t\,\mathrm{Li}_2(abt)}{1-at},
\label{eq:li22}
\end{equation}
where the integral representation is valid for
the arguments used here~\cite{Frellesvig:2016ske}.
As $ab\to1^-$, a branch point approaches the
integration endpoint. We resolve this behaviour
by subdividing the interval at successively
closer points to $t=1$ and using
\hhLiGuard{} guard digits.
Comparison with a separate integral representation
gives a maximum relative difference of
$\hhLiValueDev$ over all arguments in the main
scans, precision controls and additional
kinematic tests.
We check Bose symmetry under $y\leftrightarrow z$
at $v=1/3$ and $y=1/2$, finding relative
differences below $\boseDevMone$ for $\M_1$
and $\boseDevMtwo$ for $\M_2$.
The four colour structures of
Eq.~(\ref{eq:colourdec}) are evaluated separately.
Their $L$ and $L^2$ coefficients reproduce
the one-loop and tree amplitudes, respectively
(Sec.~\ref{sec:hhoneloop}).
Direct evaluation of the full expressions at
$N_c=3$ and $\mu_R^2=m_H^2/2$ agrees with
the reconstructed colour decomposition to
$\storedMaxDiff$ over \storedCount{} retained
kinematic points.

For $gg\to AA$, we use the streaming evaluator
of Ref.~\cite{P1} to process the two published
expression files of Ref.~\cite{Bhattacharya:2019oun},
of sizes 71 MB and 133 MB.
The $N^2$, $Nn_f$, $n_f^2$ and $n_f/N$
contributions are extracted in a single pass.
We set $m_A=\mu_R=1$ GeV and use the
one-loop normalisation of Ref.~\cite{P1};
the published two-loop expressions include
an overall factor of $64$.
The cancellations near the forward limit
cause a loss of approximately \aaLossFour{}
decimal digits at $\delta=10^{-4}$
and \aaLossEight{} at $\delta=10^{-8}$
for $\beta=1/2$. We therefore evaluate
the amplitudes with $50+9k$ digits at
$\delta=10^{-k}$ for $\beta\ge1/2$,
with 15 additional digits for $\beta=1/4$
and 30 additional digits for $\beta=1/10$.
Repeating the $k=10$, $\beta=1/2$
calculation with 15 additional digits
changes the colour contributions by less
than $\rzAActrl$.
The main scans use $\beta=\scanBetas$,
and the additional tests use
$\beta=\aaBlindBetas$, with
$\delta=\aaDeltaFirst,\ldots,\aaDeltaLast$.
Both helicity components are evaluated,
giving \aaPointCount{} amplitude evaluations.
We use the same fit degrees as for $gg\to HH$.

\subsection{\texorpdfstring{$gg\to ZZ$}{gg -> ZZ}}

We evaluate the massless quark loop amplitudes
of Ref.~\cite{vonManteuffel:2015msa} by
translating the published C++ expressions
into arbitrary precision Python.
Nine independent tensor coefficients determine
the full set of twenty through crossing relations.
The \texttt{p0002} special-function expressions
use the ordered $\mathrm{Li}_{2,2}$ arguments
of Ref.~\cite{Gehrmann:2015ora}.
At the benchmark point, all \zzLiCalls{}
calls satisfy $|a|<1$ and $|ab|<1$
in the specified order; only
\zzLiSwapped{} would satisfy these conditions
after interchanging the arguments.
Across the main scans, precision controls
and additional kinematic tests, the
\zzLiScanArgs{} arguments requiring the
integral representation agree with the
implementation used for $gg\to HH$,
with a loss of at most \zzLiScanDigits{}
digits relative to the working precision.
A separate transcription reproduces all
\zzPzeroStatements{} statements of
\texttt{p0002}.
We also reconstruct the tensor projectors
by inverting the $d$-dimensional
tensor-contraction matrix. At $d=6$,
the result agrees with the published
projectors to $\zzTensorDev$.

We perform two further checks of the
amplitude algebra.

First, the published C++ expressions,
compiled in quadruple precision,
agree with the Python translation to
$\zzCxxBenchOneLoop$ at one loop
and $\zzCxxBenchTwo$ at two loops
at the benchmark point.
Near the forward limit, at
$\beta=\zzCxxFwdBeta$ and
$\delta=\zzCxxFwdDelta$, the differences
are $\zzCxxFwdOneLoop$ at one loop,
and $\zzCxxFwdCF$ and $\zzCxxFwdCA$
for the two-loop $C_F$ and $C_A$
contributions, respectively.
The C++ comparison uses polylogarithm
values supplied by the Python implementation.
It therefore checks the algebraic translation
and coefficient assignments, while the
special functions are tested separately.

Second, an independent FORM-master integral
reader reproduces all \zzBenchOneLoopCount{}
one-loop tensor coefficients to
$\zzBenchOneLoop$.
For the selected two-loop coefficients
\zzBenchCFList{} and
\zzBenchCFListX{} after interchanging
$t$ and $u$, the differences are
$\zzBenchCF$ in the $C_F$ part
and $\zzBenchCA$ in the $C_A$ part.
At $\beta=\zzForwardCFBeta$ and
$\delta=\zzForwardCFDelta$, all
\zzForwardCFCount{} helicity amplitudes
agree to $\zzForwardCF$ for $C_F$
and $\zzForwardCA$ for $C_A$.
This second calculation uses separate
amplitude and master integral expressions,
but shares the kinematic parametrisation,
\texttt{mpmath} and the
$\mathrm{Li}_{2,2}$ implementation.
The crossed-master labels
\texttt{MN13m} and \texttt{MN13cm}
correspond to \texttt{N13x34x12}
and \texttt{N13x34}, respectively.
Interchanging these assignments spoils
the cancellation of inverse Gram
determinants in the $C_A$ contribution.

\subsection{Fits}
\label{app:fits}

\begin{figure}[t]
\centering
\includegraphics[width=0.91\textwidth]{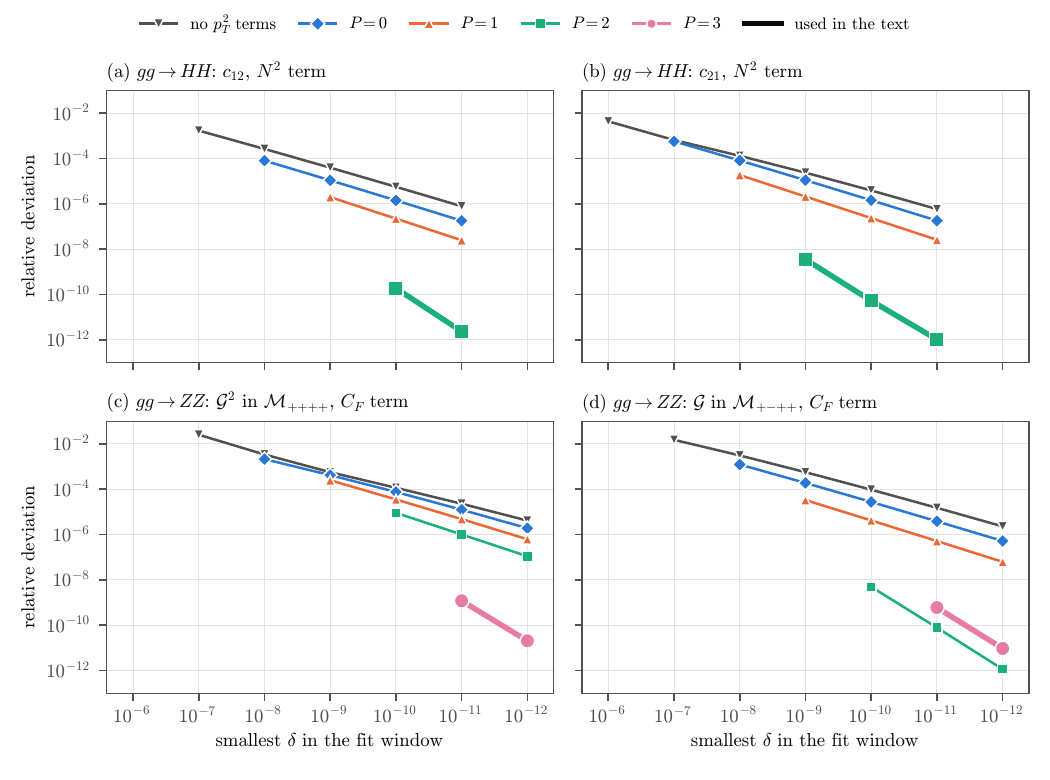}
\caption{Dependence of the extracted leading logarithmic
coefficients on the fit window and the degree of the
power correction in $gg\to HH$ and $gg\to ZZ$.
The plotted deviations are relative to the analytic
predictions; the fit windows contain the retained
amplitude points.}
\label{fig:fit_windows}
\end{figure}

For the $HH$ scans, $p_T^2/m_H^2$
ranges from $\pTsqMax$ to $\pTsqMin$
along lines of fixed $\beta$.
We extract the forward coefficients
using Eq.~(\ref{eq:fit}), including
terms proportional to $p_T^2$.
For each velocity, colour structure
and power of $L$, we use $P=2$
and solve the linear system defined
by the $J+P+2$ points closest to
$\delta=0$. The fitted polynomial
interpolates these points exactly.
Its residual therefore cannot estimate
the neglected higher-power terms.
Instead, we compare the fitted coefficients
with those obtained after shifting the
window one point towards larger $\delta$.
This change provides a numerical stability
test. The retained data and plotting scripts reproduce
Fig.~\ref{fig:fit_windows} with \texttt{make figure}
in the manuscript directory.
For the coefficients with known predictions,
the deviations are smaller than the
corresponding fit-window changes.
For the extracted finite matching coefficients
$r_0$, the largest ratios of these deviations
to the window changes are approximately
$\rzRatioHH$ for $gg\to HH$ and
$\rzRatioZZ$ for $gg\to ZZ$.
These comparisons test numerical stability;
they are not rigorous bounds on neglected
power corrections or systematic effects
common to neighbouring windows.

For the leading logarithmic coefficients
of Table~\ref{tab:colour}, the largest
window change is $\shiftMax$ relative
to the leading coefficient, compared with
a maximum deviation of $\devLeadMax$
from the prediction.
The deviations are calculated from the
modulus of the complex difference.
The imaginary parts, which vanish in
the predictions, are below $\imagMax$
relative to the corresponding coefficients.
We also allow one additional logarithmic
power, taking $J=4$ for $\M_1$
and $J=3$ for $\M_2$.
The additional coefficients remain below
$\extraPowerMax$ of the leading ones.
Omitting the $p_T^2$ corrections
in Eq.~(\ref{eq:fit}) increases the
deviations to approximately $\plainDevMax$,
demonstrating their importance in the
forward extrapolation.

To check the renormalisation scale
dependence, we fit the coefficient of
$L$ with $J=3$ for $\M_1$ and
$J=2$ for $\M_2$. The coefficients
of $\G^3$ and $\G^2$ in $\M_1$
and of $\G^2$ and $\G$ in $\M_2$
are consistent with zero, as required
by Sec.~\ref{sec:scale}.
The one-loop logarithmic coefficient
in $\M_1$ is then extracted with
$J=1$. For $\M_2$, we determine
the one-loop forward constant by fitting
a constant plus a term linear in $p_T^2$
through the two smallest values of
$\delta$. The coefficient of $L^2$
is checked against $3\beta_0^2\M^{(0)}$
at each kinematic point, without fitting.
The $gg\to ZZ$ fit parameters are given
in Sec.~\ref{sec:zz}.

The fit-window plots compare the results
as the window approaches $\delta=0$
and as the degree $P$ of the power-correction
polynomial is varied.
The tests cover the four leading
logarithmic coefficients in the $N^2$
part of $gg\to HH$, and the separate
$C_F$ and $C_A$ contributions for the
eight selected $gg\to ZZ$ helicity
configurations, at all \velocityCount{}
velocities.
Moving the window towards smaller
$\delta$ improves the agreement for
all fit models. Including the
$p_T^2$ corrections reduces the
deviations by several orders of magnitude.
Selecting the model by the smallest
window shift, without using the
theoretical predictions, favours
$P=2$ for $HH$, $P=3$ for the first
$ZZ$ helicity group and $P=2$
for the second.
We nevertheless use $P=3$ for both
$ZZ$ groups in the reported results.
For the second group, the differences
from the $P=2$ coefficients are smaller
than the $P=3$ window changes.
The reported coefficients agree with
the predictions within these changes.
Including more distant points in
least-squares fits increases the
influence of neglected power corrections.

\partheading{Part III. Further checks and higher-order results}

\section{Independent checks of the DPS pinch}
\label{sec:pinchchecks}

We also examine the DPS pinch and its local power counting
in Feynman parameter space. The calculations verify the
Symanzik polynomial identities and local pinch structure
for candidate scalings obtained from a separate momentum-space
survey. They provide independent algebraic checks, but do not
constitute a complete classification of momentum regions or
replace the regulated calculations of
Secs.~\ref{app:boundary}--\ref{app:regions}.

\input{pinch_checks.tex}

\section{Planar \texorpdfstring{$\mathcal N=4$}{N=4} amplitude checks}
\label{sec:n4}

The self-crossing singularity of planar six-gluon
amplitudes provides another setting in which to test
the DPS description~\cite{Dixon:2016epj}.
We compare the logarithmic coefficients, normalisation
and finite matching constants with the known
Sudakov/Bessel representation and its angular
components.
At the crossing,
$\delta_*=p_T^2/Q_*^2$, with
$Q_*^2=sxy(1-x)(1-y)$ and
$m_1^2=sxy$, $m_2^2=s(1-x)(1-y)$.
We compare the BDS-like normalised amplitude
$\mathcal E=A_6/A_6^{\rm BDS\mbox{-}like}$,
where the denominator is the modified
Bern-Dixon-Smirnov factor.
This differs from the remainder function $R_6$;
the two are related through exponentiation
and the function $Y$ of the BDS-like
normalisation~\cite{Dixon:2016epj}.

\paragraph{MHV input constants: }
In planar $\mathcal N=4$ super Yang-Mills theory,
the beta function vanishes, while the
non-cusp contribution to the kernel anomalous
dimension vanishes at one loop.
With $a=g^2N_c/(8\pi^2)$, the lowest-order
kernel coefficient satisfies
$a_s\DD=a/2=\gamma_K^{(1)}a/8$
in Eq.~(\Lref{eq:n4exp}).
Here $\gamma_K$ denotes the planar cusp
anomalous dimension, distinct from the
QCD kernel anomalous dimension
$\gamma_{\mathcal K}$.
The constants $\kappa_n$ include contributions
from the hard amplitudes and the BDS-like
normalisation as well as the kernel
(Sec.~\ref{sec:n4regions}).
In $3\to3$ kinematics, the phase cancels.
The singular MHV component has $\nu=0$
and one-loop coefficient $C_1=2\pi i$.
At fixed cross ratio $v$, the coefficients
of $\ln^k|\delta_*|$ are independent of $v$.

Using the MHV results summarised in
Table~\Lref{tab:n4letter}, we determine
$\kappa_1$ at two loops and
$B^{(L-1)},\kappa_{L-1}$ from the
double and single logarithms at $L\ge3$:
\begin{equation}
\begin{aligned}
\nFourConstants
\end{aligned}
\label{eq:n4const}
\end{equation}
These seven constants agree with the published
expressions of Ref.~\cite{CaronHuot:2019vjl}
after removing the additional cosmic normalisation
factor used in that reference.
In that reference, $a=2g^2$ and
$\gamma_K=2\Gamma_{\rm cusp}$.
Converting $\Gamma_{\rm virt}$ and $\Gamma_3$
to the BDS-like convention of
Ref.~\cite{Dixon:2016epj} reproduces
all seven constants.

\paragraph{MHV consistency checks: }

The leading logarithms reproduce the
Gaussian integral representation
\[
C_1\G\int_0^1dt\,e^{-2a\G^2t^2},
\]
obtained by Fourier transforming the
quadratic Sudakov exponent in $X$.
This agrees with the all-order result of
Ref.~\cite{Dixon:2016epj}.
At three loops, the coefficient of
$\G^3$ is
$-2(\kappa_1+\gamma_K^{(2)}/4)C_1/3$.
It vanishes for $\kappa_1=\zeta_2$,
consistent with the two-loop cusp
coefficient $\gamma_K^{(2)}=-4\zeta_2$.
The $\zeta_2$ and $\zeta_4$ terms at
four and five loops provide further
checks of the known cusp anomalous
dimension. The five-loop cubic logarithm
also determines $\gamma_K^{(4)}$,
in agreement with its known value.
The constants have uniform transcendental weights
$2n$ for $\kappa_n$ and $2n-1$ for $B^{(n)}$.
A linear one-loop boundary term would modify
the next-to-leading logarithms.
Of the 25 MHV coefficients through five loops,
one fixes the normalisation, seven determine the
constants in Eq.~(\ref{eq:n4const}), and one fixes
the four-loop cusp coefficient.
The remaining 16 independently test the
resulting representation and all agree.

We also check the analytic continuation
to $2\to4$ kinematics, where
$\G=\tfrac12\ln\delta_*+i\pi/2$
and $C_1=-2\pi i$.
The coefficients of $\ln^3\delta_*$,
$\ln^2\delta_*$ and $\ln\delta_*$
agree with Eqs.~(2.13), (2.32) and (3.10)
of Ref.~\cite{Dixon:2016epj}, using
$H^v_{\vec w}=H_{\vec w}(1-v)$
in that reference.
The quadratic coefficient $-\pi^2/2$
arises from the phase of $\G$.
The single-logarithm coefficient,
$-2i\pi^3/3$, is independent of $v$
and again gives $\kappa_1=\zeta_2$.

\paragraph{NMHV angular component: }
The transcendental functions entering the
NMHV amplitude remain finite at self-crossing~\cite{Dixon:2016epj}.
Nevertheless, the BDS-like ratio
$\rho=A_6^{\rm NMHV}/A_6^{\rm BDS\mbox{-}like}$
contains logarithms of $|\delta_*|$
from two loops onwards in the $v\to0^-$
overlap of $3\to3$ self-crossing and
multi-Regge kinematics.
These logarithms multiply
$(1+w)/(1+w^*)$, an angular factor
with $|\Delta J_z|=2$.
The corresponding coefficients through
four loops are given in Appendix D
of Ref.~\cite{Dixon:2016epj}, including
its erratum.
Using the same MHV matching constants,
the $\nu=2$ transform with one-loop
normalisation $K$ gives
$-\tfrac12\ln^2|\delta_*|+\ln|\delta_*|$
at two loops, in units of $K$.
The \nFourNmhvChecked{} coefficients
through four loops agree with the
polynomials in Eqs.~(D.15)--(D.17)
(Table~\Lref{tab:n4letter}).
Terms proportional to $\ln|v|$
are not included in this comparison.
The agreement includes the constant terms:
$\zeta_2-1$ at two loops,
$\zeta_3+\zeta_4/2+3$ at three loops,
and the weight-six combination at four loops.
No additional matching constant is introduced
for the NMHV component.

The relation between the two angular
components is also apparent from their
Fourier-transform generators.
Using
\[
e^{2\gamma_E\alpha}
\frac{\Gamma(1+\alpha)}{\Gamma(1-\alpha)}
=\exp\left[
-\sum_{k=3,5,\ldots}\frac{2\zeta_k}{k}\alpha^k
\right],
\]
both generators contain the common factor
\[
\exp\left[
-2\G\alpha
-\sum_{k=3,5,\ldots}\frac{2\zeta_k}{k}\alpha^k
\right],
\]
with prefactors $1/(2\alpha)$ for
$\nu=0$ and $1/(1-\alpha)$ for $\nu=2$.
The absence of the pole in the latter
gives the leading NMHV logarithms
$Ke^{-2a\G^2}$, without the
$1/(2n+1)$ factors of the MHV component.
Differentiating the $\nu=0$ generator
with respect to $\G$ removes its pole.
The identity
$1/(1-\alpha)=\int_0^\infty ds\,
e^{-s(1-\alpha)}$
then shifts $\G$ to $\G-s/2$,
yielding the integral relation quoted
in the Letter.

\paragraph{Known helicity map and five-loop comparison: }
This relation is consistent with the
established multi-Regge helicity map~\cite{Basso:2014bfkl,CaronHuot:2019vjl}.
Let $\xi=1+w$ and take $\xi\to0$
at fixed $v<0$.
The spectral momentum is $P=(2\nu,n)$,
with $P_\pm=2\nu\pm in$.
Here $\nu$ is the continuous
multi-Regge spectral variable, not
the angular index of the DPS transform.
At large $|P|$, the ratio of NMHV
to MHV impact factors approaches
$P_+/P_-$. For the selected leading
endpoint contribution, this changes
the angular dependence without changing
the leading radial behaviour.
It gives
$\partial_\xi A_{\rm NMHV}
=\partial_{\bar\xi}A_{\rm MHV}$.

Writing $\ell=\ln|\delta_*|$ and
$A_{\rm MHV}=2\pi i f(\ell)$,
$A_{\rm NMHV}=\pi i(\xi/\bar\xi)g(\ell)$,
we obtain
\begin{equation}
g(\ell)+g'(\ell)=2f'(\ell),
\qquad
\left(1+\tfrac12\partial_{\G}\right)
\frac{\M_2(\G)}K=
\partial_{\G}\frac{\M_1(\G)}{C_1}.
\label{eq:n4knownflip}
\end{equation}
The normalisation factors relating
the two descriptions are independent
of $\ell$ at fixed $v$.
This identity agrees with the
Bessel-transform relation and applies
to the leading endpoint component,
not to the complete NMHV amplitude
at general kinematics.
We further compare the nine five-loop
coefficients, including the constant
and odd zeta terms, with an independent
endpoint expansion of the published
multi-Regge amplitude~\cite{CaronHuot:2019vjl}.
This calculation uses the primary
multi-Regge expressions without
input from the DPS kernel.

Set $z=-w$, $t=1-z$,
$\bar t=1-\bar z$ and
$u=\ln(t\bar t)$.
The single-valued harmonic polylogarithms
are constructed using the
shuffle-regularised Drinfeld associator~\cite{Dixon:2012yy}.
Writing
$A=Z(x_0,x_1)\widetilde Z(y_0,y_1)$,
their generating series has the
endpoint expansion
$\mathcal L=e^{-u x_1}A+
t\mathcal L_{10}+\bar t\mathcal L_{01}
+\cdots$, where
\begin{equation}
(1+\partial_u)\mathcal L_{10}
=-x_1\mathcal L_{10}-x_0e^{-u x_1}A.
\label{eq:n4endpoint}
\end{equation}
The expansion through weight nine
requires the associator through weight
eight. After selecting the
$(1+w)/(1+w^*)$ component,
multiplication by
$\exp(\gamma_K\zeta_2/2)$ converts
the result to the BDS-like convention.
The coefficients independent of
$\ln|v|$ reproduce all
\nFourNmhvLower{} lower-loop and
\nFourNmhvFiveCount{} five-loop
entries without fitted parameters.
These comparisons test the known
multi-Regge representation and
helicity map in the self-crossing
overlap, providing an additional
check of their consistency with
the DPS angular decomposition.

\section{Finite matching in planar \texorpdfstring{$\mathcal N=4$}{N=4} theory}
\label{sec:n4regions}

We determine $\kappa_1$ from the momentum regions,
hard amplitudes and BDS-like normalisation, independently
of its extraction from the two-loop self-crossing
logarithm~\cite{Dixon:2016epj}. The calculation differs
from QCD in its particle content, regularisation scheme
and hard subprocesses.

\paragraph{Field content and regularisation: }
The four adjoint Weyl fermions contribute
$T_Fn_f=2C_A$ in the fermion-loop diagrams.
A complex adjoint scalar contributes
$-C_A/(3-2\epsilon)$ to the gluon self-energy
bracket of Eq.~(\ref{eq:sigma}) and adds a
scalar triangle to the collinear function.
Its vertex is $g_sf^{abc}(p+p')^\mu$, rather
than the ghost vertex $g_sf^{abc}p^\mu$,
and carries no closed-ghost-loop sign.
The seagull diagrams vanish after colour
contraction.
We use the four-dimensional helicity scheme
appropriate to the BDS-like normalisation.
Relative to conventional dimensional
regularisation, this introduces
$2\epsilon$ real adjoint scalars.
The scalar-loop multiplicity is therefore
$3+\epsilon$ complex scalars.
At $x=1/2$, explicit four-dimensional
metric traces reproduce this conversion
to $\nFourFDH$ for each tensor structure
through the required orders in $\epsilon$.
The total parton self-energy contribution
reduces to $-2C_A$, and $\beta_0=0$.

The two incoming splittings have independent
momentum fractions $x_1,x_2$.
We evaluate all \nFourRegPairs{} distinct
pairs formed from the five values
\nFourRegFracs.
At $\mu=\mu_b$, the kernel poles are
$C_A[2/\epsilon^2+2L_*/\epsilon]$, where
$L_*=-\ln[x_1(1-x_1)x_2(1-x_2)]$
in the units of Sec.~\ref{app:regions}.
Restoring the scales adds
$-\ln(|s_{36}|/\mu_b^2)$, so that
$L_*=-X$.
The pole coefficients agree with the
expected values to $\rzNfourPole$.
The rapidity poles cancel to
$\rzNfourEta$, the imaginary parts
are below $\rzNfourImag$, and the
two angular components agree to
$\rzNfourSpread$.
The numerical finite terms agree to
$\rzNfourClosed$ with
\begin{equation}
r_0^{\mathcal N=4}=C_A\big[-5\zeta_2-2\ln x_1\ln(1-x_1)-2\ln x_2\ln(1-x_2)\big].
\label{eq:r0n4}
\end{equation}
The scalar contributions cancel the
rational constant and helicity-dependent
terms present in the QCD result of
Eq.~(\Lref{eq:r0g}).
The latter has the supersymmetric
multiplicity $1-4+3=0$.
The remaining terms contain only
$\zeta_2$ and the logarithms of the
splitting fractions.
For $x_1=x$ and $x_2=1-x$,
these logarithms reduce to the
corresponding gluon-splitting expression
in Eq.~(\Lref{eq:r0g}).

\paragraph{Hard amplitudes and BDS-like normalisation: }
In the six-gluon self-crossing configuration,
the two hard subprocesses are $2\to2$
scatterings. In planar $\mathcal N=4$
super Yang--Mills theory, the one-loop
four-point amplitude divided by its
tree amplitude is universal for the
relevant external states~\cite{Bern:1994zx,Bern:2005iz}:
\begin{equation}
M_4^{(1)}(s,t)=-\frac1{\epsilon^2}\Big[\Big(\frac{\mu^2}{-s}\Big)^\epsilon+\Big(\frac{\mu^2}{-t}\Big)^\epsilon\Big]
+\frac12\ln^2\frac st+4\zeta_2,
\label{eq:m4}
\end{equation}
where $s,t$ are adjacent planar invariants
and the expansion parameter is
$a=g^2N_c/(8\pi^2)$.
For the final-state pairs $(1,2)$ and
$(4,5)$, the corresponding invariants
are $s_{12}$ and
$t_1=(p_2+(1-x)p_3)^2$, and
$s_{45}$ and $t_2=(p_4+xp_3)^2$,
respectively. The incoming gluons
$p_3,p_6$ split with fractions $x,y$
as in Fig.~1 and Eq.~(2.19) of
Ref.~\cite{Dixon:2016epj}.
The BDS-like normalised amplitude
$\mathcal E=A_6/A_6^{\rm BDS\mbox{-}like}$
includes a subtraction
$-\hat M_6/2$ at one loop.
The combination
$\hat M_6=M_6^{(1)}+Y$
depends only on two-particle invariants~\cite{Dixon:2016epj}.
Combining the one-loop box representation~\cite{Bern:1994zx} with $Y$ gives
\begin{equation}
\hat M_6=\sum_{i=1}^6\Big[-\frac1{\epsilon^2}\Big(\frac{\mu^2}{-s_{i,i+1}}\Big)^\epsilon
+\frac12\ell_i^2-\ell_i\ell_{i+1}+\frac12\ell_i\ell_{i+3}\Big]+9\zeta_2,\qquad
\ell_i=\ln\frac{-s_{i,i+1}}{\mu^2},
\label{eq:m6hat}
\end{equation}
where the dependence on the
three-particle invariants cancels.
In $3\to3$ kinematics, the relevant
two-particle invariants are negative.
The logarithm is
$X=\ln(|Q_*^2|b^2/b_0^2)$, with
$Q_*^2=s_{36}xy(1-x)(1-y)$,
and the terms above are real.

At two loops, the singular contribution
to $\mathcal E$, normalised to the
lowest-order DPS term
$a\widetilde{\mathcal K}_0$, is
\begin{equation}
h_1+h_2-\tfrac12\hat M_6+\frac{\widetilde{\mathcal K}^{(1)}}{\widetilde{\mathcal K}_0},
\label{eq:e2}
\end{equation}
where $h_1,h_2$ are the one-loop
hard corrections. The kernel is
evaluated at $\mu_b$, with
$a=2C_Aa_s$.

\paragraph{Pole cancellation and finite matching: }
The double poles cancel between the
two hard amplitudes, the BDS-like
subtraction and the kernel:
$-4+3+1=0$.
The kinematic identities
$s_{12}s_{45}
=xy(1-x)(1-y)s_{36}^2$
and
$s_{23}s_{34}s_{56}s_{61}
=t_1^2t_2^2/[xy(1-x)(1-y)]$
combine the remaining single poles
into $X/\epsilon$ from the hard
amplitudes and BDS-like subtraction,
and $L_*/\epsilon=-X/\epsilon$
from the kernel. These poles cancel.
The finite contribution is
\begin{equation}
-\frac{X^2}2+\ln x\ln(1-x)+\ln y\ln(1-y)+\frac72\zeta_2+\frac{r_0^{\mathcal N=4}}{2C_A}.
\label{eq:kappaone}
\end{equation}
The quadratic term is the cusp
contribution of Eq.~(\Lref{eq:n4exp}),
and the coefficient linear in $X$
vanishes, consistent with $B^{(1)}=0$.
The dependence on the hard scattering
angles, including the cross ratio $v$,
cancels.

Identifying $x_1=1-x$ and $x_2=1-y$
in Eq.~(\ref{eq:r0n4}), the logarithms
of the momentum fractions also cancel.
The remaining finite constant is
\begin{equation}
\kappa_1=\frac72\zeta_2-\frac52\zeta_2=\zeta_2.
\label{eq:kappaoneval}
\end{equation}
The numerical region calculation
reproduces this result to
$\kappaOneReg$ for all
\nFourRegPairs{} fraction pairs
and both angular components.
Thus $\kappa_1$ is obtained
without using the two-loop
self-crossing single logarithm
as input. The hard-amplitude
combination is evaluated analytically;
the finite kernel terms are checked
by numerical integration.

The calculation explicitly uses
gluon daughters. For the MHV
configuration with two negative-helicity
incoming gluons, the gluons in each
outgoing pair have equal helicities.
Quark or scalar daughters do not
contribute to these tree-level
hard subprocesses.
Other external-state configurations
are related by supersymmetric
Ward identities. The universality
of the four-point amplitude ratio
and the one-loop splitting correction~\cite{Bern:1994zx} is consistent
with the same finite matching
constant across the supermultiplet.

\paragraph{Continuation to $2\to4$: }
When the relevant invariants become
positive, the hard amplitudes and
$\hat M_6$ acquire phases through
$(\mu^2/(-s))^\epsilon$.
Continuing
$\ell_{12},\ell_{45}\to\ell-i\pi$
produces imaginary single poles.
These cancel only after including
the kernel's $i\pi/\epsilon$
contribution of Eq.~(\Lref{eq:imK}).
The finite result retains the form
of Eq.~(\ref{eq:kappaone}), now with
$X=\ln(Q_*^2b^2/b_0^2)-i\pi$.
It remains independent of $v$
and reproduces the two-loop
single-logarithm coefficient
$-2i\pi^3/3$ of
Ref.~\cite{Dixon:2016epj}.
This agreement follows with
$\kappa_1=\zeta_2$ and no
additional finite Glauber matching
constant. It provides a separate
check of the phase after IR
subtraction.

\section{Conditional three-loop logarithms}
\label{sec:threeloop}

The evolution equation gives predictions
for the leading three-loop DPS logarithms
under two conditions. First, the
factorised description must remain
valid and the boundary at relative
order $a_s^2$ must contain no powers
of $X$ higher than $X^2$.
Then the two highest logarithmic
coefficients are fixed in each
angular component.
If the coefficient of $X^2$
also vanishes, the third logarithmic
coefficient is determined.
These conditions concern the
higher-order boundary and are
not implied by renormalisation
group evolution alone.

\paragraph{Boundary sensitivity: }
To identify the possible contributions
not constrained by the evolution
equation, consider the additional
boundary term
\begin{equation}
\delta\widetilde{\mathcal K}(\mu_b)
=a_s^3\widetilde{\mathcal K}_0
(\rho_0+\rho_1X+\rho_2X^2+\rho_3X^3+\rho_4X^4).
\label{eq:boundaryhigher}
\end{equation}
This polynomial illustrates the
dependence on higher-order matching
coefficients; it is not an evaluated
three-loop QCD contribution.
A boundary term proportional to $X^4$
changes the highest logarithmic power:
$\G^5$ in $\M_1^{(3)}$ and
$\G^4$ in $\M_2^{(3)}$.
A term proportional to $X^3$ changes
the next-to-highest power:
$\G^4$ and $\G^3$, respectively.
A term $\rho_2X^2$ leaves these two
highest powers unchanged but shifts
the third logarithmic coefficients
by $4\rho_2/3$ for $\G^3$ in
$\M_1^{(3)}/C_1$, and by $4\rho_2$
for $\G^2$ in $\M_2^{(3)}/K$.
Under both boundary conditions,
the correction relative to the
lowest-order kernel reduces to
$a_sr_0+a_s^2(r_0^{(3)}+r_1^{(3)}X)$,
where $r_0^{(3)}$ and
$r_1^{(3)}$ remain unknown.

A sufficient condition for the
absence of a quadratic boundary
term is local rapidity evolution
with an $X$-independent generator,
whose one-loop contribution
vanishes as a colour operator
at $\mu_b$, together with
finite matching independent of $X$.
For a scalar evolution equation,
this has the form
\begin{equation}
\exp\left[X\sum_{n\ge1}a_s^n(\mu_b)
\gamma_{\rm rap}^{(n)}\right]
\left(1+\text{terms independent of }X\right),
\qquad \gamma_{\rm rap}^{(1)}=0.
\label{eq:rapboundary}
\end{equation}
The same argument applies to
matrix-valued evolution under
these assumptions. The generator
must vanish as an operator;
cancellation after contraction
with a particular hard amplitude
is insufficient. These conditions
have not been established for the
complete three-loop region system.
In particular, strict space-like
collinear factorisation can be
violated by spectator colour
correlations~\cite{Catani:2011spacelike,Forshaw:2012breaking}.
This does not rule out a joint
two-splitting kernel, but the
colour-singlet nature of the
hard vertices does not eliminate
intermediate colour mixing.
Glauber contributions must also
be treated with consistent
overlap subtractions~\cite{Rothstein:2016glauber}.

\paragraph{Evolution, hard functions and scheme: }
We evolve the kernel as in
Sec.~\ref{sec:rg}, including
the two-loop anomalous dimension
and coupling evolution through
$\beta_0$ and $\beta_1$.
Under the stated boundary conditions,
\[
\widetilde{\mathcal K}(\mu_b)
=a_s(\mu_b)\widetilde{\mathcal K}_0
\left[1+a_s(\mu_b)r_0
+a_s^2(\mu_b)
(r_0^{(3)}+r_1^{(3)}X)\right].
\]
The three highest logarithmic
coefficients in each angular
component then follow from
Eq.~(\Lref{eq:threeloop}).
For nonidentical hard subprocesses,
the term $8\DD h/3$ in that
equation becomes
$4\DD(h_1+h_2)/3$.
The third logarithmic coefficients
depend on the renormalisation scale
through $L_*$ and the hard functions,
and on the IR subtraction scheme.
At the displayed logarithmic powers,
conversion from minimal subtraction
to Catani's convention includes
$-[2I^{(1)}]_{\epsilon^0}$
times the two-loop amplitude.
Using Eq.~(\ref{eq:I1finite}),
this increases the coefficient of
$\G^3$ in $\M_1^{(3)}/C_1$
by
$4\DD[2I^{(1)}]_{\epsilon^0}/3$,
and that of $\G^2$ in
$\M_2^{(3)}/K$ by
$4\DD[2I^{(1)}]_{\epsilon^0}$.
Table~\Lref{tab:threeloop}
gives the minimal-subtraction
values for $gg\to HH$.
The next logarithmic powers,
$\G^2$ in $\M_1^{(3)}$
and $\G$ in $\M_2^{(3)}$,
depend on the two-loop non-cusp
anomalous dimensions, $\beta_1$
and the unknown boundary
coefficient $r_1^{(3)}$.
They are not fixed by the
present calculation.

For quark splittings, the
corresponding predictions use
$\DD=2C_F-C_A$, $B=6C_F$
and the finite coefficient
$r_0$ of Eq.~(\Lref{eq:r0q}).
Explicit three-loop $gg\to HH$
and $gg\to ZZ$ calculations
would test the predicted logarithms
and the higher-order boundary
conditions.

The released \texttt{three\_loop.py}
implements the evolution and Fourier
transforms. It reproduces
Eqs.~(\Lref{eq:rg}) and
~(\ref{eq:r0read}) at two loops
and agrees with the exponentiated
conformal limit.

\paragraph{Conclusion: }
The two-loop amplitude comparisons establish
the logarithmic coefficients in the channels
studied, while the separate momentum-region
calculations reproduce the finite matching
functions and Glauber phase.
The tests with off-shell vector bosons of different
invariant masses and the planar $\mathcal N=4$
comparisons provide further checks of the same structure.
Its evolution then fixes the stated
higher-order logarithms under the
specified boundary conditions.

\paragraph{Use of AI tools: }
ChatGPT (OpenAI) and Claude (Anthropic) assisted
with research notes,
code development and debugging, and
repository organisation. The implementations
were validated through symbolic checks,
separate amplitude-algebra calculations,
numerical precision controls and variations
of the fitting procedure. Shared dependencies
between checks are identified in the text.
The author reviewed the results and takes
responsibility for their accuracy and
scientific interpretation.

\bibliography{references}

%% file: numbers.tex
\newcommand{\devLeadMax}{\ensuremath{7\times10^{-12}}}

\newcommand{\shiftMax}{\ensuremath{6\times10^{-10}}}
\newcommand{\plainDevMax}{\ensuremath{10^{-6}}}

\newcommand{\scanBetas}{\ensuremath{\frac14,\frac12,\frac23}}

\newcommand{\zzDevMax}{\ensuremath{4\times10^{-10}}}
\newcommand{\zzShiftMax}{\ensuremath{3\times10^{-8}}}

\newcommand{\rzFitHH}{\ensuremath{5.4\times10^{-10}}}
\newcommand{\rzFitZZ}{\ensuremath{1.6\times10^{-9}}}
\newcommand{\rzBlindHH}{\ensuremath{1.7\times10^{-9}}}
\newcommand{\rzBlindZZ}{\ensuremath{4.3\times10^{-9}}}

\newcommand{\rzRegQ}{\ensuremath{8.4\times10^{-8}}}
\newcommand{\rzRegG}{\ensuremath{5.9\times10^{-8}}}

\newcommand{\kappaOneReg}{\ensuremath{2.1\times10^{-7}}}

\newcommand{\glauberImag}{\ensuremath{3.4\times10^{-9}}}

\newcommand{\rzFitAA}{\ensuremath{6.4\times10^{-10}}}

\newcommand{\rzAAlead}{\ensuremath{8.2\times10^{-12}}}
\newcommand{\rzAALogShift}{\ensuremath{7.3\times10^{-10}}}

\newcommand{\rzBlindAA}{\ensuremath{2.0\times10^{-9}}}

\newcommand{\aaBlindBetas}{\ensuremath{\frac{1}{10},\frac45}}

\newcommand{\nFourPass}{16}

\newcommand{\nFourNmhvChecked}{15}

\newcommand{\nFourRegPairs}{fifteen}

\newcommand{\glauberConfigs}{eighteen}

\newcommand{\glauberValues}{104}
\newcommand{\nFourNmhvFiveCount}{nine}

%% file: sm_nocite.tex
\nocite{Ahmed:2015qpa,Bauer:2001yt,Bauer:2002nz,Becher:2009qa,Bern:1994zx,Bern:2005iz,Binoth:2000ps,Buffing:2017mqm,Caola:2015ila,Catani:2011spacelike,Catani:2013tia,Coleman:1965xm,Dixon:1996wi,Dixon:2012yy,Echevarria:2011epo,Eden:1966dnq,Forshaw:2012breaking,Frellesvig:2016ske,Gardi:2009qi,Gardi:2024dissecting,Gehrmann:2015ora,Henn:2014lfa,Kosower:1999xi,Landau:1959fi,Mangano:1990by,Passarino:1978jh,Rothstein:2016glauber,mpmath}

%% file: tables/table_n4_letter.tex
\begin{table*}[!htb]
\centering
\small
\setlength{\tabcolsep}{5pt}
\begin{tabular}{@{}r@{\hspace{4pt}}r@{\hspace{6pt}}l@{\hspace{12pt}}r@{\hspace{4pt}}r@{\hspace{6pt}}l@{}}
\toprule
\multicolumn{3}{c}{\textbf{MHV}: $2\pi i\,\ell^k$} & \multicolumn{3}{c}{\textbf{NMHV}: $\pi i\,\frac{1+w}{1+w^*}\,\ell^k$}\\
\cmidrule(lr){1-3}\cmidrule(l){4-6}
$L$ & $k$ & coefficient & $L$ & $k$ & coefficient\\
\midrule
1 & 1 & $\tfrac{1}{2}$ & 1 & 0 & $1$\\
\addlinespace[3pt]
2 & 3 & $- \tfrac{1}{12}$ & 2 & 2 & $- \tfrac{1}{2}$\\
 & 2 & $0$ &  & 1 & $1$\\
 & 1 & $\tfrac{\zeta_{2}}{2}\,[\kappa_1]$ &  & 0 & $\zeta_{2} - 1$\\
\addlinespace[3pt]
3 & 5 & $\tfrac{1}{80}$ & 3 & 4 & $\tfrac{1}{8}$\\
 & 4 & $0$ &  & 3 & $- \tfrac{1}{2}$\\
 & 3 & $0$ &  & 2 & $\tfrac{3}{2}$\\
 & 2 & $- \tfrac{\zeta_{3}}{4}\,[B^{(2)}]$ &  & 1 & $- \zeta_{3} - 3$\\
 & 1 & $\tfrac{\zeta_{4}}{4}\,[\kappa_2]$ &  & 0 & $\zeta_{3} + \tfrac{\zeta_{4}}{2} + 3$\\
\addlinespace[3pt]
4 & 7 & $- \tfrac{1}{672}$ & 4 & 6 & $- \tfrac{1}{48}$\\
 & 6 & $0$ &  & 5 & $\tfrac{1}{8}$\\
 & 5 & $- \tfrac{\zeta_{2}}{80}$ &  & 4 & $- \tfrac{\zeta_{2}}{8} - \tfrac{5}{8}$\\
 & 4 & $- \tfrac{\zeta_{3}}{48}$ &  & 3 & $\tfrac{\zeta_{2}}{2} - \tfrac{\zeta_{3}}{6} + \tfrac{5}{2}$\\
 & 3 & $- \tfrac{7 \zeta_{4}}{24}$ &  & 2 & $- \tfrac{3 \zeta_{2}}{2} + \tfrac{\zeta_{3}}{2} - \tfrac{7 \zeta_{4}}{4} - \tfrac{15}{2}$\\
 & 2 & $- \tfrac{3 \zeta_{2} \zeta_{3}}{4} + \zeta_{5}\,[B^{(3)}]$ &  & 1 & $- 3 \zeta_{2} \zeta_{3} + 3 \zeta_{2} - \zeta_{3} + \tfrac{7 \zeta_{4}}{2} + 4 \zeta_{5} + 15$\\
 & 1 & $- \zeta_{3}^{2} - \tfrac{13 \zeta_{6}}{48}\,[\kappa_3]$ &  & 0 & $3 \zeta_{2} \zeta_{3} - 3 \zeta_{2} - 2 \zeta_{3}^{2} + \zeta_{3} - \tfrac{7 \zeta_{4}}{2} - 4 \zeta_{5} - \tfrac{13 \zeta_{6}}{24} - 15$\\
\addlinespace[3pt]
5 & 9 & $\tfrac{1}{6912}$ & & & \\
 & 8 & $0$ & & & \\
 & 7 & $\tfrac{\zeta_{2}}{336}$ & & & \\
 & 6 & $\tfrac{5 \zeta_{3}}{288}$ & & & \\
 & 5 & $\tfrac{9 \zeta_{4}}{80}$ & & & \\
 & 4 & $\tfrac{7 \zeta_{2} \zeta_{3}}{24} + \tfrac{\zeta_{5}}{4}$ & & & \\
 & 3 & $\tfrac{2 \zeta_{3}^{2}}{3} + \tfrac{115 \zeta_{6}}{72}\,[\gamma_K^{(4)}]$ & & & \\
 & 2 & $\tfrac{17 \zeta_{2} \zeta_{5}}{4} + \tfrac{11 \zeta_{3} \zeta_{4}}{4} - \tfrac{55 \zeta_{7}}{16}\,[B^{(4)}]$ & & & \\
 & 1 & $\tfrac{\zeta_{2} \zeta_{3}^{2}}{4} + \tfrac{45 \zeta_{3} \zeta_{5}}{4} + \tfrac{257 \zeta_{8}}{72}\,[\kappa_4]$ & & & \\
\bottomrule
\end{tabular}
\caption{Planar $\mathcal N=4$: the coefficients of the powers of $\ell=\ln|\delta_*|$ in the $L$-loop MHV amplitude $\mathcal E^{(L)}$, Eqs.~(3.17)--(3.21) of Ref.~\cite{Dixon:2016epj}, and in the next-to-MHV amplitude $\rho^{(L)}$, Eqs.~(D.14)--(D.17) of its erratum, in $3\to3$ kinematics, against Eqs.~(\ref{eq:kernel}), (\ref{eq:Fm}) and~(\ref{eq:n4exp}) with $\nu=0$ and $\nu=2$. The one-loop coefficients define $C_1$ and $K$; a coefficient followed by a constant in brackets fixes that constant, and the remaining coefficients check the known radial representation and helicity map. All agree.}
\label{tab:n4letter}
\end{table*}

%% file: tables/table_threeloop_letter.tex
\begin{table}[!htb]
\centering
\footnotesize
\setlength{\tabcolsep}{7pt}
\renewcommand{\arraystretch}{1.35}
\begin{tabular}{lccc}
\toprule
$x$ & $N_c^2$ & $N_cn_f$ & $n_f^2$\\
\midrule
\multicolumn{4}{@{}l}{$[\G^3]\,\M_1^{(3)}/C_1$}\\
$\frac58$ & $-0.008-58.231\,i$ & $-11.055+8.378\,i$ & $1.778$\\
$\frac34$ & $-4.510-60.101\,i$ & $-10.491+8.378\,i$ & $1.778$\\
$\frac56$ & $-10.719-62.615\,i$ & $-9.728+8.378\,i$ & $1.778$\\
\midrule
\multicolumn{4}{@{}l}{$[\G^2]\,\M_2^{(3)}/K$}\\
$\frac58$ & $-96.420-174.693\,i$ & $-8.767+25.133\,i$ & $5.333$\\
$\frac34$ & $-107.957-180.302\,i$ & $-9.047+25.133\,i$ & $5.333$\\
$\frac56$ & $-125.901-187.844\,i$ & $-7.441+25.133\,i$ & $5.333$\\
\bottomrule
\end{tabular}
\caption{Conditional three-loop DPS coefficients for $gg\to HH$ at $\mu=m_H$ in minimal subtraction; the boundary assumptions are stated in the text.}
\label{tab:threeloop}
\end{table}

%% file: tables/table_colour.tex
\begin{table}[!htb]
\centering
\small
\begin{tabular}{llrrc}
\toprule
coefficient & colour factor & prediction & extracted & deviation\\
\midrule
$c_{13}$, $\G^3$ in $A_1$ & $N^2$ & $47.8220215046$ & $47.8220215046$ & $3.7\times10^{-13}$\\
 & $Nn_f$ & $0$ & $2.8\times10^{-12}$ & ($5.9\times10^{-14}$)\\
 & $n_f/N$ & $0$ & $5.5\times10^{-14}$ & ($1.1\times10^{-15}$)\\
 & $n_f^2$ & $0$ & $1.6\times10^{-13}$ & ($3.2\times10^{-15}$)\\
\midrule
$c_{12}$, $\G^2$ in $A_1$ & $N^2$ & $263.0211182755$ & $263.0211182752$ & $2.3\times10^{-12}$\\
 & $Nn_f$ & $-47.8220215046$ & $-47.8220215046$ & $2.0\times10^{-12}$\\
 & $n_f/N$ & $0$ & $1.9\times10^{-12}$ & ($7.2\times10^{-15}$)\\
 & $n_f^2$ & $0$ & $5.4\times10^{-12}$ & ($2.0\times10^{-14}$)\\
\midrule
$c_{22}$, $\G^2$ in $A_2$ & $N^2$ & $-21.4675497995$ & $-21.4675497995$ & $1.1\times10^{-13}$\\
 & $Nn_f$ & $0$ & $3.7\times10^{-13}$ & ($1.7\times10^{-14}$)\\
 & $n_f/N$ & $0$ & $2.0\times10^{-14}$ & ($9.1\times10^{-16}$)\\
 & $n_f^2$ & $0$ & $1.4\times10^{-15}$ & ($6.5\times10^{-17}$)\\
\midrule
$c_{21}$, $\G$ in $A_2$ & $N^2$ & $-57.2467994654$ & $-57.2467994654$ & $1.0\times10^{-12}$\\
 & $Nn_f$ & $14.3116998664$ & $14.3116998664$ & $6.3\times10^{-13}$\\
 & $n_f/N$ & $0$ & $4.7\times10^{-13}$ & ($8.3\times10^{-15}$)\\
 & $n_f^2$ & $0$ & $3.4\times10^{-14}$ & ($5.9\times10^{-16}$)\\
\bottomrule
\end{tabular}
\caption{Coefficients of the leading logarithms of the two-loop $gg\to HH$ remainders in the forward limit, by colour factor, at $\beta=1/2$ and $\mu_R=m_H$, compared with Eq.~(\ref{eq:colourpred}). The deviations are relative. For terms predicted to vanish, the modulus of the extracted coefficient is given, with its size relative to the $N^2$ coefficient in brackets. Moving the fit window one step up in $\delta$ changes the coefficients by at most $6\times10^{-10}$ of the leading coefficient at all three velocities.}
\label{tab:colour}
\end{table}

%% file: tables/table_stage2a.tex
\begin{table}[!htb]
\centering
\small
\begin{tabular}{lccc}
\toprule
coefficient & finite & added by renormalisation & two-loop amplitude\\
& remainder & and $-2I^{(1)}\M^{(1)}$ & at $O(\epsilon^0)$\\
\midrule
$\G^3$ in $\M_1$, units of $C_1$ & $-\frac43C_A$ & $+\frac43C_A$ & $-\frac83C_A$\\
$\beta_0\G^2$ in $\M_1$, units of $C_1$ & $-2$ & $+2$ & $-4$\\
$\G^2$ in $\M_2$, units of $K$ & $-4C_A$ & $+4C_A$ & $-8C_A$\\
$\beta_0\G$ in $\M_2$, units of $K$ & $-4$ & $+4$ & $-8$\\
\bottomrule
\end{tabular}
\caption{Leading coefficients of the two-loop finite remainder, the contribution of ultraviolet renormalisation and infrared subtraction acting on the one-loop DPS term, and the resulting coefficients of the unrenormalised, unsubtracted two-loop amplitude at $O(\epsilon^0)$.}
\label{tab:stage2a}
\end{table}

%% file: tables/table_zz.tex
\begin{table}[!htb]
\centering
\small
\begin{tabular}{lllllll}
\toprule
&&&\multicolumn{2}{c}{$\beta=\frac12$}&\multicolumn{2}{c}{largest}\\
\cmidrule(lr){4-5}\cmidrule(lr){6-7}
coefficient & part & prediction & predicted & extracted & deviation & shift\\
\midrule
\multicolumn{7}{l}{amplitudes $\hel++++$, $\hel++--$, $\hel----$, $\hel--++$, in units of $C_1$}\\
$\G^3$ & $C_F$ & $-\frac83$ & $-2.6666666667$ & $-2.6666666667$ & $3.7\times10^{-13}$ & $2.5\times10^{-11}$\\
 & $C_A$ & $\frac43$ & $1.3333333333$ & $1.3333333333$ & $3.6\times10^{-13}$ & $2.7\times10^{-11}$\\
$\G^2$ & $C_F$ & $-4L_x-6$ & $0.6959057343$ & $0.6959057343$ & $1.5\times10^{-11}$ & $9.0\times10^{-10}$\\
 & $C_A$ & $4L_x$ & $-6.6959057343$ & $-6.6959057343$ & $1.4\times10^{-11}$ & $9.0\times10^{-10}$\\
\midrule
\multicolumn{7}{l}{amplitudes $\hel+-++$, $\hel+---$, $\hel-+++$, $\hel-+--$, in units of $K$}\\
$\G^3$ & $C_F$ & $0$ & $0$ & $1.8\times10^{-13}$ & $2.8\times10^{-13}$ & $2.0\times10^{-11}$\\
 & $C_A$ & $0$ & $0$ & $4.5\times10^{-13}$ & $8.1\times10^{-13}$ & $6.8\times10^{-11}$\\
$\G^2$ & $C_F$ & $-8$ & $-8.0000000000$ & $-8.0000000000$ & $1.1\times10^{-11}$ & $7.3\times10^{-10}$\\
 & $C_A$ & $4$ & $4.0000000000$ & $4.0000000000$ & $3.0\times10^{-11}$ & $2.3\times10^{-9}$\\
$\G$ & $C_F$ & $-8L_x-4$ & $9.3918114686$ & $9.3918114685$ & $1.5\times10^{-10}$ & $8.8\times10^{-9}$\\
 & $C_A$ & $8L_x-4$ & $-17.3918114686$ & $-17.3918114684$ & $3.8\times10^{-10}$ & $2.6\times10^{-8}$\\
\bottomrule
\end{tabular}
\caption{Coefficients of the leading logarithms of the two-loop $gg\to ZZ$ amplitudes in the forward limit, by colour part, in units of the one-loop coefficients, both expanded in $a_s$, compared with Eq.~(\ref{eq:zzpred}); $L_x=\ln[x(1-x)]$. The extracted values are those of the first amplitude of each group at $\beta=\frac12$: the real part where the prediction is nonzero, the modulus where it vanishes. The last two columns give the largest modulus of the difference from the prediction and the largest change when the fit window is moved one step up in $\delta$, over the four amplitudes of each group and the three velocities.}
\label{tab:zz}
\end{table}

%% file: tables/table_r0_zz.tex
\begin{table}[!htb]
\centering
\footnotesize
\setlength{\tabcolsep}{5pt}
\begin{tabular}{llcccc}
\toprule
& & \multicolumn{2}{c}{$\nu=0$: $\M_{\hel++++}$} & \multicolumn{2}{c}{$\nu=2$: $\M_{\hel+-++}$}\\
\cmidrule(lr){3-4}\cmidrule(lr){5-6}
& & $C_F$ & $C_A$ & $C_F$ & $C_A$\\
\midrule
$x=\frac58$ & amplitudes & $2.65228123$ & $1.27907520$ & $2.65228123$ & $1.27907520$\\
 & closed form & $2.65228123$ & $1.27907520$ & $2.65228123$ & $1.27907520$\\
$x=\frac34$ & amplitudes & $4.54429352$ & $-0.36421199$ & $4.54429352$ & $-0.36421199$\\
 & closed form & $4.54429352$ & $-0.36421199$ & $4.54429352$ & $-0.36421199$\\
$x=\frac56$ & amplitudes & $7.31097639$ & $-2.84235222$ & $7.31097639$ & $-2.84235222$\\
 & closed form & $7.31097639$ & $-2.84235222$ & $7.31097639$ & $-2.84235222$\\
\midrule
\multicolumn{2}{l}{largest difference} & $2.0\times10^{-10}$ & $1.7\times10^{-10}$ & $6.5\times10^{-10}$ & $1.6\times10^{-9}$\\
\multicolumn{2}{l}{window shift} & $1.1\times10^{-8}$ & $1.0\times10^{-8}$ & $3.6\times10^{-8}$ & $1.0\times10^{-7}$\\
\bottomrule
\end{tabular}
\caption{The constant $r_0$ of the one-loop kernel read off the two-loop $gg\to ZZ$ amplitudes with Eq.~(\ref{eq:r0read}), by colour factor, at the three velocities of the scans, against Eq.~(\Lref{eq:r0q}). The values shown are those of the first amplitude of each group; the other three amplitudes of each group agree with them to $4.8\times10^{-11}$. The row marked largest difference gives, for each column, the largest difference between the amplitudes and the closed form over the four amplitudes of the group and the three velocities, and the row marked window shift the largest change of the fitted value when the fit window is moved one step up in $\delta$ (Sec.~\ref{app:fits}).}
\label{tab:r0zz}
\end{table}

%% file: tables/table_r0_amp.tex
\begin{table}[!htb]
\centering
\footnotesize
\setlength{\tabcolsep}{5pt}
\begin{tabular}{llcccc}
\toprule
& & \multicolumn{2}{c}{$C_A$} & \multicolumn{2}{c}{$T_Fn_f$}\\
\cmidrule(lr){3-4}\cmidrule(lr){5-6}
& & $HH$ & $AA$ & $HH$ & $AA$\\
\midrule
\multicolumn{6}{l}{$\nu=0$}\\
$x=\frac58$ & amplitudes & $4.96102843$ & $4.99578015$ & $-4.72601066$ & $-4.79551411$\\
 & closed form & $4.96102843$ & $4.99578015$ & $-4.72601066$ & $-4.79551411$\\
$x=\frac34$ & amplitudes & $5.18575874$ & $5.20342420$ & $-4.67802109$ & $-4.71335201$\\
 & closed form & $5.18575874$ & $5.20342420$ & $-4.67802109$ & $-4.71335201$\\
$x=\frac56$ & amplitudes & $5.44666610$ & $5.45382124$ & $-4.62275054$ & $-4.63706082$\\
 & closed form & $5.44666610$ & $5.45382124$ & $-4.62275054$ & $-4.63706082$\\
\multicolumn{2}{l}{largest difference} & $5.4\times10^{-10}$ & $6.4\times10^{-10}$ & $2.1\times10^{-10}$ & $2.5\times10^{-10}$\\
\multicolumn{2}{l}{window shift} & $4.1\times10^{-8}$ & $4.9\times10^{-8}$ & $1.7\times10^{-8}$ & $2.0\times10^{-8}$\\
\midrule
\multicolumn{6}{l}{$\nu=2$}\\
$x=\frac58$ & amplitudes & $5.72676090$ & $3.91372974$ & $-6.25747561$ & $-2.63141328$\\
 & closed form & $5.72676090$ & $3.91372974$ & $-6.25747561$ & $-2.63141328$\\
$x=\frac34$ & amplitudes & $5.45921432$ & $4.67872651$ & $-5.22493225$ & $-3.66395664$\\
 & closed form & $5.45921432$ & $4.67872651$ & $-5.22493225$ & $-3.66395664$\\
$x=\frac56$ & amplitudes & $5.54920634$ & $5.16581976$ & $-4.82783103$ & $-4.06105786$\\
 & closed form & $5.54920634$ & $5.16581976$ & $-4.82783103$ & $-4.06105786$\\
\multicolumn{2}{l}{largest difference} & $1.5\times10^{-10}$ & $1.0\times10^{-10}$ & $5.6\times10^{-11}$ & $6.8\times10^{-11}$\\
\multicolumn{2}{l}{window shift} & $6.6\times10^{-9}$ & $4.4\times10^{-9}$ & $4.5\times10^{-9}$ & $5.5\times10^{-9}$\\
\bottomrule
\end{tabular}
\caption{The constant $r_0$ of the one-loop kernel read off the two-loop $gg\to HH$ and $gg\to AA$ amplitudes with Eq.~(\ref{eq:r0read}), for the two components and the two colour factors of gluon splittings, at the three velocities of the scans, against the closed forms: Eq.~(\Lref{eq:r0g}) with $F_0=F_S$ for $HH$, and Eq.~(\ref{eq:r0aa}) for $AA$, which has $F_0=F_A$ for $\nu=0$ and the helicity-flip term with the opposite sign for $\nu=2$. The two processes differ by the helicity term alone in both components. The rows marked largest difference give, for each component, the largest difference between the amplitudes and the closed forms over the three velocities, and the rows marked window shift the largest change of the fitted value when the fit window is moved one step up in $\delta$ (Sec.~\ref{app:fits}).}
\label{tab:r0amp}
\end{table}

%% file: tables/table_r0.tex
\begin{table}[!htb]
\centering
\small
\setlength{\tabcolsep}{4pt}
\begin{tabular}{lcccccc}
\toprule
& \multicolumn{2}{c}{$g\to q\bar q$} & \multicolumn{2}{c}{$g\to gg$, $\nu=0$} & \multicolumn{2}{c}{$g\to gg$, $\nu=2$}\\
\cmidrule(lr){2-3}\cmidrule(lr){4-5}\cmidrule(lr){6-7}
$x$ & $C_F$ & $C_A$ & $C_A$ & $T_Fn_f$ & $C_A$ & $T_Fn_f$\\
\midrule
0.1 & $11.78598720$ & $-6.98106379$ & $5.75285579$ & $-4.56253144$ & $5.78524783$ & $-4.62731552$\\
0.3 & $3.56622290$ & $0.49139902$ & $5.07516361$ & $-4.70175009$ & $5.51057203$ & $-5.57256693$\\
0.5 & $2.13039556$ & $1.72312203$ & $4.89055459$ & $-4.74074074$ & $6.07573977$ & $-7.11111111$\\
\midrule
difference & $8.4\times10^{-8}$ & $4.1\times10^{-8}$ & $5.4\times10^{-8}$ & $3.5\times10^{-9}$ & $5.9\times10^{-8}$ & $1.8\times10^{-9}$\\
\bottomrule
\end{tabular}
\caption{The constant $r_0$ of the one-loop kernel from the regions (Sec.~\ref{app:regions}), coefficient by colour factor, for quark splittings ($gg\to ZZ$) and for the two components of gluon splittings ($gg\to HH$); by the symmetry $x\to1-x$, $x=0.7$ and $0.9$ give the same. The last row gives the largest difference between numerical region integrations and the analytic expressions of Eqs.~(\Lref{eq:r0g}) and~(\Lref{eq:r0q}) over the three values of $x$.}
\label{tab:r0}
\end{table}

%% file: pinch_checks.tex
\subsection{Pinch structure and scalar cross-checks}
\label{sec:pinch}

Feynman parameter space provides a complementary
description of the DPS pinch and its local power
counting. We verify the scalar polynomial identities
and local pinch powers independently of the fitted
amplitude coefficients. The candidate scalings are
taken from a separate momentum-space survey; the
checks here do not repeat that enumeration or replace
the regulated region integrations of
Secs.~\ref{app:regions} and~\ref{sec:r0regions}.

\paragraph{The one-loop pinch: }
For $p_3=xp_1+yp_2+p_{3\perp}$, with
$p_{3\perp}^2=-p_T^2$ and $0<x,y<1$,
define $\lambda^2=p_T^2/s$.
The four daughter propagators of the scalar
box have Feynman parameters
$\alpha_k,\alpha_{\bar k},\alpha_r,
\alpha_{\bar r}$.
The second Symanzik polynomial is exactly
\begin{equation}
\frac{F}{s}=-f_1f_2+\lambda^2(\alpha_k+\alpha_{\bar k})(\alpha_r+\alpha_{\bar r}),
\qquad
f_1=x\alpha_k-(1-x)\alpha_{\bar k},\quad
f_2=y\alpha_r-(1-y)\alpha_{\bar r}.
\label{eq:pinchparameter}
\end{equation}
At $p_T=0$, both $F$ and its gradient
vanish on $f_1=f_2=0$ within the
positive Feynman parameter domain.
These conditions are the two light-cone
components of the Landau loop equation.
They correspond to two hard vertices
coinciding at the DPS pinch~\cite{Coleman:1965xm,Gaunt:2011xd}.
The singularity is therefore an
interior pinch rather than an endpoint
singularity in the original parameters.
In massless $2\to2$ scattering,
Ref.~\cite{Gardi:2024dissecting}
identified analogous parameter-space
pinches beginning at three loops
in particular nonplanar graphs.
With two massive final-state particles,
the DPS box develops such a pinch
already at one loop.
It arises from cancellations between
terms of opposite sign in $F$,
associated with two distinct
hard subgraphs.
Taking the normal coordinates
$f_1,f_2\sim\lambda$ gives a
measure proportional to $\lambda^2$,
while $F^{-2}\sim\lambda^{-4}$.
The scalar box therefore scales
as $\lambda^{-2}$ in four dimensions.
The QCD splitting numerators provide
two powers of transverse momentum,
changing this to $\lambda^0$,
up to logarithms.
The scalar singularity alone does not
determine the power of the QCD amplitude.

\paragraph{Input from momentum space: }
The candidate scalings come from a
separate survey of 54 two-loop graphs
containing the DPS box, together with
16 additional loop-momentum
parametrisations.
The box momentum is assigned its
DPS pinch or a hard, collinear, soft
or intermediate scaling.
The additional loop momentum is
assigned hard, collinear, soft,
Glauber, ultrasoft or intermediate
longitudinal scalings.
Each record specifies the scaling
$|p_e^2|\sim\lambda^{w_e}s$
of the squared momentum of each
internal line $e$, the momentum
routing and the type of reduced
diagram. The supplied labels
distinguish DPS-type configurations,
with two hard vertices, from
single parton scattering (SPS)
configurations, with one hard
vertex producing both final-state
particles.
The pointwise power $P$ includes
the integration measure, propagator
denominators and QCD numerators,
evaluated at generic points of the
assigned scaling. The survey uses
Feynman gauge and the axial gauges
$n=p_1+p_2$ and $n=3p_1+2p_2$.
The scalar power $P_{\rm den}$
instead sets the numerators to unity.

The released \texttt{vectors.json.xz}
contains the squared-momentum scalings,
momentum assignments and original
surface labels. The polynomial
checks below are reproducible from
these inputs, but the programs used
to generate the original momentum-space
survey are not included.
Consequently, neither the candidate
scalings nor their DPS/SPS labels
are independently determined by the
parameter-space calculation.

\paragraph{Two-loop polynomials and local powers: }
For an $L$-loop scalar graph with
$E$ propagators, the Feynman parameter
integrand, apart from the projective
measure and overall factors, is
$U^a/(F-i0)^b$, where
$a=E-2(L+1)$ and $b=E-2L$
in four dimensions.
These expressions describe the
integrand without assuming that
the scalar integral is finite.
We construct the first and second
Symanzik polynomials from spanning
trees and two-forests and independently
from the determinant and adjugate
of the loop-momentum quadratic form.
The two constructions agree
identically in $x,y,\lambda^2$
for all 54 canonical two-loop
graphs containing the DPS box.
For a supplied parameter scaling
$\alpha_e\sim\lambda^{-w_e}$,
write
$U\sim\lambda^{\mu_U}U_0$
and $F\sim\lambda^{\mu_F}F_0$.

The survey contains 2823 records,
representing 2363 distinct
graph-parameter vectors in
70 parametrisations.
Of these, 1709 records, corresponding
to 1305 distinct vectors, have
a leading polynomial $F_0$ with
exactly two factors that change
sign within the positive parameter
domain. The factors are associated
with the two incoming parton families.
This algebraic property does not
by itself identify the reduced
diagram. In particular, the same
two-factor structure occurs in
678 records labelled SPS.
The parameter-space check verifies
the factors without using the
supplied DPS/SPS labels.
For the 142 selected DPS-type
scalings with the box at its
DPS pinch and the additional
loop hard, collinear or ultrasoft,
the local scalar powers agree
with the supplied $P_{\rm den}$.
Of these, 127 have
$P_{\rm den}\le0$; the remaining
15 have $P_{\rm den}=2$.
This agreement concerns the
measure and propagators.
It does not test the
numerator-inclusive power $P$.

To determine the local power
near an interior pinch, consider
two normal coordinates $z_1,z_2$.
By multiaffinity, the polynomial
takes the form
$F=Az_1z_2+Bz_1+Cz_2+D$,
with the remaining parameters
held positive.
Its stationary point is
$z_1=-C/A$, $z_2=-B/A$,
and the critical value is
$F_{\rm crit}=(AD-BC)/A$.
Suppose the first nonzero term
in $F_{\rm crit}$ occurs at
relative order $\lambda^\delta$
with respect to $\lambda^{\mu_F}$.
For $b\ge2$, the resulting
local scalar power is
\begin{equation}
\omega=-\sum_e w_e+a\mu_U-b\mu_F+\delta(1-b).
\label{eq:pinchlocalpower}
\end{equation}
For $b=1$, the normal integration
is logarithmic.
This local analysis applies away
from zeros of the critical value
along the tangent directions and
away from tangent boundaries.
Such configurations require
additional analysis.

A complete region classification
also requires dissecting the
parameter domain at the pinch,
turning the interior singularity
into endpoint singularities.
The regions can then be identified
from the Newton polytopes of
the dissected integrals~\cite{Gardi:2024dissecting}.
The present calculation determines
the stated local powers without
performing this full dissection.

\paragraph{Interjet exchange and gauge dependence: }
For the 16 graphs in which the
additional gluon connects the two
incoming jets, we examine the
common squared-momentum assignments
at the Glauber scaling.
The leading polynomial has no
stationary point with all Feynman
parameters strictly positive.
This follows from exact algebra
at arbitrary interior $x,y$:
either a derivative has a fixed
nonzero sign, or every solution
of a necessary stationary equation
forces another derivative to have
a fixed nonzero sign.
The result excludes a stationary
point in the positive interior
for these scalar configurations.
It does not exclude boundary
pinches or imply that the regulated
Glauber contribution vanishes.
In particular, it does not
eliminate the Glauber phase
found in Sec.~\ref{app:regions}.

In other graphs, a Glauber
assignment can give the same
squared-momentum scalings as a
collinear or soft assignment.
These possibilities cannot be
distinguished by the scalar
polynomial check alone.
The momentum-space survey also
includes intermediate longitudinal
scalings between the soft,
collinear and Glauber regions.
For $gg\to HH$, $gg\to AA$
and $gg\to ZZ$, the only
negative pointwise power found
at these intermediate scalings
is $P=-\tfrac12$, occurring
in individual daughter
self-energy graphs in the
two axial gauges.
No negative power is found
in Feynman gauge at these
scalings.
The half-integer powers arise
from gauge-vector terms in
the exchanged-gluon numerator.
They characterise individual
integrands rather than a
demonstrated singularity of
the gauge-invariant amplitude.
Their cancellation in the
complete axial-gauge sum,
including axial-pole
prescriptions and regulated
overlaps, is not established
by the pointwise survey.

\paragraph{Reproduction and scope: }
The released package
\texttt{checks/pinch\_regions/}
contains the supplied scaling
vectors, exact Symanzik
polynomial identities, algebraic
sign checks and regression tests.
The scalar polynomial and
local-power calculations are
reproducible from these data.
The original momentum-space
enumeration that selected the
candidate vectors is not
regenerated by this package.
The checks establish the
one-loop interior pinch,
the stated two-loop polynomial
identities, the local scalar
powers and the absence of
positive-interior stationary
points in the selected
interjet-exchange configurations.
They provide complementary
evidence for the DPS singular
structure alongside the
two-loop amplitude comparisons
and regulated momentum-region
calculations.
A complete classification of
the momentum regions and
renormalised matching of
all overlaps are separate
extensions of these results.

%% file: letter.bbl
\begin{thebibliography}{57}%
\makeatletter
\providecommand \@ifxundefined [1]{%
 \@ifx{#1\undefined}
}%
\providecommand \@ifnum [1]{%
 \ifnum #1\expandafter \@firstoftwo
 \else \expandafter \@secondoftwo
 \fi
}%
\providecommand \@ifx [1]{%
 \ifx #1\expandafter \@firstoftwo
 \else \expandafter \@secondoftwo
 \fi
}%
\providecommand \natexlab [1]{#1}%
\providecommand \enquote  [1]{``#1''}%
\providecommand \bibnamefont  [1]{#1}%
\providecommand \bibfnamefont [1]{#1}%
\providecommand \citenamefont [1]{#1}%
\providecommand \href@noop [0]{\@secondoftwo}%
\providecommand \href [0]{\begingroup \@sanitize@url \@href}%
\providecommand \@href[1]{\@@startlink{#1}\@@href}%
\providecommand \@@href[1]{\endgroup#1\@@endlink}%
\providecommand \@sanitize@url [0]{\catcode `\\12\catcode `\$12\catcode
  `\&12\catcode `\#12\catcode `\^12\catcode `\_12\catcode `\%12\relax}%
\providecommand \@@startlink[1]{}%
\providecommand \@@endlink[0]{}%
\providecommand \url  [0]{\begingroup\@sanitize@url \@url }%
\providecommand \@url [1]{\endgroup\@href {#1}{\urlprefix }}%
\providecommand \urlprefix  [0]{URL }%
\providecommand \Eprint [0]{\href }%
\providecommand \doibase [0]{https://doi.org/}%
\providecommand \selectlanguage [0]{\@gobble}%
\providecommand \bibinfo  [0]{\@secondoftwo}%
\providecommand \bibfield  [0]{\@secondoftwo}%
\providecommand \translation [1]{[#1]}%
\providecommand \BibitemOpen [0]{}%
\providecommand \bibitemStop [0]{}%
\providecommand \bibitemNoStop [0]{.\EOS\space}%
\providecommand \EOS [0]{\spacefactor3000\relax}%
\providecommand \BibitemShut  [1]{\csname bibitem#1\endcsname}%
\let\auto@bib@innerbib\@empty
\bibitem [{\citenamefont {Gaunt}\ and\ \citenamefont
  {Stirling}(2011)}]{Gaunt:2011xd}%
  \BibitemOpen
  \bibfield  {author} {\bibinfo {author} {\bibfnamefont {J.~R.}\ \bibnamefont
  {Gaunt}}\ and\ \bibinfo {author} {\bibfnamefont {W.~J.}\ \bibnamefont
  {Stirling}},\ }\bibfield  {title} {\bibinfo {title} {{Double parton
  scattering singularity in one-loop integrals}},\ }\href
  {https://doi.org/10.1007/JHEP06(2011)048} {\bibfield  {journal} {\bibinfo
  {journal} {JHEP}\ }\textbf {\bibinfo {volume} {06}}\bibfield  {number}
  {\bibinfo  {number} { (2011)},\ \bibinfo {pages} {048}},\ }\Eprint
  {https://arxiv.org/abs/1103.1888} {arXiv:1103.1888 [hep-ph]} \BibitemShut
  {NoStop}%
\bibitem [{\citenamefont {Diehl}\ \emph {et~al.}(2012)\citenamefont {Diehl},
  \citenamefont {Ostermeier},\ and\ \citenamefont
  {Sch{\"a}fer}}]{Diehl:2011yj}%
  \BibitemOpen
  \bibfield  {author} {\bibinfo {author} {\bibfnamefont {M.}~\bibnamefont
  {Diehl}}, \bibinfo {author} {\bibfnamefont {D.}~\bibnamefont {Ostermeier}},\
  and\ \bibinfo {author} {\bibfnamefont {A.}~\bibnamefont {Sch{\"a}fer}},\
  }\bibfield  {title} {\bibinfo {title} {{Elements of a theory for multiparton
  interactions in QCD}},\ }\href {https://doi.org/10.1007/JHEP03(2012)089}
  {\bibfield  {journal} {\bibinfo  {journal} {JHEP}\ }\textbf {\bibinfo
  {volume} {03}}\bibfield  {number} {\bibinfo  {number} { (2012)},\ \bibinfo
  {pages} {089}},\ }\bibinfo {note} {[Erratum ibid. 03 (2016) 001]},\ \Eprint
  {https://arxiv.org/abs/1111.0910} {arXiv:1111.0910 [hep-ph]} \BibitemShut
  {NoStop}%
\bibitem [{\citenamefont {Glover}\ and\ \citenamefont {van~der
  Bij}(1989)}]{Glover:1988rg}%
  \BibitemOpen
  \bibfield  {author} {\bibinfo {author} {\bibfnamefont {E.~W.~N.}\
  \bibnamefont {Glover}}\ and\ \bibinfo {author} {\bibfnamefont {J.~J.}\
  \bibnamefont {van~der Bij}},\ }\bibfield  {title} {\bibinfo {title} {{Z boson
  pair production via gluon fusion}},\ }\href
  {https://doi.org/10.1016/0550-3213(89)90262-9} {\bibfield  {journal}
  {\bibinfo  {journal} {Nucl. Phys. B}\ }\textbf {\bibinfo {volume} {321}},\
  \bibinfo {pages} {561} (\bibinfo {year} {1989})}\BibitemShut {NoStop}%
\bibitem [{\citenamefont {Banerjee}\ \emph {et~al.}(2018)\citenamefont
  {Banerjee}, \citenamefont {Borowka}, \citenamefont {Dhani}, \citenamefont
  {Gehrmann},\ and\ \citenamefont {Ravindran}}]{Banerjee:2018lfq}%
  \BibitemOpen
  \bibfield  {author} {\bibinfo {author} {\bibfnamefont {P.}~\bibnamefont
  {Banerjee}}, \bibinfo {author} {\bibfnamefont {S.}~\bibnamefont {Borowka}},
  \bibinfo {author} {\bibfnamefont {P.~K.}\ \bibnamefont {Dhani}}, \bibinfo
  {author} {\bibfnamefont {T.}~\bibnamefont {Gehrmann}},\ and\ \bibinfo
  {author} {\bibfnamefont {V.}~\bibnamefont {Ravindran}},\ }\bibfield  {title}
  {\bibinfo {title} {{Two-loop massless QCD corrections to the $g+g\to H+H$
  four-point amplitude}},\ }\href {https://doi.org/10.1007/JHEP11(2018)130}
  {\bibfield  {journal} {\bibinfo  {journal} {JHEP}\ }\textbf {\bibinfo
  {volume} {11}}\bibfield  {number} {\bibinfo  {number} { (2018)},\ \bibinfo
  {pages} {130}},\ }\Eprint {https://arxiv.org/abs/1809.05388}
  {arXiv:1809.05388 [hep-ph]} \BibitemShut {NoStop}%
\bibitem [{\citenamefont {von Manteuffel}\ and\ \citenamefont
  {Tancredi}(2015)}]{vonManteuffel:2015msa}%
  \BibitemOpen
  \bibfield  {author} {\bibinfo {author} {\bibfnamefont {A.}~\bibnamefont {von
  Manteuffel}}\ and\ \bibinfo {author} {\bibfnamefont {L.}~\bibnamefont
  {Tancredi}},\ }\bibfield  {title} {\bibinfo {title} {{The two-loop helicity
  amplitudes for $gg\to V_1V_2\to4$ leptons}},\ }\href
  {https://doi.org/10.1007/JHEP06(2015)197} {\bibfield  {journal} {\bibinfo
  {journal} {JHEP}\ }\textbf {\bibinfo {volume} {06}}\bibfield  {number}
  {\bibinfo  {number} { (2015)},\ \bibinfo {pages} {197}},\ }\Eprint
  {https://arxiv.org/abs/1503.08835} {arXiv:1503.08835 [hep-ph]} \BibitemShut
  {NoStop}%
\bibitem [{\citenamefont {Manohar}\ and\ \citenamefont
  {Waalewijn}(2012)}]{Manohar:2012pe}%
  \BibitemOpen
  \bibfield  {author} {\bibinfo {author} {\bibfnamefont {A.~V.}\ \bibnamefont
  {Manohar}}\ and\ \bibinfo {author} {\bibfnamefont {W.~J.}\ \bibnamefont
  {Waalewijn}},\ }\bibfield  {title} {\bibinfo {title} {{What is double parton
  scattering?}},\ }\href {https://doi.org/10.1016/j.physletb.2012.05.044}
  {\bibfield  {journal} {\bibinfo  {journal} {Phys. Lett. B}\ }\textbf
  {\bibinfo {volume} {713}},\ \bibinfo {pages} {196} (\bibinfo {year}
  {2012})},\ \Eprint {https://arxiv.org/abs/1202.5034} {arXiv:1202.5034
  [hep-ph]} \BibitemShut {NoStop}%
\bibitem [{\citenamefont {Gaunt}(2013)}]{Gaunt:2012dd}%
  \BibitemOpen
  \bibfield  {author} {\bibinfo {author} {\bibfnamefont {J.~R.}\ \bibnamefont
  {Gaunt}},\ }\bibfield  {title} {\bibinfo {title} {{Single perturbative
  splitting diagrams in double parton scattering}},\ }\href
  {https://doi.org/10.1007/JHEP01(2013)042} {\bibfield  {journal} {\bibinfo
  {journal} {JHEP}\ }\textbf {\bibinfo {volume} {01}}\bibfield  {number}
  {\bibinfo  {number} { (2013)},\ \bibinfo {pages} {042}},\ }\Eprint
  {https://arxiv.org/abs/1207.0480} {arXiv:1207.0480 [hep-ph]} \BibitemShut
  {NoStop}%
\bibitem [{\citenamefont {Diehl}\ \emph {et~al.}(2017)\citenamefont {Diehl},
  \citenamefont {Gaunt},\ and\ \citenamefont {Sch{\"o}nwald}}]{Diehl:2017kgu}%
  \BibitemOpen
  \bibfield  {author} {\bibinfo {author} {\bibfnamefont {M.}~\bibnamefont
  {Diehl}}, \bibinfo {author} {\bibfnamefont {J.~R.}\ \bibnamefont {Gaunt}},\
  and\ \bibinfo {author} {\bibfnamefont {K.}~\bibnamefont {Sch{\"o}nwald}},\
  }\bibfield  {title} {\bibinfo {title} {{Double hard scattering without double
  counting}},\ }\href {https://doi.org/10.1007/JHEP06(2017)083} {\bibfield
  {journal} {\bibinfo  {journal} {JHEP}\ }\textbf {\bibinfo {volume}
  {06}}\bibfield  {number} {\bibinfo  {number} { (2017)},\ \bibinfo {pages}
  {083}},\ }\Eprint {https://arxiv.org/abs/1702.06486} {arXiv:1702.06486
  [hep-ph]} \BibitemShut {NoStop}%
\bibitem [{\citenamefont {Blok}\ \emph {et~al.}(2011)\citenamefont {Blok},
  \citenamefont {Dokshitzer}, \citenamefont {Frankfurt},\ and\ \citenamefont
  {Strikman}}]{BDFS2011}%
  \BibitemOpen
  \bibfield  {author} {\bibinfo {author} {\bibfnamefont {B.}~\bibnamefont
  {Blok}}, \bibinfo {author} {\bibfnamefont {{\relax{Yu}}.}~\bibnamefont
  {Dokshitzer}}, \bibinfo {author} {\bibfnamefont {L.}~\bibnamefont
  {Frankfurt}},\ and\ \bibinfo {author} {\bibfnamefont {M.}~\bibnamefont
  {Strikman}},\ }\bibfield  {title} {\bibinfo {title} {The four jet production
  at {LHC} and {Tevatron} in {QCD}},\ }\href
  {https://doi.org/10.1103/PhysRevD.83.071501} {\bibfield  {journal} {\bibinfo
  {journal} {Phys. Rev. D}\ }\textbf {\bibinfo {volume} {83}},\ \bibinfo
  {pages} {071501} (\bibinfo {year} {2011})},\ \Eprint
  {https://arxiv.org/abs/1009.2714} {arXiv:1009.2714 [hep-ph]} \BibitemShut
  {NoStop}%
\bibitem [{\citenamefont {Blok}\ \emph {et~al.}(2012)\citenamefont {Blok},
  \citenamefont {Dokshitzer}, \citenamefont {Frankfurt},\ and\ \citenamefont
  {Strikman}}]{BDFS2012}%
  \BibitemOpen
  \bibfield  {author} {\bibinfo {author} {\bibfnamefont {B.}~\bibnamefont
  {Blok}}, \bibinfo {author} {\bibfnamefont {{\relax{Yu}}.}~\bibnamefont
  {Dokshitzer}}, \bibinfo {author} {\bibfnamefont {L.}~\bibnamefont
  {Frankfurt}},\ and\ \bibinfo {author} {\bibfnamefont {M.}~\bibnamefont
  {Strikman}},\ }\bibfield  {title} {\bibinfo {title} {{pQCD} physics of
  multiparton interactions},\ }\href
  {https://doi.org/10.1140/epjc/s10052-012-1963-8} {\bibfield  {journal}
  {\bibinfo  {journal} {Eur. Phys. J. C}\ }\textbf {\bibinfo {volume} {72}},\
  \bibinfo {pages} {1963} (\bibinfo {year} {2012})},\ \Eprint
  {https://arxiv.org/abs/1106.5533} {arXiv:1106.5533 [hep-ph]} \BibitemShut
  {NoStop}%
\bibitem [{\citenamefont {Blok}\ \emph {et~al.}(2014)\citenamefont {Blok},
  \citenamefont {Dokshitzer}, \citenamefont {Frankfurt},\ and\ \citenamefont
  {Strikman}}]{BDFS2014}%
  \BibitemOpen
  \bibfield  {author} {\bibinfo {author} {\bibfnamefont {B.}~\bibnamefont
  {Blok}}, \bibinfo {author} {\bibfnamefont {{\relax{Yu}}.}~\bibnamefont
  {Dokshitzer}}, \bibinfo {author} {\bibfnamefont {L.}~\bibnamefont
  {Frankfurt}},\ and\ \bibinfo {author} {\bibfnamefont {M.}~\bibnamefont
  {Strikman}},\ }\bibfield  {title} {\bibinfo {title} {Perturbative {QCD}
  correlations in multi-parton collisions},\ }\href
  {https://doi.org/10.1140/epjc/s10052-014-2926-z} {\bibfield  {journal}
  {\bibinfo  {journal} {Eur. Phys. J. C}\ }\textbf {\bibinfo {volume} {74}},\
  \bibinfo {pages} {2926} (\bibinfo {year} {2014})},\ \Eprint
  {https://arxiv.org/abs/1306.3763} {arXiv:1306.3763 [hep-ph]} \BibitemShut
  {NoStop}%
\bibitem [{\citenamefont {Bhattacharya}(2026)}]{P1}%
  \BibitemOpen
  \bibfield  {author} {\bibinfo {author} {\bibfnamefont {A.}~\bibnamefont
  {Bhattacharya}},\ }\href@noop {} {\bibinfo {title} {{Two-loop $gg\to AA$
  amplitudes near threshold and the double parton scattering singularity}}}
  (\bibinfo {year} {2026}),\ \Eprint {https://arxiv.org/abs/2610.19024}
  {arXiv:2610.19024 [hep-ph]} \BibitemShut {NoStop}%
\bibitem [{\citenamefont {Diehl}\ \emph {et~al.}(2019)\citenamefont {Diehl},
  \citenamefont {Gaunt}, \citenamefont {Pl{\"o}{\ss}l},\ and\ \citenamefont
  {Sch{\"a}fer}}]{Diehl:2019rdh}%
  \BibitemOpen
  \bibfield  {author} {\bibinfo {author} {\bibfnamefont {M.}~\bibnamefont
  {Diehl}}, \bibinfo {author} {\bibfnamefont {J.~R.}\ \bibnamefont {Gaunt}},
  \bibinfo {author} {\bibfnamefont {P.}~\bibnamefont {Pl{\"o}{\ss}l}},\ and\
  \bibinfo {author} {\bibfnamefont {A.}~\bibnamefont {Sch{\"a}fer}},\
  }\bibfield  {title} {\bibinfo {title} {{Two-loop splitting in double parton
  distributions}},\ }\href {https://doi.org/10.21468/SciPostPhys.7.2.017}
  {\bibfield  {journal} {\bibinfo  {journal} {SciPost Phys.}\ }\textbf
  {\bibinfo {volume} {7}},\ \bibinfo {pages} {017} (\bibinfo {year} {2019})},\
  \Eprint {https://arxiv.org/abs/1902.08019} {arXiv:1902.08019 [hep-ph]}
  \BibitemShut {NoStop}%
\bibitem [{\citenamefont {Diehl}\ \emph {et~al.}(2021)\citenamefont {Diehl},
  \citenamefont {Gaunt},\ and\ \citenamefont {Plossl}}]{Diehl:2021non}%
  \BibitemOpen
  \bibfield  {author} {\bibinfo {author} {\bibfnamefont {M.}~\bibnamefont
  {Diehl}}, \bibinfo {author} {\bibfnamefont {J.~R.}\ \bibnamefont {Gaunt}},\
  and\ \bibinfo {author} {\bibfnamefont {P.}~\bibnamefont {Plossl}},\
  }\bibfield  {title} {\bibinfo {title} {{Two-loop splitting in double parton
  distributions: the colour non-singlet case}},\ }\href
  {https://doi.org/10.1007/JHEP08(2021)040} {\bibfield  {journal} {\bibinfo
  {journal} {JHEP}\ }\textbf {\bibinfo {volume} {08}}\bibfield  {number}
  {\bibinfo  {number} { (2021)},\ \bibinfo {pages} {040}},\ }\Eprint
  {https://arxiv.org/abs/2105.08425} {arXiv:2105.08425 [hep-ph]} \BibitemShut
  {NoStop}%
\bibitem [{\citenamefont {Diehl}\ and\ \citenamefont
  {Plossl}(2026)}]{Diehl:2026quant}%
  \BibitemOpen
  \bibfield  {author} {\bibinfo {author} {\bibfnamefont {M.}~\bibnamefont
  {Diehl}}\ and\ \bibinfo {author} {\bibfnamefont {P.}~\bibnamefont {Plossl}},\
  }\bibfield  {title} {\bibinfo {title} {{A quantitative study of two-loop
  splitting in double parton distributions}},\ }\href
  {https://doi.org/10.1007/JHEP05(2026)158} {\bibfield  {journal} {\bibinfo
  {journal} {JHEP}\ }\textbf {\bibinfo {volume} {05}}\bibfield  {number}
  {\bibinfo  {number} { (2026)},\ \bibinfo {pages} {158}},\ }\Eprint
  {https://arxiv.org/abs/2602.18233} {arXiv:2602.18233 [hep-ph]} \BibitemShut
  {NoStop}%
\bibitem [{\citenamefont {Diehl}\ \emph {et~al.}(2016)\citenamefont {Diehl},
  \citenamefont {Gaunt}, \citenamefont {Ostermeier}, \citenamefont
  {Pl{\"o}{\ss}l},\ and\ \citenamefont {Sch{\"a}fer}}]{Diehl:2015bca}%
  \BibitemOpen
  \bibfield  {author} {\bibinfo {author} {\bibfnamefont {M.}~\bibnamefont
  {Diehl}}, \bibinfo {author} {\bibfnamefont {J.~R.}\ \bibnamefont {Gaunt}},
  \bibinfo {author} {\bibfnamefont {D.}~\bibnamefont {Ostermeier}}, \bibinfo
  {author} {\bibfnamefont {P.}~\bibnamefont {Pl{\"o}{\ss}l}},\ and\ \bibinfo
  {author} {\bibfnamefont {A.}~\bibnamefont {Sch{\"a}fer}},\ }\bibfield
  {title} {\bibinfo {title} {{Cancellation of Glauber gluon exchange in the
  double Drell-Yan process}},\ }\href {https://doi.org/10.1007/JHEP01(2016)076}
  {\bibfield  {journal} {\bibinfo  {journal} {JHEP}\ }\textbf {\bibinfo
  {volume} {01}}\bibfield  {number} {\bibinfo  {number} { (2016)},\ \bibinfo
  {pages} {076}},\ }\Eprint {https://arxiv.org/abs/1510.08696}
  {arXiv:1510.08696 [hep-ph]} \BibitemShut {NoStop}%
\bibitem [{\citenamefont {Caola}\ \emph
  {et~al.}(2015{\natexlab{a}})\citenamefont {Caola}, \citenamefont {Melnikov},
  \citenamefont {R{\"o}ntsch},\ and\ \citenamefont {Tancredi}}]{Caola:2015zz}%
  \BibitemOpen
  \bibfield  {author} {\bibinfo {author} {\bibfnamefont {F.}~\bibnamefont
  {Caola}}, \bibinfo {author} {\bibfnamefont {K.}~\bibnamefont {Melnikov}},
  \bibinfo {author} {\bibfnamefont {R.}~\bibnamefont {R{\"o}ntsch}},\ and\
  \bibinfo {author} {\bibfnamefont {L.}~\bibnamefont {Tancredi}},\ }\bibfield
  {title} {\bibinfo {title} {{QCD corrections to $ZZ$ production in gluon
  fusion at the LHC}},\ }\href {https://doi.org/10.1103/PhysRevD.92.094028}
  {\bibfield  {journal} {\bibinfo  {journal} {Phys. Rev. D}\ }\textbf {\bibinfo
  {volume} {92}},\ \bibinfo {pages} {094028} (\bibinfo {year}
  {2015}{\natexlab{a}})},\ \Eprint {https://arxiv.org/abs/1509.06734}
  {arXiv:1509.06734 [hep-ph]} \BibitemShut {NoStop}%
\bibitem [{SM()}]{SM}%
  \BibitemOpen
  \href@noop {} {}\bibinfo {note} {See Supplemental Material accompanying this arXiv submission for the two-loop DPS derivation, matching
  calculations, numerical validation, and higher-order consistency
  checks.}\BibitemShut {Stop}%
\bibitem [{\citenamefont {Becher}\ and\ \citenamefont
  {Neubert}(2009{\natexlab{a}})}]{Becher:2009cu}%
  \BibitemOpen
  \bibfield  {author} {\bibinfo {author} {\bibfnamefont {T.}~\bibnamefont
  {Becher}}\ and\ \bibinfo {author} {\bibfnamefont {M.}~\bibnamefont
  {Neubert}},\ }\bibfield  {title} {\bibinfo {title} {{Infrared singularities
  of scattering amplitudes in perturbative QCD}},\ }\href
  {https://doi.org/10.1103/PhysRevLett.102.162001} {\bibfield  {journal}
  {\bibinfo  {journal} {Phys. Rev. Lett.}\ }\textbf {\bibinfo {volume} {102}},\
  \bibinfo {pages} {162001} (\bibinfo {year} {2009}{\natexlab{a}})},\ \bibinfo
  {note} {[Erratum ibid. 111 (2013) 199905]},\ \Eprint
  {https://arxiv.org/abs/0901.0722} {arXiv:0901.0722 [hep-ph]} \BibitemShut
  {NoStop}%
\bibitem [{\citenamefont {Korchemskaya}\ and\ \citenamefont
  {Korchemsky}(1995)}]{Korchemskaya:1994qp}%
  \BibitemOpen
  \bibfield  {author} {\bibinfo {author} {\bibfnamefont {I.~A.}\ \bibnamefont
  {Korchemskaya}}\ and\ \bibinfo {author} {\bibfnamefont {G.~P.}\ \bibnamefont
  {Korchemsky}},\ }\bibfield  {title} {\bibinfo {title} {{High-energy
  scattering in QCD and cross singularities of Wilson loops}},\ }\href
  {https://doi.org/10.1016/0550-3213(94)00553-Q} {\bibfield  {journal}
  {\bibinfo  {journal} {Nucl. Phys. B}\ }\textbf {\bibinfo {volume} {437}},\
  \bibinfo {pages} {127} (\bibinfo {year} {1995})},\ \Eprint
  {https://arxiv.org/abs/hep-ph/9409446} {arXiv:hep-ph/9409446} \BibitemShut
  {NoStop}%
\bibitem [{\citenamefont {Catani}(1998)}]{Catani:1998bh}%
  \BibitemOpen
  \bibfield  {author} {\bibinfo {author} {\bibfnamefont {S.}~\bibnamefont
  {Catani}},\ }\bibfield  {title} {\bibinfo {title} {{The singular behaviour of
  QCD amplitudes at two-loop order}},\ }\href
  {https://doi.org/10.1016/S0370-2693(98)00332-3} {\bibfield  {journal}
  {\bibinfo  {journal} {Phys. Lett. B}\ }\textbf {\bibinfo {volume} {427}},\
  \bibinfo {pages} {161} (\bibinfo {year} {1998})},\ \Eprint
  {https://arxiv.org/abs/hep-ph/9802439} {arXiv:hep-ph/9802439} \BibitemShut
  {NoStop}%
\bibitem [{\citenamefont {Bhattacharya}\ \emph {et~al.}(2020)\citenamefont
  {Bhattacharya}, \citenamefont {Mahakhud}, \citenamefont {Mathews},\ and\
  \citenamefont {Ravindran}}]{Bhattacharya:2019oun}%
  \BibitemOpen
  \bibfield  {author} {\bibinfo {author} {\bibfnamefont {A.}~\bibnamefont
  {Bhattacharya}}, \bibinfo {author} {\bibfnamefont {M.}~\bibnamefont
  {Mahakhud}}, \bibinfo {author} {\bibfnamefont {P.}~\bibnamefont {Mathews}},\
  and\ \bibinfo {author} {\bibfnamefont {V.}~\bibnamefont {Ravindran}},\
  }\bibfield  {title} {\bibinfo {title} {{Two loop QCD amplitudes for di-pseudo
  scalar production in gluon fusion}},\ }\href
  {https://doi.org/10.1007/JHEP02(2020)121} {\bibfield  {journal} {\bibinfo
  {journal} {JHEP}\ }\textbf {\bibinfo {volume} {02}}\bibfield  {number}
  {\bibinfo  {number} { (2020)},\ \bibinfo {pages} {121}},\ }\Eprint
  {https://arxiv.org/abs/1909.08993} {arXiv:1909.08993 [hep-ph]} \BibitemShut
  {NoStop}%
\bibitem [{\citenamefont {Bern}\ \emph {et~al.}(1999)\citenamefont {Bern},
  \citenamefont {Del~Duca}, \citenamefont {Kilgore},\ and\ \citenamefont
  {Schmidt}}]{Bern:1999ry}%
  \BibitemOpen
  \bibfield  {author} {\bibinfo {author} {\bibfnamefont {Z.}~\bibnamefont
  {Bern}}, \bibinfo {author} {\bibfnamefont {V.}~\bibnamefont {Del~Duca}},
  \bibinfo {author} {\bibfnamefont {W.~B.}\ \bibnamefont {Kilgore}},\ and\
  \bibinfo {author} {\bibfnamefont {C.~R.}\ \bibnamefont {Schmidt}},\
  }\bibfield  {title} {\bibinfo {title} {{The infrared behavior of one-loop QCD
  amplitudes at next-to-next-to-leading order}},\ }\href
  {https://doi.org/10.1103/PhysRevD.60.116001} {\bibfield  {journal} {\bibinfo
  {journal} {Phys. Rev. D}\ }\textbf {\bibinfo {volume} {60}},\ \bibinfo
  {pages} {116001} (\bibinfo {year} {1999})},\ \Eprint
  {https://arxiv.org/abs/hep-ph/9903516} {arXiv:hep-ph/9903516} \BibitemShut
  {NoStop}%
\bibitem [{\citenamefont {Dixon}\ and\ \citenamefont
  {Esterlis}(2016)}]{Dixon:2016epj}%
  \BibitemOpen
  \bibfield  {author} {\bibinfo {author} {\bibfnamefont {L.~J.}\ \bibnamefont
  {Dixon}}\ and\ \bibinfo {author} {\bibfnamefont {I.}~\bibnamefont
  {Esterlis}},\ }\bibfield  {title} {\bibinfo {title} {{All orders results for
  self-crossing Wilson loops mimicking double parton scattering}},\ }\href
  {https://doi.org/10.1007/JHEP07(2016)116} {\bibfield  {journal} {\bibinfo
  {journal} {JHEP}\ }\textbf {\bibinfo {volume} {07}}\bibfield  {number}
  {\bibinfo  {number} { (2016)},\ \bibinfo {pages} {116}},\ }\bibinfo {note}
  {[Erratum ibid. 08 (2016) 131]},\ \Eprint {https://arxiv.org/abs/1602.02107}
  {arXiv:1602.02107 [hep-th]} \BibitemShut {NoStop}%
\bibitem [{\citenamefont {Caron-Huot}\ \emph {et~al.}(2019)\citenamefont
  {Caron-Huot}, \citenamefont {Dixon}, \citenamefont {Dulat}, \citenamefont
  {von Hippel}, \citenamefont {McLeod},\ and\ \citenamefont
  {Papathanasiou}}]{CaronHuot:2019vjl}%
  \BibitemOpen
  \bibfield  {author} {\bibinfo {author} {\bibfnamefont {S.}~\bibnamefont
  {Caron-Huot}}, \bibinfo {author} {\bibfnamefont {L.~J.}\ \bibnamefont
  {Dixon}}, \bibinfo {author} {\bibfnamefont {F.}~\bibnamefont {Dulat}},
  \bibinfo {author} {\bibfnamefont {M.}~\bibnamefont {von Hippel}}, \bibinfo
  {author} {\bibfnamefont {A.~J.}\ \bibnamefont {McLeod}},\ and\ \bibinfo
  {author} {\bibfnamefont {G.}~\bibnamefont {Papathanasiou}},\ }\bibfield
  {title} {\bibinfo {title} {{Six-Gluon Amplitudes in Planar ${\cal N}=4$
  Super-Yang-Mills Theory at Six and Seven Loops}},\ }\href
  {https://doi.org/10.1007/JHEP08(2019)016} {\bibfield  {journal} {\bibinfo
  {journal} {JHEP}\ }\textbf {\bibinfo {volume} {08}}\bibfield  {number}
  {\bibinfo  {number} { (2019)},\ \bibinfo {pages} {016}},\ }\Eprint
  {https://arxiv.org/abs/1903.10890} {arXiv:1903.10890 [hep-th]} \BibitemShut
  {NoStop}%
\bibitem [{\citenamefont {Beisert}\ \emph {et~al.}(2007)\citenamefont
  {Beisert}, \citenamefont {Eden},\ and\ \citenamefont
  {Staudacher}}]{Beisert:2006ez}%
  \BibitemOpen
  \bibfield  {author} {\bibinfo {author} {\bibfnamefont {N.}~\bibnamefont
  {Beisert}}, \bibinfo {author} {\bibfnamefont {B.}~\bibnamefont {Eden}},\ and\
  \bibinfo {author} {\bibfnamefont {M.}~\bibnamefont {Staudacher}},\ }\bibfield
   {title} {\bibinfo {title} {{Transcendentality and Crossing}},\ }\href
  {https://doi.org/10.1088/1742-5468/2007/01/P01021} {\bibfield  {journal}
  {\bibinfo  {journal} {J. Stat. Mech.}\ }\textbf {\bibinfo {volume} {0701}},\
  \bibinfo {pages} {P01021} (\bibinfo {year} {2007})},\ \Eprint
  {https://arxiv.org/abs/hep-th/0610251} {arXiv:hep-th/0610251} \BibitemShut
  {NoStop}%
\bibitem [{\citenamefont {Basso}\ \emph {et~al.}(2015)\citenamefont {Basso},
  \citenamefont {Caron-Huot},\ and\ \citenamefont {Sever}}]{Basso:2014bfkl}%
  \BibitemOpen
  \bibfield  {author} {\bibinfo {author} {\bibfnamefont {B.}~\bibnamefont
  {Basso}}, \bibinfo {author} {\bibfnamefont {S.}~\bibnamefont {Caron-Huot}},\
  and\ \bibinfo {author} {\bibfnamefont {A.}~\bibnamefont {Sever}},\ }\bibfield
   {title} {\bibinfo {title} {{Adjoint BFKL at finite coupling: a short-cut
  from the collinear limit}},\ }\href {https://doi.org/10.1007/JHEP01(2015)027}
  {\bibfield  {journal} {\bibinfo  {journal} {JHEP}\ }\textbf {\bibinfo
  {volume} {01}}\bibfield  {number} {\bibinfo  {number} { (2015)},\ \bibinfo
  {pages} {027}},\ }\Eprint {https://arxiv.org/abs/1407.3766} {arXiv:1407.3766
  [hep-th]} \BibitemShut {NoStop}%
\bibitem [{\citenamefont {Ahmed}\ \emph {et~al.}(2015)\citenamefont {Ahmed},
  \citenamefont {Gehrmann}, \citenamefont {Mathews}, \citenamefont {Rana},\
  and\ \citenamefont {Ravindran}}]{Ahmed:2015qpa}%
  \BibitemOpen
  \bibfield  {author} {\bibinfo {author} {\bibfnamefont {T.}~\bibnamefont
  {Ahmed}}, \bibinfo {author} {\bibfnamefont {T.}~\bibnamefont {Gehrmann}},
  \bibinfo {author} {\bibfnamefont {P.}~\bibnamefont {Mathews}}, \bibinfo
  {author} {\bibfnamefont {N.}~\bibnamefont {Rana}},\ and\ \bibinfo {author}
  {\bibfnamefont {V.}~\bibnamefont {Ravindran}},\ }\bibfield  {title} {\bibinfo
  {title} {{Pseudo-scalar form factors at three loops in QCD}},\ }\href
  {https://doi.org/10.1007/JHEP11(2015)169} {\bibfield  {journal} {\bibinfo
  {journal} {JHEP}\ }\textbf {\bibinfo {volume} {11}}\bibfield  {number}
  {\bibinfo  {number} { (2015)},\ \bibinfo {pages} {169}},\ }\Eprint
  {https://arxiv.org/abs/1510.01715} {arXiv:1510.01715 [hep-ph]} \BibitemShut
  {NoStop}%
\bibitem [{\citenamefont {Bauer}\ \emph
  {et~al.}(2002{\natexlab{a}})\citenamefont {Bauer}, \citenamefont {Pirjol},\
  and\ \citenamefont {Stewart}}]{Bauer:2001yt}%
  \BibitemOpen
  \bibfield  {author} {\bibinfo {author} {\bibfnamefont {C.~W.}\ \bibnamefont
  {Bauer}}, \bibinfo {author} {\bibfnamefont {D.}~\bibnamefont {Pirjol}},\ and\
  \bibinfo {author} {\bibfnamefont {I.~W.}\ \bibnamefont {Stewart}},\
  }\bibfield  {title} {\bibinfo {title} {{Soft collinear factorization in
  effective field theory}},\ }\href
  {https://doi.org/10.1103/PhysRevD.65.054022} {\bibfield  {journal} {\bibinfo
  {journal} {Phys. Rev. D}\ }\textbf {\bibinfo {volume} {65}},\ \bibinfo
  {pages} {054022} (\bibinfo {year} {2002}{\natexlab{a}})},\ \Eprint
  {https://arxiv.org/abs/hep-ph/0109045} {arXiv:hep-ph/0109045} \BibitemShut
  {NoStop}%
\bibitem [{\citenamefont {Bauer}\ \emph
  {et~al.}(2002{\natexlab{b}})\citenamefont {Bauer}, \citenamefont {Fleming},
  \citenamefont {Pirjol}, \citenamefont {Rothstein},\ and\ \citenamefont
  {Stewart}}]{Bauer:2002nz}%
  \BibitemOpen
  \bibfield  {author} {\bibinfo {author} {\bibfnamefont {C.~W.}\ \bibnamefont
  {Bauer}}, \bibinfo {author} {\bibfnamefont {S.}~\bibnamefont {Fleming}},
  \bibinfo {author} {\bibfnamefont {D.}~\bibnamefont {Pirjol}}, \bibinfo
  {author} {\bibfnamefont {I.~Z.}\ \bibnamefont {Rothstein}},\ and\ \bibinfo
  {author} {\bibfnamefont {I.~W.}\ \bibnamefont {Stewart}},\ }\bibfield
  {title} {\bibinfo {title} {{Hard scattering factorization from effective
  field theory}},\ }\href {https://doi.org/10.1103/PhysRevD.66.014017}
  {\bibfield  {journal} {\bibinfo  {journal} {Phys. Rev. D}\ }\textbf {\bibinfo
  {volume} {66}},\ \bibinfo {pages} {014017} (\bibinfo {year}
  {2002}{\natexlab{b}})},\ \Eprint {https://arxiv.org/abs/hep-ph/0202088}
  {arXiv:hep-ph/0202088} \BibitemShut {NoStop}%
\bibitem [{\citenamefont {Becher}\ and\ \citenamefont
  {Neubert}(2009{\natexlab{b}})}]{Becher:2009qa}%
  \BibitemOpen
  \bibfield  {author} {\bibinfo {author} {\bibfnamefont {T.}~\bibnamefont
  {Becher}}\ and\ \bibinfo {author} {\bibfnamefont {M.}~\bibnamefont
  {Neubert}},\ }\bibfield  {title} {\bibinfo {title} {{On the structure of
  infrared singularities of gauge-theory amplitudes}},\ }\href
  {https://doi.org/10.1088/1126-6708/2009/06/081} {\bibfield  {journal}
  {\bibinfo  {journal} {JHEP}\ }\textbf {\bibinfo {volume} {06}}\bibfield
  {number} {\bibinfo  {number} { (2009)},\ \bibinfo {pages} {081}},\ }\bibinfo
  {note} {[Erratum ibid. 11 (2013) 024]},\ \Eprint
  {https://arxiv.org/abs/0903.1126} {arXiv:0903.1126 [hep-ph]} \BibitemShut
  {NoStop}%
\bibitem [{\citenamefont {Bern}\ \emph {et~al.}(1994)\citenamefont {Bern},
  \citenamefont {Dixon}, \citenamefont {Dunbar},\ and\ \citenamefont
  {Kosower}}]{Bern:1994zx}%
  \BibitemOpen
  \bibfield  {author} {\bibinfo {author} {\bibfnamefont {Z.}~\bibnamefont
  {Bern}}, \bibinfo {author} {\bibfnamefont {L.~J.}\ \bibnamefont {Dixon}},
  \bibinfo {author} {\bibfnamefont {D.~C.}\ \bibnamefont {Dunbar}},\ and\
  \bibinfo {author} {\bibfnamefont {D.~A.}\ \bibnamefont {Kosower}},\
  }\bibfield  {title} {\bibinfo {title} {{One loop n point gauge theory
  amplitudes, unitarity and collinear limits}},\ }\href
  {https://doi.org/10.1016/0550-3213(94)90179-1} {\bibfield  {journal}
  {\bibinfo  {journal} {Nucl. Phys. B}\ }\textbf {\bibinfo {volume} {425}},\
  \bibinfo {pages} {217} (\bibinfo {year} {1994})},\ \Eprint
  {https://arxiv.org/abs/hep-ph/9403226} {arXiv:hep-ph/9403226} \BibitemShut
  {NoStop}%
\bibitem [{\citenamefont {Bern}\ \emph {et~al.}(2005)\citenamefont {Bern},
  \citenamefont {Dixon},\ and\ \citenamefont {Smirnov}}]{Bern:2005iz}%
  \BibitemOpen
  \bibfield  {author} {\bibinfo {author} {\bibfnamefont {Z.}~\bibnamefont
  {Bern}}, \bibinfo {author} {\bibfnamefont {L.~J.}\ \bibnamefont {Dixon}},\
  and\ \bibinfo {author} {\bibfnamefont {V.~A.}\ \bibnamefont {Smirnov}},\
  }\bibfield  {title} {\bibinfo {title} {{Iteration of planar amplitudes in
  maximally supersymmetric Yang-Mills theory at three loops and beyond}},\
  }\href {https://doi.org/10.1103/PhysRevD.72.085001} {\bibfield  {journal}
  {\bibinfo  {journal} {Phys. Rev. D}\ }\textbf {\bibinfo {volume} {72}},\
  \bibinfo {pages} {085001} (\bibinfo {year} {2005})},\ \Eprint
  {https://arxiv.org/abs/hep-th/0505205} {arXiv:hep-th/0505205} \BibitemShut
  {NoStop}%
\bibitem [{\citenamefont {Binoth}\ and\ \citenamefont
  {Heinrich}(2000)}]{Binoth:2000ps}%
  \BibitemOpen
  \bibfield  {author} {\bibinfo {author} {\bibfnamefont {T.}~\bibnamefont
  {Binoth}}\ and\ \bibinfo {author} {\bibfnamefont {G.}~\bibnamefont
  {Heinrich}},\ }\bibfield  {title} {\bibinfo {title} {{An automatized
  algorithm to compute infrared divergent multiloop integrals}},\ }\href
  {https://doi.org/10.1016/S0550-3213(00)00429-6} {\bibfield  {journal}
  {\bibinfo  {journal} {Nucl. Phys. B}\ }\textbf {\bibinfo {volume} {585}},\
  \bibinfo {pages} {741} (\bibinfo {year} {2000})},\ \Eprint
  {https://arxiv.org/abs/hep-ph/0004013} {arXiv:hep-ph/0004013} \BibitemShut
  {NoStop}%
\bibitem [{\citenamefont {Buffing}\ \emph {et~al.}(2018)\citenamefont
  {Buffing}, \citenamefont {Diehl},\ and\ \citenamefont
  {Kasemets}}]{Buffing:2017mqm}%
  \BibitemOpen
  \bibfield  {author} {\bibinfo {author} {\bibfnamefont {M.~G.~A.}\
  \bibnamefont {Buffing}}, \bibinfo {author} {\bibfnamefont {M.}~\bibnamefont
  {Diehl}},\ and\ \bibinfo {author} {\bibfnamefont {T.}~\bibnamefont
  {Kasemets}},\ }\bibfield  {title} {\bibinfo {title} {{Transverse momentum in
  double parton scattering: factorisation, evolution and matching}},\ }\href
  {https://doi.org/10.1007/JHEP01(2018)044} {\bibfield  {journal} {\bibinfo
  {journal} {JHEP}\ }\textbf {\bibinfo {volume} {01}}\bibfield  {number}
  {\bibinfo  {number} { (2018)},\ \bibinfo {pages} {044}},\ }\bibinfo {note}
  {[Erratum ibid. 07 (2021) 046]},\ \Eprint {https://arxiv.org/abs/1708.03528}
  {arXiv:1708.03528 [hep-ph]} \BibitemShut {NoStop}%
\bibitem [{\citenamefont {Caola}\ \emph
  {et~al.}(2015{\natexlab{b}})\citenamefont {Caola}, \citenamefont {Henn},
  \citenamefont {Melnikov}, \citenamefont {Smirnov},\ and\ \citenamefont
  {Smirnov}}]{Caola:2015ila}%
  \BibitemOpen
  \bibfield  {author} {\bibinfo {author} {\bibfnamefont {F.}~\bibnamefont
  {Caola}}, \bibinfo {author} {\bibfnamefont {J.~M.}\ \bibnamefont {Henn}},
  \bibinfo {author} {\bibfnamefont {K.}~\bibnamefont {Melnikov}}, \bibinfo
  {author} {\bibfnamefont {A.~V.}\ \bibnamefont {Smirnov}},\ and\ \bibinfo
  {author} {\bibfnamefont {V.~A.}\ \bibnamefont {Smirnov}},\ }\bibfield
  {title} {\bibinfo {title} {{Two-loop helicity amplitudes for the production
  of two off-shell electroweak bosons in gluon fusion}},\ }\href
  {https://doi.org/10.1007/JHEP06(2015)129} {\bibfield  {journal} {\bibinfo
  {journal} {JHEP}\ }\textbf {\bibinfo {volume} {06}}\bibfield  {number}
  {\bibinfo  {number} { (2015)},\ \bibinfo {pages} {129}},\ }\Eprint
  {https://arxiv.org/abs/1503.08759} {arXiv:1503.08759 [hep-ph]} \BibitemShut
  {NoStop}%
\bibitem [{\citenamefont {Catani}\ \emph {et~al.}(2012)\citenamefont {Catani},
  \citenamefont {de~Florian},\ and\ \citenamefont
  {Rodrigo}}]{Catani:2011spacelike}%
  \BibitemOpen
  \bibfield  {author} {\bibinfo {author} {\bibfnamefont {S.}~\bibnamefont
  {Catani}}, \bibinfo {author} {\bibfnamefont {D.}~\bibnamefont {de~Florian}},\
  and\ \bibinfo {author} {\bibfnamefont {G.}~\bibnamefont {Rodrigo}},\
  }\bibfield  {title} {\bibinfo {title} {{Space-like (versus time-like)
  collinear limits in QCD: Is factorization violated?}},\ }\href
  {https://doi.org/10.1007/JHEP07(2012)026} {\bibfield  {journal} {\bibinfo
  {journal} {JHEP}\ }\textbf {\bibinfo {volume} {07}}\bibfield  {number}
  {\bibinfo  {number} { (2012)},\ \bibinfo {pages} {026}},\ }\Eprint
  {https://arxiv.org/abs/1112.4405} {arXiv:1112.4405 [hep-ph]} \BibitemShut
  {NoStop}%
\bibitem [{\citenamefont {Catani}\ \emph {et~al.}(2014)\citenamefont {Catani},
  \citenamefont {Cieri}, \citenamefont {de~Florian}, \citenamefont {Ferrera},\
  and\ \citenamefont {Grazzini}}]{Catani:2013tia}%
  \BibitemOpen
  \bibfield  {author} {\bibinfo {author} {\bibfnamefont {S.}~\bibnamefont
  {Catani}}, \bibinfo {author} {\bibfnamefont {L.}~\bibnamefont {Cieri}},
  \bibinfo {author} {\bibfnamefont {D.}~\bibnamefont {de~Florian}}, \bibinfo
  {author} {\bibfnamefont {G.}~\bibnamefont {Ferrera}},\ and\ \bibinfo {author}
  {\bibfnamefont {M.}~\bibnamefont {Grazzini}},\ }\bibfield  {title} {\bibinfo
  {title} {{Universality of transverse-momentum resummation and hard factors at
  the NNLO}},\ }\href {https://doi.org/10.1016/j.nuclphysb.2014.02.011}
  {\bibfield  {journal} {\bibinfo  {journal} {Nucl. Phys. B}\ }\textbf
  {\bibinfo {volume} {881}},\ \bibinfo {pages} {414} (\bibinfo {year}
  {2014})},\ \Eprint {https://arxiv.org/abs/1311.1654} {arXiv:1311.1654
  [hep-ph]} \BibitemShut {NoStop}%
\bibitem [{\citenamefont {Coleman}\ and\ \citenamefont
  {Norton}(1965)}]{Coleman:1965xm}%
  \BibitemOpen
  \bibfield  {author} {\bibinfo {author} {\bibfnamefont {S.}~\bibnamefont
  {Coleman}}\ and\ \bibinfo {author} {\bibfnamefont {R.~E.}\ \bibnamefont
  {Norton}},\ }\bibfield  {title} {\bibinfo {title} {{Singularities in the
  physical region}},\ }\href {https://doi.org/10.1007/BF02750472} {\bibfield
  {journal} {\bibinfo  {journal} {Nuovo Cim.}\ }\textbf {\bibinfo {volume}
  {38}},\ \bibinfo {pages} {438} (\bibinfo {year} {1965})}\BibitemShut
  {NoStop}%
\bibitem [{\citenamefont {Dixon}(1996)}]{Dixon:1996wi}%
  \BibitemOpen
  \bibfield  {author} {\bibinfo {author} {\bibfnamefont {L.~J.}\ \bibnamefont
  {Dixon}},\ }\bibfield  {title} {\bibinfo {title} {{Calculating scattering
  amplitudes efficiently}},\ }in\ \href@noop {} {\emph {\bibinfo {booktitle}
  {{QCD and beyond: Proceedings of TASI 95}}}}\ (\bibinfo {year} {1996})\ pp.\
  \bibinfo {pages} {539--584},\ \Eprint {https://arxiv.org/abs/hep-ph/9601359}
  {arXiv:hep-ph/9601359} \BibitemShut {NoStop}%
\bibitem [{\citenamefont {Dixon}\ \emph {et~al.}(2012)\citenamefont {Dixon},
  \citenamefont {Duhr},\ and\ \citenamefont {Pennington}}]{Dixon:2012yy}%
  \BibitemOpen
  \bibfield  {author} {\bibinfo {author} {\bibfnamefont {L.~J.}\ \bibnamefont
  {Dixon}}, \bibinfo {author} {\bibfnamefont {C.}~\bibnamefont {Duhr}},\ and\
  \bibinfo {author} {\bibfnamefont {J.}~\bibnamefont {Pennington}},\ }\bibfield
   {title} {\bibinfo {title} {{Single-valued harmonic polylogarithms and the
  multi-Regge limit}},\ }\href {https://doi.org/10.1007/JHEP10(2012)074}
  {\bibfield  {journal} {\bibinfo  {journal} {JHEP}\ }\textbf {\bibinfo
  {volume} {10}}\bibfield  {number} {\bibinfo  {number} { (2012)},\ \bibinfo
  {pages} {074}},\ }\Eprint {https://arxiv.org/abs/1207.0186} {arXiv:1207.0186
  [hep-th]} \BibitemShut {NoStop}%
\bibitem [{\citenamefont {Echevarria}\ \emph {et~al.}(2012)\citenamefont
  {Echevarria}, \citenamefont {Idilbi},\ and\ \citenamefont
  {Scimemi}}]{Echevarria:2011epo}%
  \BibitemOpen
  \bibfield  {author} {\bibinfo {author} {\bibfnamefont {M.~G.}\ \bibnamefont
  {Echevarria}}, \bibinfo {author} {\bibfnamefont {A.}~\bibnamefont {Idilbi}},\
  and\ \bibinfo {author} {\bibfnamefont {I.}~\bibnamefont {Scimemi}},\
  }\bibfield  {title} {\bibinfo {title} {{Factorization Theorem For Drell-Yan
  At Low $q_T$ And Transverse Momentum Distributions On-The-Light-Cone}},\
  }\href {https://doi.org/10.1007/JHEP07(2012)002} {\bibfield  {journal}
  {\bibinfo  {journal} {JHEP}\ }\textbf {\bibinfo {volume} {07}}\bibfield
  {number} {\bibinfo  {number} { (2012)},\ \bibinfo {pages} {002}},\ }\Eprint
  {https://arxiv.org/abs/1111.4996} {arXiv:1111.4996 [hep-ph]} \BibitemShut
  {NoStop}%
\bibitem [{\citenamefont {Eden}\ \emph {et~al.}(1966)\citenamefont {Eden},
  \citenamefont {Landshoff}, \citenamefont {Olive},\ and\ \citenamefont
  {Polkinghorne}}]{Eden:1966dnq}%
  \BibitemOpen
  \bibfield  {author} {\bibinfo {author} {\bibfnamefont {R.~J.}\ \bibnamefont
  {Eden}}, \bibinfo {author} {\bibfnamefont {P.~V.}\ \bibnamefont {Landshoff}},
  \bibinfo {author} {\bibfnamefont {D.~I.}\ \bibnamefont {Olive}},\ and\
  \bibinfo {author} {\bibfnamefont {J.~C.}\ \bibnamefont {Polkinghorne}},\
  }\href@noop {} {\emph {\bibinfo {title} {{The analytic S-matrix}}}}\
  (\bibinfo  {publisher} {Cambridge University Press},\ \bibinfo {address}
  {Cambridge, U.K.},\ \bibinfo {year} {1966})\BibitemShut {NoStop}%
\bibitem [{\citenamefont {Forshaw}\ \emph {et~al.}(2012)\citenamefont
  {Forshaw}, \citenamefont {Seymour},\ and\ \citenamefont
  {Siodmok}}]{Forshaw:2012breaking}%
  \BibitemOpen
  \bibfield  {author} {\bibinfo {author} {\bibfnamefont {J.~R.}\ \bibnamefont
  {Forshaw}}, \bibinfo {author} {\bibfnamefont {M.~H.}\ \bibnamefont
  {Seymour}},\ and\ \bibinfo {author} {\bibfnamefont {A.}~\bibnamefont
  {Siodmok}},\ }\bibfield  {title} {\bibinfo {title} {{On the breaking of
  collinear factorization in QCD}},\ }\href
  {https://doi.org/10.1007/JHEP11(2012)066} {\bibfield  {journal} {\bibinfo
  {journal} {JHEP}\ }\textbf {\bibinfo {volume} {11}}\bibfield  {number}
  {\bibinfo  {number} { (2012)},\ \bibinfo {pages} {066}},\ }\Eprint
  {https://arxiv.org/abs/1206.6363} {arXiv:1206.6363 [hep-ph]} \BibitemShut
  {NoStop}%
\bibitem [{\citenamefont {Frellesvig}\ \emph {et~al.}(2016)\citenamefont
  {Frellesvig}, \citenamefont {Tommasini},\ and\ \citenamefont
  {Wever}}]{Frellesvig:2016ske}%
  \BibitemOpen
  \bibfield  {author} {\bibinfo {author} {\bibfnamefont {H.}~\bibnamefont
  {Frellesvig}}, \bibinfo {author} {\bibfnamefont {D.}~\bibnamefont
  {Tommasini}},\ and\ \bibinfo {author} {\bibfnamefont {C.}~\bibnamefont
  {Wever}},\ }\bibfield  {title} {\bibinfo {title} {{On the reduction of
  generalized polylogarithms to $\mathrm{Li}_n$ and $\mathrm{Li}_{2,2}$ and on
  the evaluation thereof}},\ }\href {https://doi.org/10.1007/JHEP03(2016)189}
  {\bibfield  {journal} {\bibinfo  {journal} {JHEP}\ }\textbf {\bibinfo
  {volume} {03}}\bibfield  {number} {\bibinfo  {number} { (2016)},\ \bibinfo
  {pages} {189}},\ }\Eprint {https://arxiv.org/abs/1601.02649}
  {arXiv:1601.02649 [hep-ph]} \BibitemShut {NoStop}%
\bibitem [{\citenamefont {Gardi}\ and\ \citenamefont
  {Magnea}(2009)}]{Gardi:2009qi}%
  \BibitemOpen
  \bibfield  {author} {\bibinfo {author} {\bibfnamefont {E.}~\bibnamefont
  {Gardi}}\ and\ \bibinfo {author} {\bibfnamefont {L.}~\bibnamefont {Magnea}},\
  }\bibfield  {title} {\bibinfo {title} {{Factorization constraints for soft
  anomalous dimensions in QCD scattering amplitudes}},\ }\href
  {https://doi.org/10.1088/1126-6708/2009/03/079} {\bibfield  {journal}
  {\bibinfo  {journal} {JHEP}\ }\textbf {\bibinfo {volume} {03}}\bibfield
  {number} {\bibinfo  {number} { (2009)},\ \bibinfo {pages} {079}},\ }\Eprint
  {https://arxiv.org/abs/0901.1091} {arXiv:0901.1091 [hep-ph]} \BibitemShut
  {NoStop}%
\bibitem [{\citenamefont {Gardi}\ \emph {et~al.}(2024)\citenamefont {Gardi},
  \citenamefont {Herzog}, \citenamefont {Jones},\ and\ \citenamefont
  {Ma}}]{Gardi:2024dissecting}%
  \BibitemOpen
  \bibfield  {author} {\bibinfo {author} {\bibfnamefont {E.}~\bibnamefont
  {Gardi}}, \bibinfo {author} {\bibfnamefont {F.}~\bibnamefont {Herzog}},
  \bibinfo {author} {\bibfnamefont {S.}~\bibnamefont {Jones}},\ and\ \bibinfo
  {author} {\bibfnamefont {Y.}~\bibnamefont {Ma}},\ }\bibfield  {title}
  {\bibinfo {title} {{Dissecting polytopes: Landau singularities and asymptotic
  expansions in $2\to2$ scattering}},\ }\href
  {https://doi.org/10.1007/JHEP08(2024)127} {\bibfield  {journal} {\bibinfo
  {journal} {JHEP}\ }\textbf {\bibinfo {volume} {08}}\bibfield  {number}
  {\bibinfo  {number} { (2024)},\ \bibinfo {pages} {127}},\ }\Eprint
  {https://arxiv.org/abs/2407.13738} {arXiv:2407.13738 [hep-th]} \BibitemShut
  {NoStop}%
\bibitem [{\citenamefont {Gehrmann}\ \emph {et~al.}(2015)\citenamefont
  {Gehrmann}, \citenamefont {von Manteuffel},\ and\ \citenamefont
  {Tancredi}}]{Gehrmann:2015ora}%
  \BibitemOpen
  \bibfield  {author} {\bibinfo {author} {\bibfnamefont {T.}~\bibnamefont
  {Gehrmann}}, \bibinfo {author} {\bibfnamefont {A.}~\bibnamefont {von
  Manteuffel}},\ and\ \bibinfo {author} {\bibfnamefont {L.}~\bibnamefont
  {Tancredi}},\ }\bibfield  {title} {\bibinfo {title} {{The two-loop helicity
  amplitudes for $q\overline{q}'\to V_1V_2\to4$ leptons}},\ }\href
  {https://doi.org/10.1007/JHEP09(2015)128} {\bibfield  {journal} {\bibinfo
  {journal} {JHEP}\ }\textbf {\bibinfo {volume} {09}}\bibfield  {number}
  {\bibinfo  {number} { (2015)},\ \bibinfo {pages} {128}},\ }\Eprint
  {https://arxiv.org/abs/1503.04812} {arXiv:1503.04812 [hep-ph]} \BibitemShut
  {NoStop}%
\bibitem [{\citenamefont {Henn}\ \emph {et~al.}(2014)\citenamefont {Henn},
  \citenamefont {Melnikov},\ and\ \citenamefont {Smirnov}}]{Henn:2014lfa}%
  \BibitemOpen
  \bibfield  {author} {\bibinfo {author} {\bibfnamefont {J.~M.}\ \bibnamefont
  {Henn}}, \bibinfo {author} {\bibfnamefont {K.}~\bibnamefont {Melnikov}},\
  and\ \bibinfo {author} {\bibfnamefont {V.~A.}\ \bibnamefont {Smirnov}},\
  }\bibfield  {title} {\bibinfo {title} {{Two-loop planar master integrals for
  the production of off-shell vector bosons in hadron collisions}},\ }\href
  {https://doi.org/10.1007/JHEP05(2014)090} {\bibfield  {journal} {\bibinfo
  {journal} {JHEP}\ }\textbf {\bibinfo {volume} {05}}\bibfield  {number}
  {\bibinfo  {number} { (2014)},\ \bibinfo {pages} {090}},\ }\Eprint
  {https://arxiv.org/abs/1402.7078} {arXiv:1402.7078 [hep-ph]} \BibitemShut
  {NoStop}%
\bibitem [{\citenamefont {Kosower}\ and\ \citenamefont
  {Uwer}(1999)}]{Kosower:1999xi}%
  \BibitemOpen
  \bibfield  {author} {\bibinfo {author} {\bibfnamefont {D.~A.}\ \bibnamefont
  {Kosower}}\ and\ \bibinfo {author} {\bibfnamefont {P.}~\bibnamefont {Uwer}},\
  }\bibfield  {title} {\bibinfo {title} {{One-loop splitting amplitudes in
  gauge theory}},\ }\href {https://doi.org/10.1016/S0550-3213(99)00583-0}
  {\bibfield  {journal} {\bibinfo  {journal} {Nucl. Phys. B}\ }\textbf
  {\bibinfo {volume} {563}},\ \bibinfo {pages} {477} (\bibinfo {year}
  {1999})},\ \Eprint {https://arxiv.org/abs/hep-ph/9903515}
  {arXiv:hep-ph/9903515} \BibitemShut {NoStop}%
\bibitem [{\citenamefont {Landau}(1959)}]{Landau:1959fi}%
  \BibitemOpen
  \bibfield  {author} {\bibinfo {author} {\bibfnamefont {L.~D.}\ \bibnamefont
  {Landau}},\ }\bibfield  {title} {\bibinfo {title} {{On analytic properties of
  vertex parts in quantum field theory}},\ }\href
  {https://doi.org/10.1016/0029-5582(59)90154-3} {\bibfield  {journal}
  {\bibinfo  {journal} {Nucl. Phys.}\ }\textbf {\bibinfo {volume} {13}},\
  \bibinfo {pages} {181} (\bibinfo {year} {1959})}\BibitemShut {NoStop}%
\bibitem [{\citenamefont {Mangano}\ and\ \citenamefont
  {Parke}(1991)}]{Mangano:1990by}%
  \BibitemOpen
  \bibfield  {author} {\bibinfo {author} {\bibfnamefont {M.~L.}\ \bibnamefont
  {Mangano}}\ and\ \bibinfo {author} {\bibfnamefont {S.~J.}\ \bibnamefont
  {Parke}},\ }\bibfield  {title} {\bibinfo {title} {{Multiparton amplitudes in
  gauge theories}},\ }\href {https://doi.org/10.1016/0370-1573(91)90091-Y}
  {\bibfield  {journal} {\bibinfo  {journal} {Phys. Rept.}\ }\textbf {\bibinfo
  {volume} {200}},\ \bibinfo {pages} {301} (\bibinfo {year} {1991})},\ \Eprint
  {https://arxiv.org/abs/hep-th/0509223} {arXiv:hep-th/0509223} \BibitemShut
  {NoStop}%
\bibitem [{\citenamefont {Passarino}\ and\ \citenamefont
  {Veltman}(1979)}]{Passarino:1978jh}%
  \BibitemOpen
  \bibfield  {author} {\bibinfo {author} {\bibfnamefont {G.}~\bibnamefont
  {Passarino}}\ and\ \bibinfo {author} {\bibfnamefont {M.~J.~G.}\ \bibnamefont
  {Veltman}},\ }\bibfield  {title} {\bibinfo {title} {{One-loop corrections for
  $e^+e^-$ annihilation into $\mu^+\mu^-$ in the Weinberg model}},\ }\href
  {https://doi.org/10.1016/0550-3213(79)90234-7} {\bibfield  {journal}
  {\bibinfo  {journal} {Nucl. Phys. B}\ }\textbf {\bibinfo {volume} {160}},\
  \bibinfo {pages} {151} (\bibinfo {year} {1979})}\BibitemShut {NoStop}%
\bibitem [{\citenamefont {Rothstein}\ and\ \citenamefont
  {Stewart}(2016)}]{Rothstein:2016glauber}%
  \BibitemOpen
  \bibfield  {author} {\bibinfo {author} {\bibfnamefont {I.~Z.}\ \bibnamefont
  {Rothstein}}\ and\ \bibinfo {author} {\bibfnamefont {I.~W.}\ \bibnamefont
  {Stewart}},\ }\bibfield  {title} {\bibinfo {title} {{An effective field
  theory for forward scattering and factorization violation}},\ }\href
  {https://doi.org/10.1007/JHEP08(2016)025} {\bibfield  {journal} {\bibinfo
  {journal} {JHEP}\ }\textbf {\bibinfo {volume} {08}}\bibfield  {number}
  {\bibinfo  {number} { (2016)},\ \bibinfo {pages} {025}},\ }\Eprint
  {https://arxiv.org/abs/1601.04695} {arXiv:1601.04695 [hep-ph]} \BibitemShut
  {NoStop}%
\bibitem [{\citenamefont {{The mpmath development team}}(2023)}]{mpmath}%
  \BibitemOpen
  \bibfield  {author} {\bibinfo {author} {\bibnamefont {{The mpmath development
  team}}},\ }\href@noop {} {\bibinfo {title} {{mpmath: a Python library for
  arbitrary-precision floating-point arithmetic (version 1.3.0)}}},\ \bibinfo
  {howpublished} {\url{https://mpmath.org}} (\bibinfo {year}
  {2023})\BibitemShut {NoStop}%
\bibitem [{\citenamefont {Beneke}\ and\ \citenamefont
  {Smirnov}(1998)}]{Beneke:1997zp}%
  \BibitemOpen
  \bibfield  {author} {\bibinfo {author} {\bibfnamefont {M.}~\bibnamefont
  {Beneke}}\ and\ \bibinfo {author} {\bibfnamefont {V.~A.}\ \bibnamefont
  {Smirnov}},\ }\bibfield  {title} {\bibinfo {title} {{Asymptotic expansion of
  Feynman integrals near threshold}},\ }\href
  {https://doi.org/10.1016/S0550-3213(98)00138-2} {\bibfield  {journal}
  {\bibinfo  {journal} {Nucl. Phys. B}\ }\textbf {\bibinfo {volume} {522}},\
  \bibinfo {pages} {321} (\bibinfo {year} {1998})},\ \Eprint
  {https://arxiv.org/abs/hep-ph/9711391} {arXiv:hep-ph/9711391} \BibitemShut
  {NoStop}%
\bibitem [{\citenamefont {Chiu}\ \emph {et~al.}(2012)\citenamefont {Chiu},
  \citenamefont {Jain}, \citenamefont {Neill},\ and\ \citenamefont
  {Rothstein}}]{Chiu:2012ir}%
  \BibitemOpen
  \bibfield  {author} {\bibinfo {author} {\bibfnamefont {J.-y.}\ \bibnamefont
  {Chiu}}, \bibinfo {author} {\bibfnamefont {A.}~\bibnamefont {Jain}}, \bibinfo
  {author} {\bibfnamefont {D.}~\bibnamefont {Neill}},\ and\ \bibinfo {author}
  {\bibfnamefont {I.~Z.}\ \bibnamefont {Rothstein}},\ }\bibfield  {title}
  {\bibinfo {title} {{A Formalism for the Systematic Treatment of Rapidity
  Logarithms in Quantum Field Theory}},\ }\href
  {https://doi.org/10.1007/JHEP05(2012)084} {\bibfield  {journal} {\bibinfo
  {journal} {JHEP}\ }\textbf {\bibinfo {volume} {05}}\bibfield  {number}
  {\bibinfo  {number} { (2012)},\ \bibinfo {pages} {084}},\ }\Eprint
  {https://arxiv.org/abs/1202.0814} {arXiv:1202.0814 [hep-ph]} \BibitemShut
  {NoStop}%
\end{thebibliography}%


%
